\RequirePackage{tikz}
\documentclass[sn-mathphys]{style/sn-article-template/sn-jnl}

\usepackage{amsmath}
\usepackage{amssymb}

\usepackage{amsthm}

\usepackage{silence}
\usepackage{hyperref}

\usepackage{soul}
\usepackage{graphicx}
\usepackage{enumitem}
\usepackage{tikz}
\usepackage[htt]{hyphenat}
\usepackage[most]{tcolorbox}
\usepackage{ltablex}
\usepackage{multirow}
\usepackage{lipsum}
\usepackage{changepage}
\usepackage{tabularx}
\usepackage{mathrsfs}
\usepackage{readarray}

\usepackage{thmtools}
\usepackage{thm-restate}

\usepackage{xcolor,colortbl,bm,graphicx,adjustbox}
\usepackage{afterpage}

\declaretheorem[name=Theorem, numberwithin=section]{thm}
\declaretheorem[name=Lemma, sibling=thm]{lem}
\declaretheorem[name=Proposition, sibling=thm]{prop}
\declaretheorem[name=Observation, sibling=thm]{observation}

\declaretheoremstyle[
notebraces={}{},
headpunct={},
headformat={\NAME\ \NUMBER.~\NOTE},
headfont=\bfseries, 
postheadhook = {\hspace*{0pt}},
spaceabove = 0.5cm, 
spacebelow = 0.5cm]{myproblemstyle}
\declaretheorem[style=myproblemstyle, sibling=thm, name=Problem]{myproblem}

\declaretheoremstyle[
notebraces={}{},
headpunct={},
headformat={\NAME\ \NUMBER.~\NOTE},
headfont=\bfseries, 
postheadhook = {\hspace*{0pt}},
spaceabove = 0.5cm, 
spacebelow = 0.5cm]{mydefinitionstyle}
\declaretheorem[style=mydefinitionstyle, sibling=thm, name=Definition]{mydefinition}

\newcommand{\inp}{\textit{Input: }} 

\newcommand{\ques}{\textit{Question: }}

\newcommand{\algorithmfont}[1]{\textsf{#1}}
\newcommand{\myoplus}{\oplus}

\newcommand{\NP}{\textsf{NP}}
\renewcommand{\P}{\textsf{P}}

\usepackage{eqparbox}

\algdef{S}[WHILE]{WhileNoDo}[1]{\algorithmicwhile\ #1}%
\algrenewcommand\algorithmicindent{1.0em}

\usepackage{silence}
\usepackage{linegoal}



\makeatletter
\algnewcommand{\LineComment}[1]{\State \(\triangleright\) #1
}
\makeatother

\usepackage{pifont}
\newcommand{\maynotexist}{\ding{55}}%

\usepackage{xpatch}
\makeatletter
\xpatchcmd{\algorithmic}
  {\ALG@tlm\z@}{\leftmargin\z@\ALG@tlm\z@}
  {}{}
\makeatother

\usetikzlibrary{arrows,decorations.markings,decorations.pathreplacing,patterns,matrix,calc,positioning,backgrounds,arrows.meta,shapes,decorations.markings,fadings,bending,decorations.pathmorphing}

\makeatletter
\tikzset{
  reset label anchor/.code={%
    \let\tikz@auto@anchor=\pgfutil@empty
    \def\tikz@anchor{#1}
  },
  reset label anchor/.default=center,
  every label/.append style={reset label anchor}
}
\makeatother

\makeatletter
\pgfarrowsdeclare{sss}{sss}{%
  \pgfutil@tempdima\pgflinewidth
  \pgfarrowsrightextend{+6\pgfutil@tempdima}%
  \pgfarrowsleftextend{+6\pgfutil@tempdima}%
}{
  \pgfutil@tempdima\pgflinewidth
  \pgfpathrectangle{\pgfqpoint{+4\pgfutil@tempdima}{+.5\pgfutil@tempdima}}{\pgfqpoint{-\pgfutil@tempdima}{-\pgfutil@tempdima}}
  \pgfpathrectangle{\pgfqpoint{+8\pgfutil@tempdima}{+.5\pgfutil@tempdima}}{\pgfqpoint{-\pgfutil@tempdima}{-\pgfutil@tempdima}}
  \pgfpathrectangle{\pgfqpoint{+12\pgfutil@tempdima}{+.5\pgfutil@tempdima}}{\pgfqpoint{-\pgfutil@tempdima}{-\pgfutil@tempdima}}
  \pgfusepathqfill
}
\makeatother

\tikzset{darrow/.style={decoration={
  markings,
  mark=at position .2 with {\arrowreversed{angle 90[width=2.5mm]}},
  mark=at position .8 with {\arrow{angle 90[width=2.5mm]}},
  }
  ,postaction={decorate}}}

\tikzset{darrow12/.style={decoration={
  markings,
  mark=at position {0.2*\pgfdecoratedpathlength} with {\arrowreversed{angle 90[width=2.5mm]}},
  mark=at position {0.8*\pgfdecoratedpathlength-0.4mm} with {\arrow{angle 90[width=2.5mm]}},
  mark=at position {0.8*\pgfdecoratedpathlength+0.4mm} with {\arrow{angle 90[width=2.5mm]}},
  }
  ,postaction={decorate}}}
  
\tikzset{darrow22/.style={decoration={
  markings,
  mark=at position {0.2*\pgfdecoratedpathlength-0.4mm} with {\arrowreversed{angle 90[width=2.5mm]}},
  mark=at position {0.2*\pgfdecoratedpathlength+0.4mm} with {\arrowreversed{angle 90[width=2.5mm]}},
  mark=at position {0.8*\pgfdecoratedpathlength-0.4mm} with {\arrow{angle 90[width=2.5mm]}},
  mark=at position {0.8*\pgfdecoratedpathlength+0.4mm} with {\arrow{angle 90[width=2.5mm]}},
  }
  ,postaction={decorate}}}

\tikzset{-->-/.style={decoration={
  markings,
  mark=at position .8 with {\arrow{angle 90[width=2.5mm]}}},postaction={decorate}}}
  
\tikzset{-<--/.style={decoration={
  markings,
  mark=at position .8 with {\arrow{angle 90[width=2.5mm]}}},postaction={decorate}}}
  
\tikzset{->-/.style={decoration={
  markings,
  mark=at position .5 with {\arrow{angle 90[width=2.5mm]}} },postaction={decorate}}}

\tikzset{--->/.style={decoration={
  markings,
  mark=at position 1 with {\arrow{angle 90[width=2.5mm]}} },postaction={decorate}}}

\tikzset{
  nat/.style     = {fill=white,draw=none,ellipse,minimum size=0.3cm,inner sep=1pt},
}

\usepackage[most]{tcolorbox}

\pgfdeclarelayer{background}
\pgfdeclarelayer{foreground}
\pgfsetlayers{background,main,foreground}

\makeatletter
\pgfkeys{%
  /tikz/on layer/.code={
    \def\tikz@path@do@at@end{\endpgfonlayer\endgroup\tikz@path@do@at@end}%
    \pgfonlayer{#1}\begingroup%
  }%
}
\makeatother

\tikzset{edge from parent/.append style={\solutionconceptsdiagramarrow}}

\newcommand{\arrowwithtext}[1] {%
    \begin{pgfonlayer}{foreground}
    \draw[line width=0.3mm] (0.0,0.0) -- (-0.125cm, 0.15cm);
    \draw[line width=0.3mm] (0.0, 0.0) -- (-0.125cm, -0.15cm);
    \node[draw,circle,white,fill,minimum size=0.45cm] (background) at (-0.5, 0.0) {};
    \node (anonymousnode) at (background) {#1};
    \end{pgfonlayer}
}  

\newcommand\solutionconceptsdiagramarrow{-Straight Barb[length=1.4mm]}

\newcommand{\myvaluationarrow}{%
    \draw (0.0,0.0) -- (-0.125cm, 0.15cm);
    \draw (0.0, 0.0) -- (-0.125cm, -0.15cm);
}

\newcommand{\myvaluationarrowreversed}{%
\begin{scope}[xscale=-1.0]
\myvaluationarrow
\end{scope}
}

\tikzset{darrow/.style={decoration={
  markings,
  mark=at position .2 with {\myvaluationarrowreversed},
  mark=at position .8 with {\myvaluationarrow},
  }
  ,postaction={decorate}}}

\newcommand{\convexpath}[2]{
[   
    create hullnodes/.code={
        \global\edef\namelist{#1}
        \foreach [count=\counter] \nodename in \namelist {
            \global\edef\numberofnodes{\counter}
            \node at (\nodename) [draw=none,name=hullnode\counter] {};
        }
        \node at (hullnode\numberofnodes) [name=hullnode0,draw=none] {};
        \pgfmathtruncatemacro\lastnumber{\numberofnodes+1}
        \node at (hullnode1) [name=hullnode\lastnumber,draw=none] {};
    },
    create hullnodes
]
($(hullnode1)!#2!-90:(hullnode0)$)
\foreach [
    evaluate=\currentnode as \previousnode using \currentnode-1,
    evaluate=\currentnode as \nextnode using \currentnode+1
    ] \currentnode in {1,...,\numberofnodes} {
  let
    \p1 = ($(hullnode\currentnode)!#2!-90:(hullnode\previousnode)$),
    \p2 = ($(hullnode\currentnode)!#2!90:(hullnode\nextnode)$),
    \p3 = ($(\p1) - (hullnode\currentnode)$),
    \n1 = {atan2(\y3,\x3)},
    \p4 = ($(\p2) - (hullnode\currentnode)$),
    \n2 = {atan2(\y4,\x4)},
    \n{delta} = {-Mod(\n1-\n2,360)}
  in 
    {-- (\p1) arc[start angle=\n1, delta angle=\n{delta}, radius=#2] -- (\p2)}
}
-- cycle
}
\newcommand\dashedcontainerthing[1]{%
\readlist*\nodes{#1}%
\begin{scope}[on background layer]
    \ifthenelse{\nodeslen>1}{
        \draw[thick, densely dashed] \convexpath{#1}{0.353cm};
        \fill[white] \convexpath{#1}{9.6pt};
    }{
        \node[draw, thick, circle, densely dashed, minimum size=20pt] (anonymousnode) at (\nodes[1]) {};
    }
\end{scope}
}

\definecolor{enclosure_color}{rgb}{0.6,0.6.,0.6}
\definecolor{figurecolourschemewt1_adjusted}{rgb}{0.8,0.8,0.8}
\definecolor{figurecolourschemewt1}{rgb}{0.5,0.5,0.5}
\definecolor{figurecolourschemewt2}{rgb}{0.5,0.0,0.5}
\definecolor{figurecolourschemewt3}{rgb}{0,0,0}
\definecolor{figurecolourschemewt4}{rgb}{0,0,1}
\definecolor{figurecolourschemewt5}{rgb}{0,1,0}
\definecolor{figurecolourschemewt6}{rgb}{1,0,0}

\newcommand{\figurecolorschemewtonename}{grey}
\newcommand{\figurecolorschemewttwoname}{purple}
\newcommand{\figurecolorschemewtthreename}{black}
\newcommand{\figurecolorschemewtfourname}{blue}
\newcommand{\figurecolorschemewtfivename}{green}
\newcommand{\figurecolorschemewtsixname}{red}

\def\porschenxsatvariant/{$\text{X3SAT}_{+}^{\,=3}$}
\def\iwjnomaxdegreetwofamily/{$\mathcal{I}^\star$}

\usepackage[normalem]{ulem} 

\normalbaroutside

\jyear{2021}%

\theoremstyle{thmstyleone}%
\theoremstyle{thmstyletwo}%

\begin{document}

\title[Envy-freeness in 3D hedonic games]{Envy-freeness in 3D hedonic games}


\author*[1]{\fnm{Michael} \sur{McKay}}\email{mikemckay2203@gmail.com}

\author[2,3]{\fnm{\'Agnes} \sur{Cseh}}\email{agnes.cseh@uni-bayreuth.de}

\author[1]{\fnm{David} \sur{Manlove}}\email{david.manlove@glasgow.ac.uk}

\affil*[1]{\orgdiv{School of Computing Science}, \orgname{University of Glasgow}, \orgaddress{\city{Glasgow}, \postcode{G12 8QQ}, \country{United Kingdom}}}

\affil[2]{\orgdiv{Department of Mathematics}, \orgname{University of Bayreuth}, \orgaddress{\city{Bayreuth}, \postcode{95447}, \country{Germany}}}

\affil[3]{\orgdiv{Institute of Economics}, \orgname{HUN-REN Centre for Economic and Regional Studies}, \orgaddress{\city{Budapest}, \postcode{1097}, \country{Hungary}}}


\abstract{We study the problem of fairly partitioning a set of agents into coalitions based on the agents' additively separable preferences, which can also be viewed as a hedonic game. We study three successively weaker solution concepts, related to envy, weakly justified envy, and justified envy. 

In a model in which coalitions may have any size, trivial solutions exist for these concepts, which provides a strong motivation for placing restrictions on coalition size. In this paper, we require feasible coalitions to have size three. We study the existence of partitions that are envy-free, weakly justified envy-free, and justified envy-free, and the computational complexity of finding such partitions, if they exist. 

We impose various restrictions on the agents' preferences and present a complete complexity classification in terms of these restrictions.}




\maketitle

\section{Introduction}
\label{sec:threed_efr_as_intro}

\subsection{Background and motivation}
\label{sec:threed_efr_as_background}

In this paper we study a model that involves fairly partitioning a set of agents into disjoint coalitions of size three. In terms of the literature, this model could be described as either a \emph{hedonic game with fixed size coalitions}, of size three, or a model of \emph{Three Dimensional Roommates} (3DR). We assume that agents' preferences are additively separable, or equivalently that each agent assigns a numerical valuation to every other agent. This assumption means this model is equivalent to an \emph{Additively Separable Hedonic Game} (ASHG), which is a type of cooperative game~\cite{BKS01,Olsen2007}. The ``hedonic'' aspect means that ``every agent only cares about which agents are in its coalition, but does not care how agents in other coalitions are grouped together''~\cite{hgch15,DG80}. This is a natural assumption to make in the study of multi-agent systems~\cite{Elkind:2009:HCN:1558013.1558070}. 


In this model, we assume that agents' utility in some partition is the sum of its valuations of the other agents in its coalition, and agents always prefer coalitions with greater utility. In some partition, we say that some agent $\alpha_i$ has \emph{envy} for another agent $\alpha_j$ if $\alpha_i$ prefers to swap places with $\alpha_j$. If no such $\alpha_i$ exists then we say that the partition is \emph{envy-free}. Part of our contribution regarding this model is on the existence of, and complexity of finding, partitions that are envy-free.

We also consider two other types of envy, which are called \emph{weakly justified envy}, and \emph{justified envy}. Informally, if agent $\alpha_i$ has envy for agent $\alpha_j$ then it is \emph{justified} if the other agents in the coalition of $\alpha_j$ strictly prefer $\alpha_i$ to $\alpha_j$. It is \emph{weakly justified} if the other agents in the coalition of $\alpha_j$ either prefer $\alpha_i$ to $\alpha_j$ or are indifferent.

A strong motivation exists for studying notions of envy in a model in which coalitions have a fixed size. Primarily, in a model in which coalitions have arbitrary size it can be trivial to construct partitions that are envy-free, for example by assigning all the agents to a single coalition~\cite{BY19,Ued18}. In addition, most of the previous work involving coalitions of fixed size involves finding partitions that are \emph{stable}, meaning there exists no coalition $S$ in which each agent in $S$ prefers $S$ to their assigned coalition in the partition~\cite{Bre20,Hua07,IMO07,ManloveMcKay3DSRAS2021,NH91}.

More generally, much less is known about models in which coalitions have a fixed size compared to other models, such as hedonic games. It has also been argued in the literature that the assumption made in hedonic games that coalitions may have any size is unrealistic in practice~\cite{Sless18}. In a real-life application there are likely to be constraints on the number or size of coalitions. For example, in a setting in which people are assigned to teams for specific projects it could be that the size of each team is limited~\cite{yekta2018findingjournal}, or must be exactly some fixed size. Team chess tournaments can involve teams of a fixed size, which at some amateur events~\cite{chesswiththreeplayerteams1, chesswiththreeplayerteams2} is exactly three. Moreover, in the International Chess Federation rules for team chess~\cite{chessfidehandbook}, team chess matches are played between two individuals, points awarded to a team for each match won or drawn, and the tournament won by the team with the most points. We could imagine that players' preferences over teams are therefore additively separable. 


A strong theoretical motivation also exists for considering coalitions of size three in particular. Similar models involving coalitions of size two, like the \emph{Stable Roommates}~\cite{Man13} problem, are comparatively well studied and well understood. In many such models, a stable or envy-free partition is bound to exist, or the problem of deciding whether such a partition exists can be solved in polynomial time. For example, deciding whether an envy-free partition exists can be trivial because such a partition must assign each agent to one of its ``most preferred'' partners~\cite{Coutanceetal2023}. In other words, it must be a \emph{perfect partition}~\cite{ABS13}.

In comparison, models in which coalitions have size three (or more) do not seem so straightforward. For example, many of the computational problems associated with stability are $\NP$-hard even when coalitions must have size three~\cite{ABEOMP09,IMO07,ManloveMcKay3DSRAS2021,NH91}. It seems very likely that such hardness results can be generalised to larger coalition sizes. For this reason, we consider our contribution to be a first step towards exploring the frontier between polynomial-time solvability and $\NP$-hardness for fixed-size coalitions. 

The fact that many of the related computational problems associated with stability are $\NP$-hard~\cite{ABEOMP09,IMO07,ManloveMcKay3DSRAS2021,NH91} also motivates the study of justified envy-freeness, which is a strictly weaker concept than stability (see Section~\ref{sec:prelims}).

\subsection{Preliminaries}
\label{sec:prelims}
An ASHG comprises a set of agents $N$ with \emph{additively separable pref\-er\-ences} over coalitions, which we define as follows. Each agent $\alpha_i$ has an integer \textit{valuation} function $v_{\alpha_i}: N \mapsto \mathbb{Z}$. For convenience, for each agent $\alpha_i$ let $v_{\alpha_i}(\alpha_i) = 0$. An ASHG is a pair $(N, V)$ where $V$ is a collection of all agents' valuation functions. In our model we assume that $|N|=3n$ for some integer $n$ where $n \geq 1$. We also refer to coalitions of size three as \emph{triples}. For some partition into triples $\pi$ and any set of agents $S$, let $\sigma(S, \pi)$ be the number of triples in $\pi$ that each contains at least one agent in $S$.

For some agent $\alpha_i$ and some triple $S$, let the \emph{utility} of $\alpha_i$ in the triple $S$ be $u_{\alpha_i}(S) = \sum_{\alpha_j \in S} v_{\alpha_i}(\alpha_j)$. For some set of agents $R$, let $u_{R}(S)=\sum_{\alpha_j \in R} u_{\alpha_j}(S)$. For some partition into triples $\pi$ and some agent $\alpha_i$ let $\pi(\alpha_i)$ be the coalition in $\pi$ that contains $\alpha_i$. Since preferences are hedonic, given a partition $\pi$ let $u_{\alpha_i}(\pi)$ be short for $u_{\alpha_i}(\pi(\alpha_i))$.

\begin{figure}
    \centering
    \begin{tikzpicture}
    
\def\alabeldist{0.6cm}
\def\outsidebendangle{25}
\def\insidebendangle{-1.0*\outsidebendangle}
\def\valuationlinewidth{0.3mm}
\def\bentlabeldistance{0.25}

\begin{scope}[scale=1.2]
\begin{scope}[every node/.style={circle,draw, minimum size=2.4mm}, scale=1.6]
    \begin{scope}[xscale=0.4,yscale=0.8]
        \node[thick, circle, label={[label distance=\alabeldist+1.0cm]180:$\alpha_1$}] (a1) at (0,0) {};
        \node[thick, circle, label={[label distance=\alabeldist]180:$\alpha_2$}] (a2) at (-1,1) {};
        \node[thick, circle, label={[label distance=\alabeldist]180:$\alpha_3$}] (a3) at (-1,-1) {};
    \end{scope}
    
    \begin{scope}[shift={(2.4, 0.0)}]
        \begin{scope}[xscale=1.0, yscale=1.0]
            \node[thick, circle, label={[label distance=\alabeldist]0:$\alpha_4$}] (a4) at (0,0) {};
            \node[thick, circle, label={[label distance=\alabeldist]0:$\alpha_5$}] (a5) at (0,1) {};
            \node[thick, circle, label={[label distance=\alabeldist]0:$\alpha_6$}] (a6) at (0,-1) {};
        \end{scope}
    \end{scope}
    
\end{scope}
\end{scope}

\begin{scope}
    \path [draw=none, postaction={decorate,decoration={markings,
            mark=at position 0 with {\xdef\mypgfdecoratedpathlength{\pgfdecoratedpathlength}}}},
            postaction={decorate,draw,
            decoration={markings,
                mark=at position 0 with {\arrowwithtext{$2$}},
                pre=curveto,post=curveto, 
                post length=(1.0-\bentlabeldistance)*\mypgfdecoratedpathlength, 
                pre length=\bentlabeldistance*\mypgfdecoratedpathlength, 
            amplitude=5},decorate,
            }]  (a5) to[bend right=\outsidebendangle] (a1);
            
    \path [draw=none, postaction={decorate,decoration={markings,
            mark=at position 0 with {\xdef\mypgfdecoratedpathlength{\pgfdecoratedpathlength}}}},
            postaction={decorate,draw,
            decoration={markings,
                mark=at position 0 with {\arrowwithtext{$3$}},
                pre=curveto,post=curveto, 
                post length=(1.0-\bentlabeldistance)*\mypgfdecoratedpathlength, 
                pre length=\bentlabeldistance*\mypgfdecoratedpathlength,  
            amplitude=5},decorate,
            }]  (a6) to[bend left=\outsidebendangle] (a1);
            
    \path [line width=\valuationlinewidth, postaction={decorate,decoration={markings,
            mark=at position 0 with {\xdef\mypgfdecoratedpathlength{\pgfdecoratedpathlength}}}},
            postaction={decorate,draw,
            decoration={markings,
                mark=at position 0 with {\arrowwithtext{$3$}},
                pre=curveto,post=curveto, 
                post length=(1.0-\bentlabeldistance)*\mypgfdecoratedpathlength, 
                pre length=\bentlabeldistance*\mypgfdecoratedpathlength,  
            amplitude=5},decorate,
            }]  (a1) to[bend right=\insidebendangle] (a5);
            
    \path [line width=\valuationlinewidth, postaction={decorate,decoration={markings,
            mark=at position 0 with {\xdef\mypgfdecoratedpathlength{\pgfdecoratedpathlength}}}},
            postaction={decorate,draw,
            decoration={markings,
                mark=at position 0 with {\arrowwithtext{$3$}},
                pre=curveto,post=curveto, 
                post length=(1.0-\bentlabeldistance)*\mypgfdecoratedpathlength, 
                pre length=\bentlabeldistance*\mypgfdecoratedpathlength,  
            amplitude=5},decorate,
            }]  (a1) to[bend left=\insidebendangle] (a6);
            
    \path [line width=\valuationlinewidth, postaction={decorate,decoration={markings,
            mark=at position 0 with {\xdef\mypgfdecoratedpathlength{\pgfdecoratedpathlength}}}},
            postaction={decorate,draw,
            decoration={markings,
                mark=at position 0 with {\arrowwithtext{$4$}},
                pre=curveto,post=curveto, 
                post length=0.3*\mypgfdecoratedpathlength, 
                pre length=0.7*\mypgfdecoratedpathlength, 
            amplitude=5},decorate,
            }] (a1) -- (a2);
            
    \path [line width=\valuationlinewidth, postaction={decorate,decoration={markings,
            mark=at position 0 with {\xdef\mypgfdecoratedpathlength{\pgfdecoratedpathlength}}}},
            postaction={decorate,draw,
            decoration={markings,
                mark=at position 0 with {\arrowwithtext{$1$}},
                pre=curveto,post=curveto, 
                post length=0.3*\mypgfdecoratedpathlength, 
                pre length=0.7*\mypgfdecoratedpathlength
            },decorate,
            }] (a1) -- (a3);
            
     \path [line width=\valuationlinewidth, postaction={decorate,decoration={markings,
            mark=at position 0 with {\xdef\mypgfdecoratedpathlength{\pgfdecoratedpathlength}}}},
            postaction={decorate,draw,
            decoration={markings,
                mark=at position 0 with {\arrowwithtext{$2$}},
                pre=curveto,post=curveto, 
                post length=0.45*\mypgfdecoratedpathlength, 
                pre length=0.55*\mypgfdecoratedpathlength
            },decorate,
            }] (a6) -- (a4);
            
     \path [line width=\valuationlinewidth, postaction={decorate,decoration={markings,
            mark=at position 0 with {\xdef\mypgfdecoratedpathlength{\pgfdecoratedpathlength}}}},
            postaction={decorate,draw,
            decoration={markings,
                mark=at position 0 with {\arrowwithtext{$2$}},
                pre=curveto,post=curveto, 
                post length=0.45*\mypgfdecoratedpathlength, 
                pre length=0.55*\mypgfdecoratedpathlength
            },decorate,
            }] (a5) -- (a4);
\end{scope}

\begin{scope}
    \dashedcontainerthing{a2,a1,a3};
    \dashedcontainerthing{a5,a4,a6};
\end{scope}
\end{tikzpicture}
    \vspace*{3mm}
    \caption{An example ASHG $(N, V)$ containing six agents and a partition into triples $\pi$, marked by the dashed enclosure. The weighted arcs depict the agents' valuations. Unless otherwise specified, $v_{\alpha_i}(\alpha_j) = 0$ for any $\alpha_i$ and $\alpha_j$ in $N$.}
    \label{fig:threed_efr_as_example}
\end{figure}
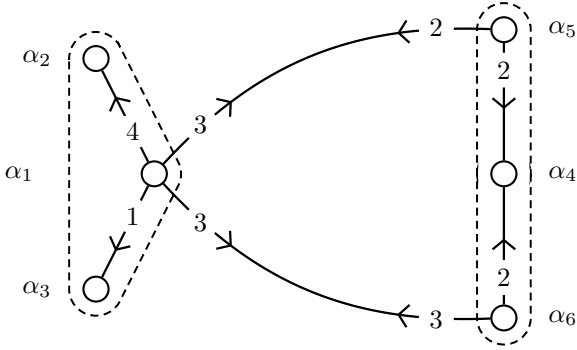
In some partition into triples $\pi$, we say that an agent $\alpha_i$ has \emph{envy} for another agent $\alpha_j$ if $u_{\alpha_i}(\pi(\alpha_j) \setminus \{ \alpha_j \}) > u_{\alpha_i}(\pi)$. In other words, $\alpha_i$ would gain a utility higher than its utility in $\pi$ if it were to swap places with $\alpha_j$. We say that a partition in which no agent has envy for another agent is \emph{envy-free}.

Suppose $\alpha_i$ has envy for $\alpha_j$ in some partition into triples $\pi$. We say that this envy is \emph{justified} (\emph{j-envy}) if $v_{\alpha_k}(\alpha_i) > v_{\alpha_k}(\alpha_j)$ for each $\alpha_k$ in $\pi(\alpha_j) \setminus \{ \alpha_j \}$. Informally, the other agents in the triple of the envied agent $\alpha_j$ all prefer the envying agent $\alpha_i$ to $\alpha_j$. 

Similarly, we say that envy is \emph{weakly justified} (\emph{wj-envy}) if $v_{\alpha_k}(\alpha_i) \geq v_{\alpha_k}(\alpha_j)$ for each $\alpha_k$ in $\pi(\alpha_j) \setminus \{ \alpha_j \}$. Now each other agent in the triple of the envied agent $\alpha_j$ would be no worse off by swapping $\alpha_j$ for the envying agent $\alpha_i$. 

A partition in which no agent has j-envy for another agent is \emph{justified envy-free} (\emph{j-envy-free}). Similarly, a partition in which no agent has wj-envy for another agent is \emph{weakly justified envy-free} (\emph{wj-envy-free}). By definition, the condition of envy-freeness is stronger than that of weakly justified envy-freeness, which is in turn stronger than justified envy-freeness.

To illustrate these solution concepts, we present an example instance in Figure~\ref{fig:threed_efr_as_example}. In this example, we can see that $\alpha_1$ has envy for $\alpha_4$, since $u_{\alpha_1}(\pi) = 4 + 1 < 3 + 3 = u_{\alpha_1}(\pi(\alpha_4) \setminus \{ \alpha_4 \})$. This envy is weakly justified since $v_{\alpha_5}(\alpha_4) = 2 \leq 2 = v_{\alpha_5}(\alpha_1)$ and $v_{\alpha_6}(\alpha_4) = 2 \leq 3 = v_{\alpha_6}(\alpha_1)$. Since $v_{\alpha_5}(\alpha_4) = v_{\alpha_5}(\alpha_1) = 2$, this envy is not justified.

We also define the solution concept of \emph{stability} in our model. Given an ASHG $(N, V)$ and a partition into triples $\pi$, we say that a triple $\{\alpha_i, \alpha_j, \alpha_k \}$ \textit{blocks} $\pi$ if $u_{\alpha_i}(\{\alpha_j, \alpha_k \}) > u_{\alpha_i}(\pi), u_{\alpha_j}(\{\alpha_i, \alpha_k \}) > u_{\alpha_j}(\pi)$, and $u_{\alpha_k}(\{\alpha_i, \alpha_j \}) > u_{\alpha_k}(\pi)$. We say that a partition $\pi$ is \emph{stable} if no triple of any three agents blocks $\pi$. In the example instance in Figure~\ref{fig:threed_efr_as_example}, $\{ \alpha_1, \alpha_5, \alpha_6 \}$ blocks $\pi$.

In this paper we consider three types of restriction on the agents' valuations. If $v_{\alpha_i}(\alpha_j)=v_{\alpha_j}(\alpha_i)$ for any pair of agents $\alpha_i, \alpha_j$ then preferences are \emph{symmetric}. If $v_{\alpha_i}(\alpha_j)\in\{0, 1\}$ for any pair of agents $\alpha_i, \alpha_j$ then preferences are \emph{binary} (also called \emph{simple}~\cite{Bilo22,ABBHOP19}). Similarly, if $v_{\alpha_i}(\alpha_j)\in\{0, 1, 2\}$ for any such $\alpha_i, \alpha_j$ then preferences are \emph{ternary}. 

If preferences are binary and symmetric then they can be represented as an undirected graph, which we call the \emph{underlying graph}. Formally, we represent the underlying graph as a pair $(N, E)$ such that $\{ \alpha_i, \alpha_j \} \in E$ if and only if $v_{\alpha_i}(\alpha_j) = 1$. The \emph{maximum degree} of such an ASHG is then the maximum degree of its underlying graph. We use standard graph-theoretic terminology when referring to the underlying graph. For example, in the context of some underlying graph $(N, E)$, for any agent $\alpha_i$ in $N$, let $\mathcal{N}(\alpha_i)$ be the \emph{open neighbourhood} of $\alpha_i$ in $(N, E)$, meaning $\alpha_j \in \mathcal{N}(\alpha_i)$ if and only if $\{ \alpha_i, \alpha_j \} \in E$ and $\alpha_j \neq \alpha_i$. Let $\deg(\alpha_i) = |\mathcal{N}(\alpha_i)|$ be the \emph{degree} of $\alpha_i$ in $(N, E)$.   We say that an agent is \emph{isolated} if it has degree $0$. A \emph{$k$-path} or \emph{$k$-cycle} is a path or cycle of $k$ agents. We shall label the consecutive agents in a path or cycle as a list. For example, if some component $P$ is a path or a cycle then the consecutive agents in $P$ will be labelled $( p_1, p_2, \dots, p_{|P|} )$, where $\{ p_i, p_{i+1} \} \in E$ for each $i$ where $1\leq i < |P|$.

\subsection{Our contribution}
In this paper we study the existence of envy-free, weakly justified envy-free, and justified envy-free partitions into triples and the complexity of the associated search problems. Specifically, we explore what happens when the agents' valuations are somehow restricted. We identify various dichotomies between polynomial-time solvability and $\NP$-hardness, shown in Table~\ref{tab:mainresults}. In the table, the symbols ``\checkmark'' and ``\maynotexist'' are short for ``must exist'' and ``may not exist'' respectively. The complexity class shown refers to the problem of finding a partition that satisfies the relevant solution concept in an instance in which the valuations are restricted as shown.

\subsubsection{Structure of the paper}

In Section~\ref{sec:threed_efr_as_intro} we provide some background, formally define the necessary concepts and terminology, and discuss some related previous work.

In Section~\ref{sec:threed_efr_as_ef} we show that an ASHG may not contain an envy-free partition into triples, even if the preferences are binary and symmetric and the maximum degree of the underlying graph is $2$. We describe a polynomial-time algorithm for this case that either constructs an envy-free partition into triples or reports that no such partition exists (Theorem~\ref{thm:threed_efr_as_ef_algorithm}). We then contrast this result by showing that the corresponding existence problem is $\NP$-complete even when the maximum degree of the underlying graph is $3$ (Theorem~\ref{thm:threed_efr_as_regularenvy_npcomplete}).

\begin{table}[b]
\begin{center}
\resizebox{\textwidth}{!}{
\begin{tabular}{ccccc}\\\noalign{\hrule}
   \multicolumn{2}{c}{input settings}            
   & \multicolumn{3}{c}{results} 
\\
 solution concept & preference restriction &  always exists? & finding & shown in\\
	      \noalign{\hrule}
	      \noalign{\hrule}
  envy & sym. and binary, $\Delta=2$  & \maynotexist & \P & Thm.~\ref{thm:threed_efr_as_ef_algorithm}\\
  envy & sym. and binary, $\Delta=3$  & \maynotexist & \NP-h. & Thm.~\ref{thm:threed_efr_as_regularenvy_npcomplete}\\[4.5pt]
  weakly justified envy & sym. and binary, $\Delta=2$  & \maynotexist & \P & Thm.~\ref{thm:threed_efr_as_wjef_algowjpathscycles}\\
  weakly justified envy & sym. and binary, $\Delta=3$  & \maynotexist & \NP-h. & Thm.~\ref{thm:threed_efr_as_wjef_npcomplete}\\[4.5pt]
  justified envy & sym. and binary & \checkmark & \P & Obs.~\ref{obs:threed_efr_as_jef_binary_symmetric_from_stability}\\
  justified envy & binary & \checkmark & \P & Thm.~\ref{thm:threed_efr_as_jef_binary_algorithm}\\
  justified envy & ternary & \maynotexist & \NP-h. & Thm.~\ref{thm:threed_efr_as_jef_terasym_npcomplete}\\
  justified envy & sym. and non-binary & \maynotexist & \NP-h. & Thm.~\ref{thm:threed_efr_as_jef_symmetric_6_npcomplete}\\
      \noalign{\hrule}
    \end{tabular}
    }
\end{center}
\vspace*{2mm}
\caption{Our existence and complexity results. ``sym.'' refers to symmetric preferences, while ``binary''and ``ternary'' refer to binary and ternary preferences, respectively.  In restrictions involving binary and symmetric preferences, $\Delta$ refers to the maximum degree of the underlying graph.}
\label{tab:mainresults}
\end{table}

In Section~\ref{sec:threed_efr_as_wjef} we identify a similar dichotomy for weakly justified envy-freeness. We first show that a weakly justified envy-free partition into triples may not exist, even when preferences are binary and symmetric and the maximum degree of the ASHG is $2$. We describe a slightly more complex polynomial-time algorithm for this case that either constructs a weakly justified envy-free partition into triples or reports that no such partition exists (Theorem~\ref{thm:threed_efr_as_wjef_algowjpathscycles}). As for envy-freeness, we show that the corresponding existence problem is also $\NP$-complete even when the maximum degree of the underlying graph is $3$ (Theorem~\ref{thm:threed_efr_as_wjef_npcomplete}). We remark that the set of ASHGs with maximum degree $2$ that do not contain a weakly justified envy-free partition into triples is a strict subset of the set of ASHGs with maximum degree $2$ that do not contain an envy-free partition into triples.

In Section~\ref{sec:threed_efr_as_jef} we consider justified envy-freeness. We first observe that if a partition into triples is stable then it is also justified envy-free. We then show that if preferences are binary but not necessarily symmetric, a justified envy-free partition into triples must exist and can be found in polynomial time, making use of a potential function~\cite{GairingSavani19} (Theorem~\ref{thm:threed_efr_as_jef_binary_algorithm}). We complement this result with two hardness results. The first is that a given ASHG may not contain a justified envy-free partition into triples even when preferences are ternary but not necessarily symmetric, and the associated existence problem is $\NP$-complete (Theorem~\ref{thm:threed_efr_as_jef_terasym_npcomplete}). The second is that a given ASHG may not contain a justified envy-free partition into triples even when preferences are symmetric but not necessarily binary, and the associated existence problem is $\NP$-complete (Theorem~\ref{thm:threed_efr_as_jef_symmetric_6_npcomplete}). We also remark that it seems unclear how to design an ASHG with symmetric preferences in which there is no justified envy-free partition into triples. The authors used an integer programming model in combination with a guided computer search. We discuss this technique further in Section~\ref{sec:threed_efr_as_jef_symmetricnonbinrary}.

In Section 6 we conclude the paper and consider some possible directions for future work.

\subsection{Related work}

We discuss previous work on:
\begin{itemize}
    \item envy-freeness and coalitions of restricted size, in Section~\ref{sec:threed_efr_as_relatedwork_efrestricted}
    \item envy-freeness and coalitions of unrestricted size, including work on hedonic games, in Section~\ref{sec:threed_efr_as_relatedwork_efgeneral}
    \item coalitions of fixed size, but not in relation to envy-freeness, including on three- and multi-dimensional roommates, in Section~\ref{sec:threed_efr_as_relatedwork_general}
\end{itemize} 

For other models of coalition formation, coalitions of unrestricted size, hedonic games, and solution concepts not involving envy, a wealth of literature exists~\cite{BJ02,ABS13,Cechlarova2008,Brandt2020FindingAR}.

\subsubsection{Envy-freeness and restricted coalition size}
\label{sec:threed_efr_as_relatedwork_efrestricted}

To our knowledge, there are four previous papers involving envy-freeness and coalitions of restricted size. 

The first paper is by Coutance et al.~\cite{Coutanceetal2023} who in 2023 considered variants of envy-freeness in a model in which coalitions must have size two and agents have ordinal preferences over possible pairs. They noted that since it is trivial to decide whether an envy-free partition exists. They therefore studied \emph{rank-envy-freeness} and some of its variants.

The second paper is by Li et al.~\cite{Li2023journal} who, also in 2023, considered a strictly more general model than ours. Rather than fixing the size of each coalition, a partition of the $n$ agents must be \emph{balanced}, meaning there is a fixed number of coalitions $k\leq n$ such that $\lfloor n/k \rfloor \leq |S| \leq \lceil n/k \rceil$ for any coalition $S$ in a feasible partition. They studied a generalisation of envy-freeness, termed \emph{envy-free up to r} (EF-$r$), in which the utility gained by any envious agent in a feasible partition may be up to $r$, for some fixed $r \geq 0$. Our definition of envy-freeness is thus EF-$0$. Interestingly, the authors applied results from discrepancy theory to show that an approximate envy-free partition with a particular fixed asymptotic bound must exist, and can be found in polynomial time. They also considered restricted types of instances, such those in which the underlying structure is a tree. They showed that in such an instance, an EF-$1$ partition must exist and can be found in polynomial time. A coalition partition in which every coalition has size three is by definition balanced, so some of the polynomial-time solvability results on approximate envy-free partitions~\cite[Theorems~9 and~10]{Li2023journal} also apply to our model.

The third paper, also published in 2023, is by Gan et al.~\cite{Ganetal2023}. The authors defined a general model in which a set of agents is assigned to a set of resources such that a single resource can be shared by one or more agents. In their model, the authors characterised the relationship between \emph{proportional} and envy-free assignments. They also defined \emph{Pareto Envy-Freeness} (PEF), which is a relaxation of envy-freeness as defined in that setting. The authors also study a restricted case of the general model in which agents are assigned to resources with the same capacity (``dorms''). Each agent has binary and symmetric preferences over all other agents, as dorm-mates, and additively separable preferences over possible dorms. The main result is that if all dorms have capacity $2$ then a PEF assignment must exist and can be found in polynomial time. It is also shown that if dorms have capacity three then a PEF assignment does not necessarily exist. The model we study in this paper can be seen as a type of special case of the latter model in which the agents are indifferent between possible dorms.

The fourth paper, published in 2020, is by Boehmer and Elkind~\cite{Boe20}. The authors considered a model in which a set of agents is to be partitioned into sets of equal size. Notably, the desired coalition size is fixed but supplied as part of the problem input. In their model, each agent's preference between possible coalitions is highly constrained. Each agent belongs to one of a number of \emph{types} and compares two possible coalitions by considering only the proportion of agents in each coalition of each type. The authors presented results for a number of solution concepts, including envy-freeness as we define it here. In particular, they showed that an envy-free partition does not necessarily exist and the associated decision problem is $\NP$-complete, even when agents' preferences are highly restricted. They also showed that the problem of deciding whether an envy-free partition exists is fixed-parameter tractable in terms of the coalition size.

\subsubsection{Envy-freeness and unrestricted coalition size}
\label{sec:threed_efr_as_relatedwork_efgeneral}

In a comprehensive study on ASHGs published in 2013, Aziz et al.~\cite{ABS13} defined envy-freeness as a solution concept in ASHGs. They noted that the \emph{singleton partition} (in which each agent belongs to a coalition of size one) is trivially envy-free and therefore considered the existence of partitions that simultaneously satisfy envy-freeness and some other solution concept. Of course, the existence of such a trivial envy-free partition forms a strong motivation for somehow bounding the size of acceptable coalitions.

Stability, envy-freeness, and justified envy-freeness were also considered by Ueda~\cite{Ued18} in 2018, in a hedonic game in which each agent has ordinal preferences over all possible coalitions. He too observed that both the singleton partition and the \emph{grand partition} (in which there is one coalition containing all the agents) are trivially envy-free. He also observed that there exist instances in which no non-trivial partition is envy-free, and additional instances in which no non-trivial partition is justified envy-free. Finally, he showed that any stable partition~\cite{hgch15} is also justified envy-free.

Barrot and Yokoo~\cite{BY19} noted Ueda's observations in 2019 and continued exploring the existence of partitions that satisfy a combination of solution concepts, including envy-freeness, weakly justified envy-freeness, and justified envy-freeness. As well as some non-existence results on ASHGs, they also considered more general systems of preference representation. Notably, they proved results on the existence of satisfactory coalition partitions in a setting in which preferences are either \emph{top responsive} and \emph{bottom responsive} (or \emph{bottom refuse}), two restrictions previously well-established in the hedonic games literature.

\subsubsection{Other solution concepts and restricted coalition size}
\label{sec:threed_efr_as_relatedwork_general}

The \emph{Stable Roommates} problem, which can be viewed as a type of hedonic game, involves assigning a set of agents into pairs such that the resulting partition is stable~\cite{Man13}. It is a classical problem of matching under preferences and has been studied extensively. Its three-dimensional case has also received some attention, and a few specific models of \emph{Three-Dimensional Stable Roommates} have been studied in the literature~\cite{Bre20,Hua07,IMO07,ManloveMcKay3DSRAS2021,NH91}.

Sless et al.~\cite{Sless18} proposed a model in 2018 that can be viewed as a type of ASHG with symmetric preferences. Arguing that there is a practical motivation for coalitions of restricted size, they focused on the existence of coalition partitions that contain exactly $k$ coalitions, for some fixed $k$ where $k \geq 1$. The authors presented both theoretical and empirical results relating to this model. They showed that the problem of finding a partition with maximum utilitarian welfare can be solved in polynomial time in the restricted case in which $k$ is fixed and there are, in a precise sense, relatively few negative edges. Otherwise, this construction problem is $\NP$-hard. They also presented other polynomial-time solvability results for a problem in which a central organiser can add some edges to the instance.

In 2019, Cseh et al.~\cite{CFH19} considered Pareto optimal partitions in a model that is similar to the model of Gan et al.~\cite{Ganetal2023} discussed previously. In the model of Cseh et al., there is a set of rooms with integer sizes and any coalition must be allocated to exactly one room where the size of the room is exactly the size of the coalition. Cseh et al.\ studied two specific variants of this model, involving so-called \emph{$\mathscr{B}$-} and \emph{$\mathscr{W}$-preferences}. They showed that if agents have $\mathscr{B}$-preferences and strict preference lists then a polynomial-time algorithm based on serial dictatorship can be used to construct a Pareto optimal partition in polynomial time. They also showed that in various other cases a Pareto optimal partition may not exist and that the associated existence problems are either $\NP$-hard or $\NP$-complete.

\begin{figure}
    \centering
    \begin{tikzpicture}[thick, every edge/.style = {draw, -to}]
\begin{scope}[yscale=0.8, xscale=0.7]
\begin{scope}[every node/.style={inner sep=8pt, align=center}, style={sibling distance=50mm, level distance=20mm}, every edge/.style = {darrow}]
  \node {perfect}
    child {node[yshift=-1.6cm, xshift=-0.1cm] (cs) {(core) stable}}
    child {node[yshift=0.2cm, xshift=2.2cm] {envy-free}
        child {node[yshift=-0.3cm] (wjef) {weakly justified envy-free}
            child {node[yshift=-0.4cm] (jef) {justified envy-free}}
        }
        child {node[xshift=1.0cm, text width=4cm] {swap stable under\\ transferable utilities}
            child {node[yshift=-0.2cm] (ss) {swap stable}
                child {node (sss) {strictly swap stable}
                }
            }
        }
    };
\end{scope}
\draw[\solutionconceptsdiagramarrow] (cs) -- ($(jef.north west)+(0.0, -0.2)$);
\begin{scope}
\end{scope}
\end{scope}
\end{tikzpicture}

    \caption[Part of the known hierarchy of solution concepts in hedonic games~\cite{BY19,Bilo22}]{Part of the known hierarchy of solution concepts in hedonic games~\cite{hgch15,BY19,Bilo22}. In the diagram, an arrow points from one concept to another if any partition that satisfies the former must also satisfy the latter. This figure is adapted from \emph{the Handbook of Computational Social Choice}~\cite[Figure~15.1]{hgch15}.}
    \label{fig:solution_concepts}
\end{figure}
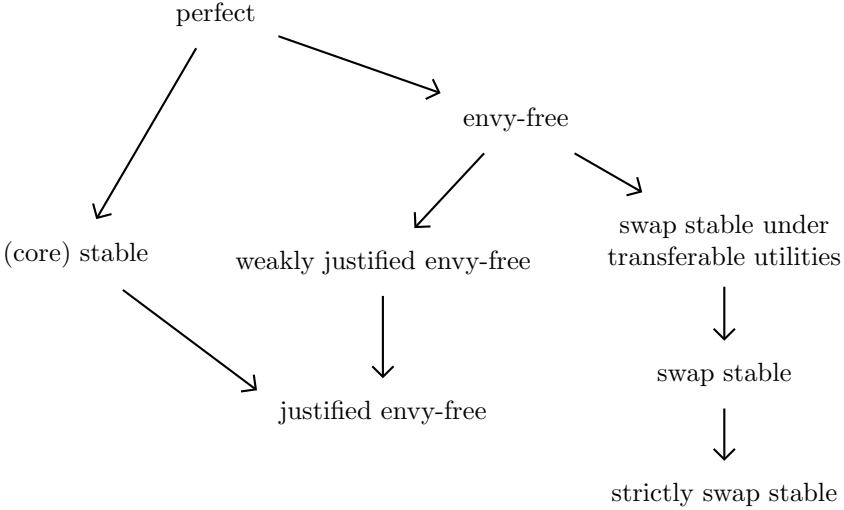

In 2022, Bil\`o et al.~\cite{Bilo22} studied ASHGs in which the coalition size is fixed and agents have binary and symmetric preferences. They considered \emph{swap stability} and two related solution concepts called \emph{strict swap stability} and \emph{swap stability under transferable utilities}. If a partition $\pi$ is \emph{strictly swap stable} then no agents $\alpha_i$ and $\alpha_j$ exist such that $\alpha_i$ could swap places with $\alpha_j$ to produce a partition $\pi'$ in which $u_{\alpha_i}(\pi') > u_{\alpha_i}(\pi)$ and $u_{\alpha_j}(\pi') > u_{\alpha_j}(\pi)$. If a partition $\pi$ is \emph{swap stable} then no two agents $\alpha_i$ and $\alpha_j$ exist such that $\alpha_i$ could swap places with $\alpha_j$ to produce a partition $\pi'$ in which $u_{\alpha_i}(\pi') > u_{\alpha_i}(\pi)$ and $u_{\alpha_j}(\pi') \geq u_{\alpha_j}(\pi)$. If a partition $\pi$ is \emph{swap stable under transferable utilities} then no two agents $\alpha_i$ and $\alpha_j$ exist such that $\alpha_i$ could swap places with $\alpha_j$ to produce a partition $\pi'$ in which $u_{\alpha_i}(\pi') + u_{\alpha_j}(\pi') > u_{\alpha_i}(\pi) + u_{\alpha_j}(\pi)$. Bil\`o et al. noted that a related solution concept had been previously studied as ``exchange stability''~\cite{Alc95}. It is straightforward to show that envy-freeness implies swap stability under transferable utilities. In fact, we illustrate in Figure~\ref{fig:solution_concepts} the hierarchical relationships between various solution concepts in ASHGs, including those studied by  Bil\`o et al. Note that a \emph{perfect} partition is one in which each agent gains its maximum possible utility~\cite{ABS13}.

\section{Envy-freeness}
\label{sec:threed_efr_as_ef}
In this section we consider envy-freeness in ASHGs with binary and symmetric preferences.

First, in Section~\ref{sec:threed_efr_as_envyfreeness_maxdeg2}, we consider ASHGs with maximum degree $2$. We show that in polynomial time we can either construct an envy-free partition into triples or report that no such partition exists (Theorem~\ref{thm:threed_efr_as_ef_algorithm}).

Next, in Section~\ref{sec:threed_efr_as_envyfreeness_maxdeg3}, we consider ASHGs with maximum degree $3$. We show that such an ASHG may not contain an envy-free partition into triples and the associated existence problem is $\NP$-complete.

\subsection{Binary and symmetric preferences with maximum degree \texorpdfstring{$2$}{2}}
\label{sec:threed_efr_as_envyfreeness_maxdeg2}
Our first result is a necessary and sufficient condition for the existence of an envy-free partition into triples in an ASHG with binary and symmetric preferences and maximum degree $2$. In other words, an ASHG in which every component of the underlying graph is either a path or a cycle.

\begin{lem}
\label{lem:threed_efr_as_ef_if_and_only_if}
Consider an ASHG with binary and symmetric preferences and maximum degree $2$. Let $P$ be the set of isolated agents, $\mathcal{Q}$ be the set of components of $3{k_1} - 2$ agents for any integer ${k_1} > 1$, and $\mathcal{R}$ be the set of components of $3{k_2} - 1$ agents for any integer ${k_2} \geq 1$. An envy-free partition into triples exists if and only if $2|\mathcal{Q}| + |\mathcal{R}| \leq |P|$.
\end{lem}
\begin{proof}
Suppose $(N, E)$ is the underlying graph of an arbitrary ASHG with binary and symmetric preferences and maximum degree $2$. Let $P=\{ p_1, p_2, \dots, p_{|P|} \}$ be the set of isolated agents, $\mathcal{Q} = \{ Q_1, Q_2, \dots, Q_{|\mathcal{Q}|} \}$ be the set of components of $3{k_1} - 2$ agents for any integer ${k_1} > 1$, and $\mathcal{R}=\{ R_1, R_2,\allowbreak \dots, R_{|\mathcal{R}|} \}$ be the set of components of $3{k_2} - 1$ agents for any integer ${k_2} \geq 1$. Let $\mathcal{S}=\{S_1,S_2,\dots,S_{|\mathcal{S}|}\}$ be the set of components of $3{k_3}$ agents for any integer ${k_3} \geq 1$.

To show the first direction, suppose $2|\mathcal{Q}| + |\mathcal{R}| \leq |P|$. We shall construct a partition into triples $\pi$ and demonstrate that it is envy-free. First, note that if any agent has utility $1$ or more then that agent is not envious, since the maximum degree of $(N, E)$ is $2$.

Construct $\pi$ as follows. First, consider each component $S$ in $\mathcal{S}$, with consecutive agents labelled $( s_1, s_2, \dots, s_{3{k_3}} )$. For each $i$ where $1\leq i\leq {k_3}$, add $\{ s_{3i-2}, s_{3i-1}, s_{3i} \}$ to $\pi$. It follows that each agent in $S$ has utility at least $1$ and is therefore not envious. 

Now consider each component $R$ in $\mathcal{R}$, with consecutive agents labelled $( r_1, r_2, \dots, r_{3{k_2}-1} )$. For each $i$ where $1\leq i \leq {k_2}-1$, add $\{ r_{3i-2}, r_{3i-1}, r_{3i} \}$ to $\pi$. Next, add $\{ r_{3{k_2}-2}, r_{3{k_2}-1}, p_{2|\mathcal{Q}|+i} \}$ to $\pi$ (recalling that $|P| \geq 2|\mathcal{Q}| + |\mathcal{R}|$). It follows that each agent in $R$ has utility at least $1$ and is therefore not envious. 

Now consider each component $Q$ in $\mathcal{Q}$, with consecutive agents labelled $( q_1, q_2, \dots, q_{3{k_1}-2} )$. For each $i$ where $1\leq i \leq {k_1}-2$, add $\{ q_{3i-2}, q_{3i-1}, q_{3i} \}$ to $\pi$. Next, add to $\pi$ the triples $\{ q_{3{k_1}-5}, q_{3{k_1}-4}, p_{i} \}$ and $\{ q_{3{k_1}-3}, q_{3{k_1}-2}, p_{2i} \}$. It follows that each agent in $Q$ has utility at least $1$ and is therefore not envious.

Finally, arbitrarily add the remaining agents in $P$ to triples in $\pi$. Since these agents are isolated they are also not envious.

To show the second direction, suppose for a contradiction that the ASHG represented by $(N, E)$ has an envy-free partition into triples $\pi$ and $2|\mathcal{Q}| + |\mathcal{R}| > |P|$. Since the degree of any agent in any component in $\mathcal{Q}$ is at least $1$, it must be that the utility of each agent in any component in $\mathcal{Q}$ is at least $1$, for otherwise such an agent is envious. Similarly, the utility of each agent in any component in $\mathcal{R}$ must also be at least $1$. It follows that any agent with utility $0$ belongs to $P$.

Now consider some component $Q$ in $\mathcal{Q}$. By definition, $Q$ has $3{k_1} - 2$ agents for some integer $k_1$ where ${k_1} > 1$. The only possibility is that there exist two triples in $\pi$ that each contains exactly two agents in $Q$ and some agent with utility $0$. Similarly, for any $R$ in $\mathcal{R}$ there must exist at least one triple in $\pi$ that contains exactly two agents in $R$ and some agent with utility $0$. It follows that there are in total at least $2|\mathcal{Q}| + |\mathcal{R}|$ agents with utility $0$. The only possibility is that there are at least $2|\mathcal{Q}| + |\mathcal{R}|$ such agents in $P$, which is a contradiction.
\end{proof}

Lemma~\ref{lem:threed_efr_as_ef_if_and_only_if} shows a necessary and sufficient condition for the existence of an envy-free partition into triples. In fact, the constructive proof of this lemma can be adapted to show that there exists an $O(|N|)$-time algorithm that either constructs an envy-free partition into triples or reports that no such partition exists. We state this as Theorem~\ref{thm:threed_efr_as_ef_algorithm} and defer the formal proof to Appendix~\ref{sec:appendix_efalgo}.

\begin{restatable}{thm}{thmthreedefrasefalgorithm}
\label{thm:threed_efr_as_ef_algorithm}
Consider an ASHG with binary and symmetric preferences and maximum degree $2$. There exists an $O(|N|)$-time algorithm that either constructs an envy-free partition into triples or reports that no such partition exists.
\end{restatable}

\subsection{Binary and symmetric preferences with maximum degree \texorpdfstring{$3$}{3}}
\label{sec:threed_efr_as_envyfreeness_maxdeg3}
We now consider ASHGs with binary and symmetric preferences and maximum degree $3$. We show that deciding if a given ASHG contains an envy-free partition into triples is $\NP$-complete even when preferences are binary and symmetric and the maximum degree is $3$.

We present a polynomial-time reduction from a variant of \emph{Exact Satisfiability} (XSAT)~\cite{GJ79}. An instance of XSAT is a boolean formula in conjunctive normal form (CNF). We represent such a formula as the set of its clauses $C=\{ c_1, c_2, \dots, c_{m} \}$. We represent each clause $c_r$ in $C$ as a set of literals. Each literal is either an occurrence of a single variable or its negation. We write $X(C)$ to mean the set of variables contained in the formula $C$. A \emph{truth assignment} $f : X(C) \mapsto \{\text{true}, \text{false}\}$ is an assignment of values to the set of variables. We say that an \emph{exact model} is a truth assignment to the variables such that each clause contains exactly one true literal (and therefore exactly two false literals). Given a formula $C$, if an exact model exists then we say that $C$ is \emph{exactly satisfiable}. 

Deciding if a given instance $C$ of XSAT is exactly satisfiable is $\NP$-complete, even in the restricted case in which each clause contains exactly three literals~\cite{SchaeferSatComplexity78}. We call this restricted case  $3$-XSAT, but it is sometimes referred to as \emph{1-in-3 SAT}. A result of Porschen et al.~\cite[Lemma~5]{PSSW14} shows that $3$-XSAT remains $\NP$-complete even in the restricted case in which each literal is positive and each variable occurs in exactly three clauses. We call this variant \porschenxsatvariant/ (Problem~\ref{pr:xsatvariant}), but it is sometimes referred to as \emph{1-in-3 Positive 3-Occurrence-SAT}~\cite{Bre20}. Note that in an instance $C$ of \porschenxsatvariant/ it must be that $|X(C)|=m$.

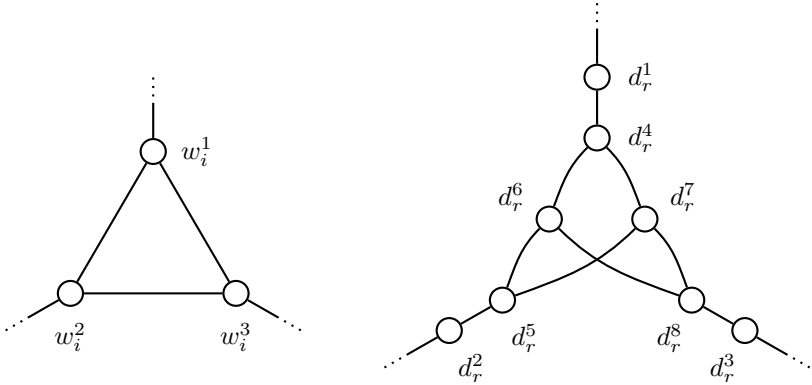
\begin{figure}
    \centering
    \begin{tikzpicture}
\begin{scope}[every node/.style={circle,draw, minimum size=2.4mm}, scale=0.9]
    \begin{scope}
        \begin{scope}[shift={(0.0, 0.0)}]
            \node[thick, circle, label={[label distance=0.4cm]0:$w_i^1$}] (v1) at (0,{1.0*1.4}) {};
            \node[thick, circle, label={[label distance=0.4cm]270:$w_i^2$}] (v2) at ({-0.866*1.4},{-0.5*1.4}) {};
            \node[thick, circle, label={[label distance=0.4cm]270:$w_i^3$}] (v3) at ({0.866*1.4},{-0.5*1.4}) {};]
            
            \node[draw=none] (v1a) at (0,{1.0*2.5}) {};
            \node[draw=none] (v2a) at ({-0.866*2.5},{-0.5*2.5}) {};
            \node[draw=none] (v3a) at ({0.866*2.5},{-0.5*2.5}) {};
            
        \end{scope}
        
        \begin{scope}[shift={(6.5, 0.0)}]
            \node[thick, circle, label={[label distance=0.4cm]0:$d_r^1$}] (dr1) at (0,{1.0*2.5}) {};
            \node[thick, circle, label={[label distance=0.4cm]300:$d_r^2$}] (dr2) at ({-0.866*2.5},{-0.5*2.5}) {};
            \node[thick, circle, label={[label distance=0.4cm]240:$d_r^3$}] (dr3) at ({0.866*2.5},{-0.5*2.5}) {};]
            
            \node[thick, circle, label={[label distance=0.4cm]0:$d_r^4$}] (dr4) at (0,{1.0*1.6}) {};
            \node[thick, circle, label={[label distance=0.4cm]300:$d_r^5$}] (dr5) at ({-0.866*1.6},{-0.5*1.6}) {};
            \node[thick, circle, label={[label distance=0.4cm]240:$d_r^8$}] (dr8) at ({0.866*1.6},{-0.5*1.6}) {};]
            
            \node[thick, circle, label={[label distance=0.4cm]150:$d_r^6$}] (dr6) at ({-0.7},{0.4}) {};]
            \node[thick, circle, label={[label distance=0.4cm]30:$d_r^7$}] (dr7) at ({0.7},{0.4}) {};]
            
            \node[draw=none] (dr1a) at (0,{1.0*3.6}) {};
            \node[draw=none] (dr2a) at ({-0.866*3.6},{-0.5*3.6}) {};
            \node[draw=none] (dr3a) at ({0.866*3.6},{-0.5*3.6}) {};
            
        \end{scope}
        
    \end{scope}

\end{scope}

\begin{scope}
    \foreach \from/\to in {v1/v2, v2/v3, v3/v1}
        \draw [thick] (\from) -- (\to);

    \path[-sss,thick] (v1) edge (v1a);
    \path[-sss,thick] (v2) edge (v2a);
    \path[-sss,thick] (v3) edge (v3a);
    
    \foreach \from/\to in {dr1/dr4, dr2/dr5, dr3/dr8}
        \draw [thick] (\from) -- (\to);

    \path[-sss,thick] (dr1) edge (dr1a);
    \path[-sss,thick] (dr2) edge (dr2a);
    \path[-sss,thick] (dr3) edge (dr3a);
        
    \draw[thick] (dr4) to[out=225, in=70] (dr6);
    \draw[thick] (dr4) to[out=315, in=110] (dr7);
    
    \draw[thick] (dr5) to[out=70, in=225] (dr6);
    \draw[thick] (dr5) to[out=20, in=225] (dr7);
    
    \draw[thick] (dr8) to[out=160, in=315] (dr6);
    \draw[thick] (dr8) to[out=110, in=315] (dr7);

\end{scope}
\end{tikzpicture}
    \caption[The reduction from \porschenxsatvariant/ to the problem of deciding if a given ASHG contains an envy-free partition into triples]{The reduction from \porschenxsatvariant/ to the problem of deciding if a given ASHG contains an envy-free partition into triples. A variable gadget $W_i$ and clause gadget $D_r$ are represented as undirected graphs.}
    \label{fig:threed_efr_as_regular_envy_free_reduction}
\end{figure}

\begin{myproblem}[\porschenxsatvariant/]
\label{pr:xsatvariant}
\begin{samepage}
\begin{adjustwidth}{8pt}{8pt}
\inp a boolean formula $C$ in conjunctive normal form, in which every literal is positive and each variable occurs in exactly three clauses\\
\ques is $C$ exactly satisfiable?
\end{adjustwidth}
\end{samepage}
\end{myproblem}
The reduction, illustrated in Figure~\ref{fig:threed_efr_as_regular_envy_free_reduction}, is as follows. Suppose $C$ is an arbitrary instance of \porschenxsatvariant/. We shall construct an ASHG represented by an underlying graph $(N, E)$.

For each variable $x_i$ in $X(C)$, there are three corresponding literals in three different clauses. For each such $x_i$, arbitrarily label each of these literals as the first, second, and third occurrences of $x_i$.
For each such $x_i$, construct a set of three agents $W_i = \{ w_i^1, w_i^2, w_i^3 \}$, which we refer to as the \emph{$i\textsuperscript{th}$ variable gadget}. Add edges $\{ w_i^1, w_i^2 \}$, $\{ w_i^2, w_i^3 \}$, and $\{ w_i^3, w_i^1 \}$. Next, for each clause $c_r$ in $C$ construct a set of eight agents $D_r = \{ d_r^1, d_r^2, \dots, d_r^8 \}$, which we refer to as the \emph{$r\textsuperscript{th}$ clause gadget}. Add edges $\{ d_r^1, d_r^4 \}$, $\{ d_r^2, d_r^5 \}$, $\{ d_r^3, d_r^8 \}$, $\{ d_r^4, d_r^6 \}$, $\{ d_r^4, d_r^7 \}$, $\{ d_r^5, d_r^6 \}$, $\{ d_r^5, d_r^7 \}$, $\{ d_r^8, d_r^6 \}$, and $\{ d_r^8, d_r^7 \}$.

Now we shall connect the variable and clause gadgets. Consider each clause $c_r = \{ x_i, x_j, x_k \}$. If $c_r$ contains the first occurrence of $x_i$ then add the edge $\{ d_r^1, w_i^1 \}$. Similarly, if $c_r$ contains the second occurrence of $x_i$ then add the edge $\{ d_r^1, w_i^2 \}$. Similarly, if $c_r$ contains the third occurrence of $x_i$ then add the edge $\{ d_r^1, w_i^3 \}$. 

In the same way, add an edge between $d_r^2$ and an agent in $W_j$ depending on the index of the occurrence of $x_j$ in $c_r$. Finally, add an edge between $d_r^3$ and an agent in $W_k$ depending on the index of the occurrence of $x_k$ in $c_r$. Let us now say that the clause gadget $D_r$ is adjacent to the variable gadgets $W_i$, $W_j$, and $W_k$, and vice-versa.

The construction of $(N, E)$ is now complete. Note that each agent in a variable gadget has degree $3$, the agents $d_r^4$, $d_r^5$, $d_r^6$, $d_r^7$, and $d_r^8$ for each $1 \leq r \leq m$ have degree $3$, and the agents $d_r^1, d_r^2, d_r^3$ for each $1 \leq r \leq m$ have degree $2$. It follows that the maximum degree of $(N, E)$ is $3$. 

It is straightforward to show that this reduction can be performed in polynomial time. To prove that the reduction is correct we show that the ASHG represented by $(N, E)$ contains an envy-free partition into triples if and only if the \porschenxsatvariant/ instance $C$ is exactly satisfiable. 

We first prove some ancillary lemmas. Recall that for any set of agents $S$, $\sigma(S, \pi)$ is the number of triples in $\pi$ that each contains at least one agent in $S$. Recall also that for any agent $\alpha_i$ we write $\mathcal{N}(\alpha_i)$ to mean the open neighbourhood of $\alpha_i$.

\begin{lem}
\label{lem:threed_efr_as_envyfree_util1_neighbourhoodthreetriples}
Suppose $\pi$ is a partition into triples in the ASHG represented by $(N, E)$. For any agent $\alpha_i$, if $u_{\alpha_i}(\pi)=1$ and $\sigma(\mathcal{N}(\alpha_i), \pi)=\deg(\alpha_i)$ then $\alpha_i$ is not envious in $\pi$.
\end{lem}
\begin{proof}
Suppose, to the contrary, that $\alpha_i$ envies some agent $\alpha_j$. Then $u_{\alpha_i}(\pi(\alpha_j) \setminus \{ \alpha_j \})=2$. It must be that two agents in $\mathcal{N}(\alpha_i)$ belong to the same triple, namely $\pi(\alpha_j)$. It follows that $\sigma(\mathcal{N}(\alpha_i), \pi) < \deg(\alpha_i)$, which is a contradiction. 
\end{proof}

\begin{lem}
\label{lem:threed_efr_as_util0}
Suppose $\pi$ is a partition into triples in the ASHG represented by $(N, E)$. For any agent $\alpha_i$, if $u_{\alpha_i}(\pi)=0$ then $\alpha_i$ is envious in $\pi$.
\end{lem}
\begin{proof}
Suppose there exists some agent $\alpha_i$ where $u_{\alpha_i}(\pi)=0$. By the construction of $(N, E)$, it must be that $\alpha_i$ has degree at least $1$ so there exists some $\alpha_j$ where $\{ \alpha_i, \alpha_j \}\in E$. It follows that $\alpha_i$ is envious of both agents in $\pi(\alpha_j)$.
\end{proof}



We now show that if the \porschenxsatvariant/ instance $C$ is exactly satisfiable then the ASHG represented by $(N, E)$ contains an envy-free partition into triples.

\begin{lem}
\label{lem:threed_efr_as_regularenvy_firstdirection}
If $C$ is exactly satisfiable then the ASHG represented by $(N, E)$ contains an envy-free partition into triples.
\end{lem}
\begin{proof}
Suppose $f$ is an exact model of $C$. We shall construct a partition into triples $\pi$ that is envy-free. For each variable $x_i$ in $X(C)$ where $f(x_i)$ is false, add $\{ w_i^1, w_i^2, w_i^3 \}$ to $\pi$. Now consider each clause $c_r = \{ x_i, x_j, x_k \}$ and the corresponding clause gadget $D_r$, labelling $i, j, k$ such that $W_i$ contains an agent adjacent to $d_r^1$, $W_j$ contains an agent adjacent to $d_r^2$, and $W_k$ contains an agent adjacent to $d_r^3$. There are three cases: $f(x_i)$ is true while both $f(x_j)$ and $f(x_k)$ are false, $f(x_j)$ is true while both $f(x_i)$ and $f(x_k)$ are false, and $f(x_k)$ is true while both $f(x_i)$ and $f(x_j)$ are false. In the first case, suppose $c_r$ contains the $u\textsuperscript{th}$ occurrence of $x_i$. Add to $\pi$ the triples $\{ w_i^u, d_r^1, d_r^4 \}$, $\{ d_r^2, d_r^5, d_r^7 \}$, and $\{ d_r^3, d_r^6, d_r^8 \}$. The constructions in the second and third cases are symmetric. In the second case, suppose $c_r$ contains the $u\textsuperscript{th}$ occurrence of $x_j$. Add to $\pi$ the triples $\{ w_j^u, d_r^2, d_r^5 \}$, $\{ d_r^1, d_r^4, d_r^6 \}$, and $\{ d_r^3, d_r^7, d_r^8 \}$. In the third case, suppose $c_r$ contains the $u\textsuperscript{th}$ occurrence of $x_k$. Add to $\pi$ the triples $\{ w_k^u, d_r^3, d_r^8 \}$, $\{ d_r^1, d_r^4, d_r^6 \}$, and $\{ d_r^2, d_r^5, d_r^7 \}$.

The construction of $\pi$ is now complete. Note that for any variable gadget $W_i$, either $\sigma(W_i, \pi) = 1$ or $\sigma(W_i, \pi) = 3$. For each clause gadget $D_r$, there exist two triples in $\pi$ that each contains three agents in $D_r$ and one triple in $\pi$ that contains two agents in $D_r$ as well as an agent in some variable gadget. There are therefore three kinds of triple in $\pi$: a triple that contains three agents belonging to the same variable gadget; a triple that contains one agent in some variable gadget as well as two agents in some clause gadget; and a triple that contains three agents in the same clause gadget. We will show that no triple of any kind contains an envious agent.

First, consider some triple $t$ in $\pi$ of the first kind, where $t = W_i$ for some variable gadget $W_i$. Since each agent in $t$ has utility $2$, no agent in $t$ is envious.

Second, consider some triple $t$ in $\pi$ of the second kind, which contains some agent $w_i^a$ in some variable gadget $W_i$ and two agents in some clause gadget $D_r$. By the construction of $\pi$, $t$ comprises either $\{ w_i^u, d_r^1, d_r^4 \}$, $\{ w_i^a, d_r^2, d_r^5 \}$, or $\{ w_i^a, d_r^3, d_r^8 \}$. Suppose $t$ comprises $\{ w_i^a, d_r^1, d_r^4 \}$. Since $u_{d_r^1}(\pi)=2$ it follows that $d_r^1$ is not envious. Since $\sigma(W_i, \pi) = 3$ and $u_{w_i^a}(\pi)\geq 1$ it follows by Lemma~\ref{lem:threed_efr_as_envyfree_util1_neighbourhoodthreetriples} that $w_i^a$ is also not envious. Similarly, since $\sigma(\mathcal{N}(d_r^4), \pi)=3$ it follows by Lemma~\ref{lem:threed_efr_as_envyfree_util1_neighbourhoodthreetriples} that $d_r^4$ is also not envious. The proof in the other cases, in which $t$ comprises either $\{ w_i^a, d_r^2, d_r^5 \}$ or $\{ w_i^a, d_r^3, d_r^8 \}$, is symmetric.

Third, consider some triple $t$ in $\pi$ of the third kind, where $t\subset D_r$ for some clause gadget $D_r$. By the construction of $\pi$, $t$ comprises either $\{ d_r^1, d_r^4, d_r^6 \}$, $\{ d_r^2, d_r^5, d_r^7 \}$, $\{ d_r^3, d_r^6, d_r^8 \}$, or $\{ d_r^3, d_r^7, d_r^8 \}$. Suppose $t$ comprises $\{ d_r^1, d_r^4, d_r^6 \}$. Since $\sigma(\mathcal{N}(d_r^1), \pi)=2$ and $u_{d_r^1}(\pi)=1$ it follows by Lemma~\ref{lem:threed_efr_as_envyfree_util1_neighbourhoodthreetriples} that $d_r^1$ is not envious. Similarly, since $\sigma(\mathcal{N}(d_r^6), \pi)=3$ and $u_{d_r^6}(\pi)=1$ it follows by Lemma~\ref{lem:threed_efr_as_envyfree_util1_neighbourhoodthreetriples} that $d_r^6$ is not envious. Since $u_{d_r^4}(\pi)=2$ it follows that $d_r^4$ is not envious. The proofs for the other three cases, in which $t$ comprises either $\{ d_r^2, d_r^5, d_r^7 \}$, $\{ d_r^3, d_r^6, d_r^8 \}$, or $\{ d_r^3, d_r^7, d_r^8 \}$, are symmetric.
\end{proof}

We now show that if the ASHG represented by $(N, E)$ contains an envy-free partition into triples then the \porschenxsatvariant/ instance $C$ is exactly satisfiable. For any partition into triples $\pi$ and any variable gadget $W_i$, if $\sigma(W_i, \pi) = 3$ then let us say that $W_i$ is \emph{open}. If $\sigma(W_i, \pi) = 1$ then let us say that $W_i$ is \emph{closed}.

\begin{lem}
\label{lem:threed_efr_as_regularenvy_seconddirection_triangle_split_stay}
If the ASHG represented by $(N, E)$ contains an envy-free partition into triples $\pi$ then any variable gadget is either open or closed.
\end{lem}
\begin{proof}
Suppose $\pi$ is an envy-free partition into triples. For each variable gadget $W_i$, since $|W_i| = 3$ by definition $1 \leq \sigma(W_i, \pi) \leq 3$. Suppose for a contradiction that there exists some variable gadget $W_i$ that is neither open nor closed, meaning $\sigma(W_i, \pi) = 2$. There must exist some triple $\{ w_i^a, w_i^b, \alpha_j \}$ in $\pi$ where $w_i^a, w_i^b \in W_i$ and $\alpha_j \notin W_i$. Label the remaining agent in $W_i$ as $w_i^c$. By the construction of $W_i$, it must be that $u_{w_i^c}(\pi)\leq 1$. It follows that $w_i^c$ envies $\alpha_j$ since $u_{w_i^c}(\pi(\alpha_j) \setminus \{ \alpha_j \}) = 2$.
\end{proof}

\begin{lem}
\label{lem:threed_efr_as_regularenvy_seconddirection}
If the ASHG represented by $(N, E)$ contains an envy-free partition into triples then $C$ is exactly satisfiable.
\end{lem}
\begin{proof}
Suppose $\pi$ is an envy-free partition into triples. By Lemma~\ref{lem:threed_efr_as_regularenvy_seconddirection_triangle_split_stay} any variable gadget is either open or closed. Construct a truth assignment $f$ in $C$ by setting $f(x_i)$ to be true if $W_i$ has an open configuration in $\pi$ and false otherwise. Each variable $x_i$ corresponds to exactly one variable gadget so it follows that $f$ is a valid truth assignment. By the construction of $(N, E)$, each clause $c_r$ corresponds to exactly one clause gadget $D_r$. Recall that each clause gadget is adjacent to three variable gadgets, which correspond to the three variables in that clause. To show that $f$ is an exact model of $C$, it is now sufficient to show that each clause gadget $D_r$ is adjacent to exactly one open variable gadget.
 
Consider an arbitrary clause gadget $D_r$ and the corresponding clause $c_r=\{ x_i, x_j, x_k \}$, labelling $i, j, k$ such that $d_r^1$ is adjacent to some agent in $W_i$, $d_r^2$ is adjacent to some agent in $W_j$ and $d_r^3$ is adjacent to some agent in $W_k$. 

First suppose for a contradiction that $D_r$ is adjacent to $0$ open variable gadgets. It follows that $W_i \in \pi$, $W_j \in \pi$, and $W_k \in \pi$. By Lemma~\ref{lem:threed_efr_as_util0}, it must be that each of the five agents in $D_r$ has utility $1$ or more. It follows that any triple that contains at least one agent in $D_r$ must either exactly two agents in $D_r$ or exactly three agents in $D_r$. Since there are five agents in $D_r$, the only possibility is that there exists some triple $\{ d_r^a, d_r^b, \alpha_m \}$ where $\alpha_m \notin D_r$. It follows that $u_{\alpha_m}(\pi) = 0$, which contradicts Lemma~\ref{lem:threed_efr_as_util0}.

Now suppose for a contradiction that either $D_r$ is adjacent to two or more open variable gadgets. Without loss of generality, assume that $W_j$ and $W_k$ are open.


Suppose $c_r$ contains the $a\textsuperscript{th}$ occurrence of $x_i$ and the $b\textsuperscript{th}$ occurrence of $x_j$. Consider $\pi(w_i^a)$. By Lemma~\ref{lem:threed_efr_as_util0}, no agent has utility $0$ in $\pi$, so $\pi(w_i^a)$ comprises either $\{ w_i^a, d_r^1, d_r^4 \}$, $\{ w_i^a, d_r^2, d_r^5 \}$, or $\{ w_i^a, d_r^3, d_r^8 \}$. Similarly, $\pi(w_i^b)$ comprises either $\{ w_i^b, d_r^1, d_r^4 \}$, $\{ w_i^b, d_r^2, d_r^5 \}$, or $\{ w_i^b, d_r^3, d_r^8 \}$. By the symmetry of the clause gadget, assume without loss of generality that $\pi(w_i^a) = \{ w_i^a, d_r^1, d_r^4 \}$ and $\pi(w_i^b) = \{ w_i^b, d_r^2, d_r^5 \}$. Now consider $d_r^6$, $d_r^7$, and $d_r^8$. Since no agent has utility $0$, the only possibility is that $\{ d_r^6, d_r^7, d_r^8 \}$ belongs to $\pi$. It then follows that $d_r^4$ envies $d_r^8$, since $u_{d_r^4}(\pi)=1 < 2 = u_{d_r^4}(\{ d_r^6, d_r^7 \})$, which is a contradiction.

\end{proof}

We have now shown that the ASHG represented by $(N, E)$ contains an envy-free partition into triples if and only if the \porschenxsatvariant/ instance $C$ is exactly satisfiable. This shows that the reduction is correct.

\begin{thm}
\label{thm:threed_efr_as_regularenvy_npcomplete}
Deciding if a given ASHG contains an envy-free partition into triples is $\NP$-complete even when preferences are binary and symmetric and the ASHG has maximum degree $3$.
\end{thm}
\begin{proof}
It is straightforward to show that this decision problem belongs to $\NP$, since for any two agents $\alpha_i, \alpha_j$ we can test if $\alpha_i$ envies $\alpha_j$ in constant time. 

We have presented a polynomial-time reduction from \porschenxsatvariant/, which is $\NP$-complete~\cite[Lemma~5]{PSSW14}. Given an arbitrary instance $C$ of \porschenxsatvariant/, the reduction constructs an ASHG represented by its underlying graph $(N, E)$ which has binary and symmetric preferences and maximum degree $3$. Together, Lemmas~\ref{lem:threed_efr_as_regularenvy_firstdirection} and~\ref{lem:threed_efr_as_regularenvy_seconddirection} show that this ASHG contains an envy-free partition into triples if and only if $C$ is exactly satisfiable, and thus that this decision problem is $\NP$-hard.
\end{proof}

\section{Weakly justified envy-freeness}
\label{sec:threed_efr_as_wjef}
In this section we consider wj-envy-freeness in ASHGs with binary and symmetric preferences.

First, in Section~\ref{sec:threed_efr_as_wjef_maxdeg2}, we consider ASHGs with maximum degree $2$. We show that in polynomial time we can either construct a wj-envy-free partition into triples or report that no such partition into triples exists (Theorem~\ref{thm:threed_efr_as_wjef_algowjpathscycles}). In fact, we present a necessary and sufficient condition for the non-existence of a wj-envy-free partition into triples in such an ASHG (in Definition~\ref{def:threed_efr_as_wjef_max_deg_2_ino}).

Next, in Section~\ref{sec:threed_efr_as_wjef_maxdeg3}, we consider ASHGs with maximum degree $3$. We show that such an ASHG may not contain a wj-envy-free partition into triples and the associated existence problem is $\NP$-complete.

\subsection{Binary and symmetric preferences with maximum degree \texorpdfstring{$2$}{2}}
\label{sec:threed_efr_as_wjef_maxdeg2}

We first define a set of ASHGs, called \iwjnomaxdegreetwofamily/, and show that membership in the set is a sufficient condition for the non-existence of a wj-envy-free partition into triples.

\begin{mydefinition}[\iwjnomaxdegreetwofamily/]
\label{def:threed_efr_as_wjef_max_deg_2_ino}
\begin{adjustwidth}{8pt}{8pt}
    An ASHG belongs to \iwjnomaxdegreetwofamily/ if and only if the underlying graph comprises a set of $k$ disjoint $4$-cycles and a single isolated agent, for any $k$ where $k \geq 2$.
    \end{adjustwidth}
\end{mydefinition}

The proof involves a sequence of lemmas. Consider an ASHG with binary and symmetric preferences that belongs to \iwjnomaxdegreetwofamily/ and its underlying graph $(\hat{N}, \hat{E})$. Label the agents in $(\hat{N}, \hat{E})$ such that $\alpha_{4k+1}$ is an isolated agent and
$(\alpha_{4i-3}, \alpha_{4i-2}, \alpha_{4i-1}, \alpha_{4i})$ is a $4$-cycle for any $i$ where $1\leq i \leq k$.

Suppose $\hat{\pi}$ is an arbitrary partition into triples.

\begin{lem}
\label{lem:threed_efr_as_wjef_maxdeg2_inoprelim}
If $\hat{\pi}$ is wj-envy-free then $u_{t}(\hat{\pi}) = 2$ for every triple $t$ in $\hat{\pi}$.
\end{lem}
\begin{proof}
Suppose for a contradiction that $\hat{\pi}$ is wj-envy-free and there exists some triple $t$ in $\hat{\pi}$ such that $u_{t}(\hat{\pi}) \neq 2$. Since preferences are binary and symmetric, it must be that $u_{t}(\hat{\pi}) \in \{ 0, 4 \}$. Without loss of generality, assume that $t$ is chosen so that $u_{t}(\hat{\pi})$ is minimised. Since $\alpha_{4k+1}$ is an isolated agent, it must be that $u_{t}(\hat{\pi}) \leq u_{\hat{\pi}(\alpha_{4k+1})}(\hat{\pi}) \leq 2$. The only possibility is that $u_{t}(\hat{\pi}) = 0$. Furthermore, without loss of generality assume that $t = \{ \alpha_1, \alpha_i, \alpha_j \}$ where $i > 4$ and $j > 4$.

Since $( \alpha_1, \alpha_2, \alpha_3, \alpha_4 )$ is a $4$-cycle, if $u_{\alpha_2}(\hat{\pi}) = 0$ then $\alpha_2$ has wj-envy for $\alpha_i$, which is a contradiction. It follows that $u_{\alpha_2}(\hat{\pi}) \geq 1$. A symmetric argument shows that $u_{\alpha_4}(\hat{\pi}) \geq 1$. The only possibility is that $\{ \alpha_2, \alpha_3, \alpha_4 \} \in \hat{\pi}$. Now $\alpha_1$ has wj-envy for $\alpha_3$, which is also a contradiction.
\end{proof}

\begin{lem}
\label{lem:threed_efr_as_wjef_maxdeg2_ino}
If an ASHG belongs to \iwjnomaxdegreetwofamily/ then it does not contain a wj-envy-free partition into triples.
\end{lem}
\begin{proof}
By Lemma~\ref{lem:threed_efr_as_wjef_maxdeg2_inoprelim} it must be that $u_{t}(\hat{\pi}) = 2$ for every triple $t$ in $\hat{\pi}$ and thus that the number of agents with utility $0$ is exactly $n$. Since $\alpha_{4k+1}$ is isolated, the number of agents in $4$-cycles with utility $0$ must be exactly $(4k + 1)/3 - 1 = (4k - 2)/3$. Since $(4k - 2)/3$ is not divisible by $4$, and the number of $4$-cycles is $k$ where $k \geq 2$, the only possibility is that there exists at least one $4$-cycle in which exactly two agents have utility $0$, and exactly two agents have utility $1$. Without loss of generality assume that this $4$-cycle is $( \alpha_1, \alpha_2, \alpha_3, \alpha_4)$ and that $u_{\alpha_1}(\hat{\pi}) = u_{\alpha_2}(\hat{\pi}) = 0$ and $\hat{\pi}(\alpha_3) = \{ \alpha_3, \alpha_4, \alpha_i \}$ where $i \geq 4$. Now $\alpha_1$ has wj-envy for $\alpha_i$, which is a contradiction.
\end{proof}

We now present an algorithm that, given an ASHG with binary and symmetric preferences and maximum degree $2$, either returns a wj-envy-free partition into triples $\pi$ or reports that the ASHG belongs to \iwjnomaxdegreetwofamily/.

In some respects the approach taken by this algorithm is straightforward. For any path or cycle that is not a $4$-cycle, it constructs as many triples as possible that each contains a path of three agents, leaving at most one or two \emph{surplus} agents per component. More care is required in the assignment of the agents in $4$-cycles to triples. For any set of three $4$-cycles, it is relatively straightforward to assign the $12$ agents to four triples in a way that ensures none of them are wj-envied as a result. The main complexity therefore stems from the case in which the number of $4$-cycles is not divisible by $3$. In this case the algorithm either uses the surplus agents or reports that the ASHG belongs to \iwjnomaxdegreetwofamily/.

The algorithm contains calls to five subroutines. In order to simplify the overall presentation, before formally describing the main algorithm we describe each subroutine and prove some related preliminary lemmas. Four of these subroutines take as input a set agents in $N$ and construct a set of triples containing some or all of the agents in that set. The final subroutine is a helper function used to shorten the the main algorithm. In what follows, assume that $(N, E)$ is the underlying graph of an arbitrary ASHG.

The first subroutine is Subroutine~\algorithmfont{nonC4Components}, shown in Algorithm~\ref{alg:3defr_wje_subroutine_nonC4s}. This subroutine takes as input a set of components $\mathcal{C}$, none of which is a $4$-cycle. It returns a pair $(T, S)$ where $T$ is a set of triples and $S$ is a set of surplus agents. For each component in $\mathcal{C}$, the corresponding set of triples in $T$ is constructed in a straightforward way by breaking up $\mathcal{C}$ into triples that each contains a path of three agents. This leaves at most two surplus agents from each component, which are added to $S$. Note that the subgraph induced by $S$ in $(N, E)$ has maximum degree $1$.

\begin{algorithm}
\textbf{Input:} a set of connected components $\mathcal{C}$, none of which is a $4$-cycle\\
\textbf{Output:} a pair $(T, S)$ where $T$ is a set of triples and $S$ is a set of agents, such that $C' = S \cup \bigcup T$ where $C'$ is the set of agents in components in $\mathcal{C}$, and if $\pi$ is an arbitrary partition into triples then for any agent $c_j$ in $\bigcup T$ if $T(c_j) \in \pi$ then $c_j$ is neither wj-envious nor wj-envied in $\pi$
\smallskip
\begin{algorithmic}
\caption{Subroutine~\algorithmfont{nonC4Components} \label{alg:3defr_wje_subroutine_nonC4s}}
\State $T \gets \varnothing$

\For{each component $C$ in $\mathcal{C}$, with consecutive agents labelled $(c_1, c_2, \dots, c_{k})$}
    \For{$i = 1$ to $\lfloor k/3 \rfloor$}
        \State $T \gets T \cup \{ \{ c_{3i-2}, c_{3i-1}, c_{3i} \} \}$
    \EndFor
    \smallskip
    
    \For{$j = 3\lfloor k/3 \rfloor$ to $k$}
        \State $S \gets S \cup \{ c_j \}$
    \EndFor
\EndFor
\smallskip

\State \Return $(T, S)$
\end{algorithmic}
\end{algorithm}

\begin{lem}
\label{lem:threed_efr_as_wjenvy_max_degree_2_nonC4stimecomplex}
Subroutine~\algorithmfont{nonC4Components} terminates in $O(|C'|)$ time.
\end{lem}
\begin{proof}
Suppose $C'$ is the set of agents in components in $\mathcal{C}$. By definition, there are $|\mathcal{C}|$ iterations of the outer ``for'' loop. In each iteration of the outer loop, the subroutine identifies some component $C$, which has $k$ agents. It it straightforward to show that each of the inner ``for'' loops involves at most $O(k)$ iterations and each iteration of each inner ``for'' loop can be performed in constant time. It follows that each iteration of the outer ``for'' loop can be performed in $O(k)$ time. It is then straightforward to show that the running time of Subroutine~\algorithmfont{nonC4Components} is $O(|C'|)$.
\end{proof}

\begin{lem}
\label{lem:threed_efr_as_max_degree_2_subgraph_nonC4s_part0}
Suppose $\pi$ is an arbitrary partition into triples and Subroutine~$\algorithmfont{nonC4Components}$ returns $(T, S)$. For any agent $c_j$ in $\bigcup T$, if $T(c_j) \in \pi$ then $c_j$ is not wj-envious in $\pi$.
\end{lem}
\begin{proof}
Suppose some agent $c_j$ in $\bigcup T$ belongs to some component $C$ in $\mathcal{C}$, which must not be a $4$-cycle. By the construction of $T$ in Subroutine~\algorithmfont{nonC4Components}, it must be that $T(c_j)$ contains either $c_{j-1}$ or $c_{j+1}$. Since $T(c_j) \in \pi$ it must be that $u_{c_j}(\pi) \geq 1$. If $c_j$ has wj-envy in $\pi$ then it must be that $u_{c_j}(\pi) = 1$ and two agents not in $\pi(c_j)$ are adjacent to $c_j$. Since $u_{c_j}(\pi) = 1$ it follows that the degree of $c_j$ in $(N, E)$ is least $3$, which is a contradiction.
\end{proof}

\begin{lem}
\label{lem:threed_efr_as_max_degree_2_subgraph_nonC4s}
Suppose $\pi$ is an arbitrary partition into triples and Subroutine~$\algorithmfont{nonC4Components}$ returns $(T, S)$. For any agent $c_j$ in $\bigcup T$, if $T(c_j) \in \pi$ then $c_j$ is not wj-envied in $\pi$.
\end{lem}
\begin{proof}
Suppose $c_j$ is an arbitrary agent in $\bigcup T$. Let $i = \lceil j/3 \rceil$. By the pseudocode of Subroutine~$\algorithmfont{nonC4Components}$, it must be that $T(c_j) = \{ c_{3i-2}, c_{3i-1}, c_{3i} \}$ and $T(c_j)$ was added to $T$ in the $i\textsuperscript{th}$ iteration of the inner ``for'' loop, in the iteration of the outer ``for'' loop relating to component $C$. Note that by definition $\{ c_{3i-2}, c_{3i-1} \} \in E$ and $\{ c_{3i-1}, c_{3i} \} \in E$. 

There are now three possibilities: $j = 3i-2$, $j=3i-1$, and $j=3i$. Suppose for a contradiction that $T(c_j) \in \pi$ and $c_j$ is wj-envied in $\pi$ by some agent $\alpha_k$ in $N$. Note that since $T(c_j) \in \pi$ it must be that $\alpha_k \notin \{ c_{3i-2}, c_{3i-1} \}$.

First, suppose that either $j = 3i-2$ or $j=3i$. Since $\alpha_k$ has wj-envy for $c_j$ in $\pi$, it must be that $v_{c_{3i-1}}(\alpha_k) \geq v_{c_{3i-1}}(c_j) = 1$. It follows that $v_{c_{3i-1}}(\alpha_k) = v_{c_{3i-1}}(c_{3i-2}) = v_{c_{3i-1}}(c_{3i}) = 1$. Since $\alpha_k \notin \{ c_{3i-2}, c_{3i-1} \}$ it follows that that the degree of $c_{3i-1}$ in $(N, E)$ is at least $3$, which is a contradiction.

Second, suppose that $j = 3i - 1$. Since $\alpha_k$ has wj-envy for $c_j$ in $\pi$, and $T(c_j) \in \pi$, it must be that $v_{c_{3i-2}}(\alpha_k) \geq v_{c_{3i-2}}(c_{3i-1}) = 1$ and $v_{c_{3i}}(\alpha_k) \geq v_{c_{3i}}(c_{3i-1}) = 1$. It follows that $v_{c_{3i-2}}(\alpha_k) = v_{c_{3i}}(\alpha_k) = 1$. The only possibility is that $( c_{3i-2}, c_{3i-1}, c_{3i}, \alpha_k)$ is a $4$-cycle, which contradicts the fact that $\mathcal{C}$ is a valid input to Subroutine~$\algorithmfont{nonC4Components}$.
\end{proof}

\begin{lem}
\label{lem:threed_efr_as_max_degree_2_subgraph_nonC4componentsiscorrect}
Subroutine~\algorithmfont{nonC4Components} is correct.
\end{lem}
\begin{proof}
By Lemma~\ref{lem:threed_efr_as_wjenvy_max_degree_2_nonC4stimecomplex}, Subroutine~\algorithmfont{nonC4Components} eventually terminates and returns a pair $(T, S)$. By the definition of Subroutine~\algorithmfont{nonC4Components}, it is straightforward to show that $C' = S \cup \bigcup T$, where $C'$ is set of agents in components in $\mathcal{C}$. By Lemmas~\ref{lem:threed_efr_as_max_degree_2_subgraph_nonC4s_part0} and~\ref{lem:threed_efr_as_max_degree_2_subgraph_nonC4s}, if $\pi$ is an arbitrary partition into triples then for any agent $c_j$ in $\bigcup T$ if $T(c_j) \in \pi$ then $c_j$ is neither wj-envious nor wj-envied in $\pi$.
\end{proof}

The second subroutine is Subroutine~\algorithmfont{oneC4TwoSingles}. This subroutine takes as input a $4$-cycle $R$ with consecutive agents labelled $(r_1, r_2, r_3, r_4)$ and two other agents $w_1$ and $w_2$. It returns $\{ \{ w_1, r_1, r_2  \}, \{ w_2, r_3, r_4 \} \}$.

\begin{lem}
\label{lem:threed_efr_as_max_degree_2_subgraph_oneC4TwoSingles}
For any partition into triples $\pi$, if Subroutine~\algorithmfont{oneC4TwoSingles} returns $T'$ and  $T'\subseteq \pi$ then no agent in $R \cup \{ w_1, w_2 \}$ is wj-envied in $\pi$.
\end{lem}
\begin{proof}
Label $R = ( r_1, r_2, r_3, r_4 )$ so $T' = \{ \{ w_1, r_1, r_2  \}, \{ w_2, r_3, r_4 \} \}$. Suppose for a contradiction that some agent $\alpha_k$ has wj-envy for some agent in $R \cup \{ w_1, w_2 \}$. By symmetry, we need only consider two cases: either $\alpha_k$ has wj-envy for $r_1$ or $\alpha_k$ has wj-envy for $w_1$. 

If $\alpha_k$ has wj-envy for $r_1$ then consider $r_2$. Since $r_2 \in \pi(r_1)$ it must be that $v_{r_2}(\alpha_k) \geq v_{r_2}(r_1) = 1$ and thus that $v_{r_2}(\alpha_k)=1$. The only possibility is that $\alpha_k = r_3$. Since $u_{r_3}(\pi) = 1 = u_{r_3}(\{ r_1, w_2 \})$ it must be that $r_3$ does not have wj-envy for $r_1$, which is a contradiction.

If $\alpha_k$ has wj-envy for $w_1$ then it must be that $u_{\alpha_k}(\{ r_1, r_2 \}) \geq 1$. The only possibility is that either $\alpha_k = r_3$ or $\alpha_k = r_4$. Since $u_{r_3}(\pi) = u_{r_4}(\pi) = 1 = u_{r_3}(\{ r_1, r_2 \}) = u_{r_4}(\{ r_1, r_2 \})$ it must be that neither $r_3$ nor $r_4$ have wj-envy for $w_1$, which is a contradiction.
\end{proof}

The third subroutine is Subroutine~\algorithmfont{multipleOfThreeC4s}, shown in Algorithm~\ref{alg:3defr_wje_subroutine_multipleOfThreeC4s}. It takes as input a set of $4$-cycles $\mathcal{R}$ where $|\mathcal{R}|=3q$ for some integer $q \geq 1$. It returns $4q$ triples, each of which contains two agents in some $4$-cycle in $\mathcal{R}$ and one agent in a different $4$-cycle in $\mathcal{R}$. The four agents in each $4$-cycle in $\mathcal{R}$ are assigned to either two or four triples.

\begin{algorithm}
\textbf{Input:} a set of $4$-cycles $\mathcal{R} = \{ R_1, R_2, \dots, R_{3q} \}$ for some integer $q$ where $q \geq 1$, such that for each $i$ where $1\leq i \leq 3q$ the consecutive agents in the $4$-cycle $R_i$ are labelled $( r_i^1, r_i^2, r_i^3, r_i^4 )$\\
\textbf{Output:} a set of triples $T$ such that $\bigcup T$ is the set of agents in components in $\mathcal{R}$ and if $\pi$ is a partition into triples where $T \subseteq \pi$ then no agent in any component in $\mathcal{R}$ is wj-envied in $\pi$ 
\smallskip
\begin{algorithmic}
\caption{Subroutine~\algorithmfont{multipleOfThreeC4s} \label{alg:3defr_wje_subroutine_multipleOfThreeC4s}}
\State $T\gets \varnothing$
    \For{$d = 1$ to $q$}
     \State $T \gets T \cup \{ \{ r_{3d-2}^1, r_{3d-2}^2, r_{3d-1}^1 \}, \{ r_{3d-2}^3, r_{3d-2}^4, r_{3d-1}^4 \}$,\\
     \hphantom{$T \gets T \cup \{\ \ $} $\{ r_{3d-1}^2, r_{3d}^1, r_{3d}^2 \}, \{ r_{3d-1}^3, r_{3d}^3, r_{3d}^4 \} \}$
    \EndFor
\smallskip
\State \Return $T$
\end{algorithmic}
\end{algorithm}

The proof of Lemma~\ref{lem:threed_efr_as_wjenvy_max_degree_2_subroutine_multipleOfThreeC4s_running_time} is straightforward.

\begin{lem}
\label{lem:threed_efr_as_wjenvy_max_degree_2_subroutine_multipleOfThreeC4s_running_time}
Subroutine~\algorithmfont{multipleOfThreeC4s} terminates in $O(|\mathcal{R}|)$ time.
\end{lem}

\begin{lem}
\label{lem:threed_efr_as_max_degree_2_subgraph_multipleOfThreeC4s}
Subroutine~\algorithmfont{multipleOfThreeC4s} is correct.
\end{lem}
\begin{proof}
By Lemma~\ref{lem:threed_efr_as_wjenvy_max_degree_2_subroutine_multipleOfThreeC4s_running_time}, Subroutine~\algorithmfont{nonC4Components} eventually terminates and returns a set of triples $T$. By the definition of Subroutine~\algorithmfont{multipleOfThreeC4s}, it suffices to show that no agent in $R_{3d} \cup R_{3d-1} \cup R_{3d-2}$ is wj-envied in $\pi$, for any $d$ where $1\leq d \leq q$. Without loss of generality assume that $d = 1$. In fact, by the symmetry of the four triples in $T$ that involve agents in $R_1$, $R_2$, and $R_3$, it suffices to show that no agent in $\{ r_{1}^1, r_{1}^2, r_{2}^1 \}$ is wj-envied in $\pi$.

First, suppose for a contradiction that some agent $\alpha_k$ has wj-envy for $r_1^1$. Since $v_{r_1^2}(r_1^1) = 1$ it must be that $v_{r_1^2}(\alpha_k) = 1$. The only possibility is that $\alpha_k = r_1^3$. By construction, $u_{r_1^3}(\pi) = 1 = u_{r_1^3}(\{ r_1^2, r_2^1 \})$ so $r_1^3$ does not have wj-envy for $r_1^1$, which is a contradiction. A symmetric argument shows that if some $\alpha_k$ has wj-envy for $r_1^2$ then it must be that $\alpha_k = r_1^4$. Since $u_{r_1^4}(\pi) = 1 = u_{r_1^4}(\{ r_1^1, r_2^1 \})$ this also leads to a contradiction.

Finally, suppose some agent $\alpha_k$ has wj-envy for $r_2^1$. It follows that $u_{\alpha_k}(\{ r_1^1, r_1^2 \}) \geq 1$, so either $\alpha_k = r_1^3$ or $\alpha_k = r_1^4$. Suppose $\alpha_k = r_1^3$. By construction, $u_{r_1^3}(\pi) = 1 = u_{r_1^3}(\{ r_1^1, r_1^2 \})$ so $r_1^3$ does not have wj-envy for $r_2^1$, which is a contradiction. A symmetric argument also leads to a contradiction if $\alpha_k = r_1^4$.
\end{proof}

The fourth subroutine is Subroutine~\algorithmfont{configureSurplusAgents}, shown in Algorithm~\ref{alg:3defr_wje_subroutine_configureSurplusAgents}. It takes as input a set of agents $\hat{S}$ where $|\hat{S}|$ is divisible by three and the subgraph induced by $\hat{S}$ in $(N, E)$ has maximum degree $1$. It returns a set of $|\hat{S}|/3$ triples. When it is called in the main algorithm, this subroutine will be given a subset of the surplus agents returned by Subroutine~\algorithmfont{nonC4Components}. Informally, in the context of the main algorithm, the goal of this subroutine is to assign the remaining surplus agents to triples such that the number of triples with non-zero utility is maximised.

We remark that the procedure of Subroutine~\algorithmfont{configureSurplusAgents} is similar to a subroutine used in a similar model that constructs a stable partition into triples~\cite{ManloveMcKay3DSRAS2021}.

\begin{algorithm}
\textbf{Input:} a set of agents $\hat{S}\subseteq N$ such that $|\hat{S}|$ is divisible by three and the maximum degree of the subgraph induced by $\hat{S}$ in $(N, E)$ is $1$\\
\textbf{Output:} a set of triples $T'$ such that $\bigcup T' = \hat{S}$ and if $\pi$ is a partition into triples where $T' \subseteq \pi$ then no agent in $\hat{S}$ has wj-envy in $\pi$ for any other agent in $\hat{S}$
\smallskip
\begin{algorithmic}
\caption{Subroutine~\algorithmfont{configureSurplusAgents} \label{alg:3defr_wje_subroutine_configureSurplusAgents}}
\State $P \gets $ the set of agents with degree $0$ in the subgraph induced by $\hat{S}$ in $(N, E)$, labelling $P = \{ p_1, p_2, \dots, p_{|P|} \}$
\State $\mathcal{Q} \gets $ a set containing each pair of agents $\{ q_i, q_j \} \subset \hat{S}$ where $\{ q_i, q_j \} \in E$, labelling $\mathcal{Q} = \{ Q_1, Q_2, \dots, Q_{|\mathcal{Q}|} \}$
\State $\mathcal{X} \gets \varnothing$
\If{$|\mathcal{Q}| \geq |\hat{S}|/3$}
    \State $\mathcal{X} \gets \{ Q_1, Q_2, \dots, Q_{|\hat{S}|/3} \}$
\Else
    \LineComment{note that $|P| > 2(|\hat{S}|/3 - |\mathcal{Q}|)$ since by definition $|P| = |\hat{S}| - 2|\mathcal{Q}|$}
    \State $\mathcal{W} \gets \{ \{ p_i, p_{2i} \} : 1 \leq i \leq |\hat{S}|/3 - |\mathcal{Q}| \}$
    \State $\mathcal{X} \gets \mathcal{Q} \cup \mathcal{W}$
\EndIf
\smallskip
\State $Y \gets \hat{S} \setminus \bigcup \mathcal{X}$
\LineComment{Label $\mathcal{X} = \{ X_1, X_2, \dots, X_{|\hat{S}|/3} \}$ and $Y = \{ y_1, y_2, \dots, y_{|\hat{S}|/3} \}$. Note that $\mathcal{X}$ is a set of pairs of agents and $Y$ is a set of individual agents. }
\State \Return $\{ X_i \cup \{ y_i \} : 1 \leq i \leq |\hat{S}|/3 \}$
\end{algorithmic}
\end{algorithm}

\begin{lem}
\label{lem:threed_efr_as_wjenvy_max_degree_2_subroutine_configureSurplusAgents_running_time}
Subroutine~\algorithmfont{configureSurplusAgents} terminates in $O(|\hat{S}|)$ time.
\end{lem}
\begin{proof}
The sets of pairs $\mathcal{X}$ and $\mathcal{W}$, the set of agents $Y$, and the returned set of triples $T$ can all be constructed in $O(|\hat{S}|)$ time.
\end{proof}

\begin{lem}
\label{lem:threed_efr_as_max_degree_2_subgraph_configureSurplusAgents}
Subroutine~\algorithmfont{configureSurplusAgents} is correct.
\end{lem}
\begin{proof}
By Lemma~\ref{lem:threed_efr_as_wjenvy_max_degree_2_subroutine_configureSurplusAgents_running_time}, Subroutine~\algorithmfont{configureSurplusAgents} eventually terminates and returns a set of triples $T'$. From the pseudocode of Subroutine~\algorithmfont{configureSurplusAgents}, it is straightforward to show that $\bigcup T' = \hat{S}$. Suppose for a contradiction that $\pi$ is an arbitrary partition into triples, $T' \subseteq \pi$, and some agent $\alpha_{j_1}$ in $\hat{S}$ has wj-envy for some agent $\alpha_{k_1}$ in $\hat{S}$, where $\pi(\alpha_{j_1}) = \{ \alpha_{j_1}, \alpha_{j_2}, \alpha_{j_3} \}$ and $\pi(\alpha_{k_1}) = \{ \alpha_{k_1}, \alpha_{k_2}, \alpha_{k_3} \}$. It follows that $u_{\alpha_{j_1}}(\{  \alpha_{k_2}, \alpha_{k_3} \}) \geq 1$. Without loss of generality assume that $v_{\alpha_{j_1}}(\alpha_{k_2})=1$, meaning $\{ \alpha_{j_1}, \alpha_{k_2} \} \in E$. By the definition of $\mathcal{Q}$, it must be that $\{ \alpha_{j_1}, \alpha_{k_2} \} \in \mathcal{Q}$.

Now, by the pseudocode of Subroutine~\algorithmfont{configureSurplusAgents}, for any pair $\{ q_a, q_b \}$ in $\mathcal{Q}$, if $|\mathcal{Q}| < |\hat{S}|/3$ then $q_a$ belongs to the same triple as $q_b$. Since $\alpha_{j_1}$ does not belong to the same triple as $\alpha_{k_2}$ it must be that $|\mathcal{Q}| \geq |\hat{S}|/3$. 

By the pseudocode, for any triple in $T'$ there exists some pair in $\mathcal{Q}$ that is a subset of that triple. It follows that either $\{ \alpha_{k_1}, \alpha_{k_2} \} \in \mathcal{Q}$, $\{ \alpha_{k_1}, \alpha_{k_3} \} \in \mathcal{Q}$, or $\{ \alpha_{k_2}, \alpha_{k_3} \} \in \mathcal{Q}$. Since $\{ \alpha_{j_1}, \alpha_{k_2} \} \in \mathcal{Q}$, and $\mathcal{Q}$ is agent-disjoint, the only possibility is that $\{ \alpha_{k_1}, \alpha_{k_3} \} \in \mathcal{Q}$. By the definition of $\mathcal{Q}$, it must be that $v_{\alpha_{k_1}}(\alpha_{k_3})=1$. Since $\alpha_{j_1}$ has wj-envy for $\alpha_{k_1}$ it must be that $v_{\alpha_{k_3}}(\alpha_{j_1}) \geq v_{\alpha_{k_3}}(\alpha_{k_1}) = 1$. In other words, $\{ \alpha_{k_3}, \alpha_{k_1} \} \in E$. Now, since $\{ \alpha_{k_3}, \alpha_{k_1} \} \in E$ and $\{ \alpha_{k_3}, \alpha_{j_1} \} \in E$ it must be that the degree of $\alpha_{k_3}$ in the subgraph induced by $\hat{S}$ in $(N, E)$ is at least $2$, which is a contradiction.
\end{proof}

The fifth subroutine is Subroutine~\algorithmfont{pickLowDegree}. This subroutine takes as input a set $S$ and integer $k\geq 1$, such that the maximum degree of the subgraph induced by $S$ in $(N, E)$ is $1$. It returns a set of $k$ agents in $S$ such that the sum of the degrees of these agents in the subgraph induced by $S$ in $(N, E)$ is minimised. Since the maximum degree of the subgraph induced by $S$ in $(N, E)$ is $1$, this subroutine can be implemented to run in $O(|N|)$ time.

We now present the main algorithm, which we call Algorithm~\algorithmfont{wjPathsCycles} and define using pseudocode in Algorithm~\ref{alg:3defr_wje_paths_cycles}. Its general procedure is as follows. First, Subroutine~\algorithmfont{nonC4Components} is used to break up components that are not $4$-cycles into a set of triples $T$ and a set of surplus agents $S$, such that each triple in $T$ contains a path of three agents. The algorithm then constructs a set $\mathcal{R}$  that contains the $4$-cycles. If $|\mathcal{R}|$ is not divisible by three then execution enters either the first or second branch of the outermost ``if'' statement. In the first branch, the algorithm either identifies that the ASHG belongs to \iwjnomaxdegreetwofamily/ or first identifies four agents, which may be surplus. Subroutine~\algorithmfont{multipleOfThreeC4s} is then used to assign both these four agents and the agents in two of the $4$-cycles, labelled $R_1$ and $R_2$, to four triples. In the second branch, the algorithm first identifies two surplus agents. Subroutine~\algorithmfont{multipleOfThreeC4s} is then used to assign both these two surplus agents and the agents in one of the $4$-cycles, labelled $R_1$, to two triples. If $|\mathcal{R}|$ is divisible by three then execution enters the third branch. 

\vspace*{4mm}
\begin{algorithm}
\caption{Algorithm~\algorithmfont{wjPathsCycles} \label{alg:3defr_wje_paths_cycles}}
\parbox[t]{\textwidth}{
\textbf{Input:} an ASHG with binary and symmetric preferences and maximum degree $2$, represented by its underlying graph $(N, E)$\\
\noindent\textbf{Output:} either a wj-envy-free partition into triples $\pi$ or ``belongs to \iwjnomaxdegreetwofamily/''}
\smallskip
\begin{algorithmic}
\State $\pi \gets \varnothing$; $\hat{T} \gets \varnothing$; $\hat{S} \gets \varnothing$
\State $\mathcal{C} \gets$ the set of all components in $(N, E)$ that are not $4$-cycles
\State $(T, S) \gets \algorithmfont{nonC4Components}(\mathcal{C})$
\State $\mathcal{R} \gets$ the set of $4$-cycles in $(N, E)$, labelling $\mathcal{R} = \{ R_1, R_2, \dots, R_{|\mathcal{R}|} \}$
\State $l \gets 0$

\If{$|\mathcal{R}| \bmod 3 = 2$}
    \If{$|S|\geq 4$}
        \State $\{ w_1, w_2, w_3, w_4 \} \gets \algorithmfont{pickLowDegree}(S, 4)$
        \State $\hat{S} \gets S \setminus \{ w_1, w_2, w_3, w_4 \}$
        \State $\hat{T} \gets T$
    \ElsIf{$|T| \geq 1$}
        \LineComment{note that $|S|=1$ by Proposition~\ref{prop:threed_efr_as_wjenvy_maxdeg2_algo_rmod3is2_specialcase}}
        \State $w_1 \gets$ the agent in $S$
        \State $\hat{t} \gets$ some triple in $T$
        \State $\{ w_2, w_3, w_4 \} \gets \hat{t}$
        \State $\hat{S} \gets \varnothing$
        \State $\hat{T} \gets T \setminus \hat{t}$
    \Else
        \State \Return ``ASHG belongs to \iwjnomaxdegreetwofamily/''
    \EndIf
    \smallskip
    
    \State $\pi \gets \pi \cup \algorithmfont{oneC4TwoSingles}(R_1, w_1, w_2) \cup \algorithmfont{oneC4TwoSingles}(R_2, w_3, w_4)$
    \State $l \gets 2$
\ElsIf{$|\mathcal{R}| \bmod 3 = 1$} 
    \LineComment{note that $|S| \geq 2$ by Proposition~\ref{prop:threed_efr_as_wjenvy_maxdeg2_algo_mod3is1}}
    \State $\{ w_1, w_2 \} \gets \algorithmfont{pickLowDegree}(S, 2)$
    \State $\hat{S} \gets S \setminus \{ w_1, w_2 \}$
    \State $\hat{T} \gets T$
    \State $\pi \gets \pi \cup \algorithmfont{oneC4TwoSingles}(R_1, w_1, w_2)$
    \State $l \gets 1$
\Else 
    \LineComment{it must be that $|\mathcal{R}| \bmod 3 = 0$} 
    \State $\hat{S} \gets S$
    \State $\hat{T} \gets T$
\EndIf
\smallskip

\LineComment{it must be that $3$ divides $(|\mathcal{R}| - l)$} 
\State $\mathcal{R}' \gets \{ R_{l+1}, R_{l+2}, \dots, R_{|\mathcal{R}|} \}$
\State $\pi \gets \pi \cup \algorithmfont{multipleOfThreeC4s}(\mathcal{R}') \cup \hat{T} \cup \algorithmfont{configureSurplusAgents}(\hat{S})$
\State \Return $\pi$
\end{algorithmic}
\end{algorithm}
\vspace*{4mm}

After execution leaves the ``if'' statement, the algorithm constructs a set of $4$-cycles $\mathcal{R}'$ where $|\mathcal{R}'|$ is divisible by three. It is straightforward to then assign the agents in $4$-cycles in $\mathcal{R}'$ to triples using Subroutine~\algorithmfont{multipleOfThreeC4s}. Finally, any remaining surplus agents, in $\hat{S}$, are assigned to triples using Subroutine~\algorithmfont{configureSurplusAgents}.

Proposition~\ref{prop:threed_efr_as_wjenvy_maxdeg2_algo_ucupbigcuppi} follows immediately by Lemma~\ref{lem:threed_efr_as_max_degree_2_subgraph_nonC4componentsiscorrect}.

\begin{prop}
\label{prop:threed_efr_as_wjenvy_maxdeg2_algo_ucupbigcuppi}
In Algorithm~\algorithmfont{wjPathsCycles}, $S \cup \bigcup T$ is the set of agents that do not belong to $4$-cycles in $(N, E)$.
\end{prop}

We now prove two propositions that show that, in two specific cases, $S$ is large enough to extract the number of surplus agents required.

\begin{prop}
\label{prop:threed_efr_as_wjenvy_maxdeg2_algo_rmod3is2_specialcase}
In Algorithm~\algorithmfont{wjPathsCycles}, after initialising $\mathcal{R}$, if $|\mathcal{R}| \bmod 3 = 2$ and $|S| < 4$ then $|S| = 1$.
\end{prop}
\begin{proof}
Suppose $|\mathcal{R}| \bmod 3 = 2$ and $|S| < 4$ after initialising $\mathcal{R}$. Then there exists some constant $k_1 \geq 0$ such that $|\mathcal{R}| = 3{k_1} + 2$, so the number of agents in $N$ that belong to $4$-cycles is $4|\mathcal{R}| = 12{k_1} + 8$. It follows that the number of agents in $N$ that do not belong to $4$-cycles is $3n - 12{k_1} - 8$. Since $(3n - 12{k_1} - 8) \bmod 3 = 1$ there exists some constant $k_2 \geq 0$ such that the number of agents in $N$ that do not belong to $4$-cycles is $3{k_2} + 1$. By Proposition~\ref{prop:threed_efr_as_wjenvy_maxdeg2_algo_ucupbigcuppi}, $S \cup \bigcup T$ is the set of agents in $N$ that do not belong to $4$-cycles. Since $|S \cup \bigcup T| = 3{k_2} + 1$, $T$ is a set of disjoint triples, and $|S| < 4$, it must be that ${k_2}=0$ and $|S|=1$.
\end{proof}

\begin{prop}
\label{prop:threed_efr_as_wjenvy_maxdeg2_algo_mod3is1}
In Algorithm~\algorithmfont{wjPathsCycles}, after initialising $\mathcal{R}$, if $|\mathcal{R}| \bmod 3 = 1$ then $|S| \geq 2$.
\end{prop}
\begin{proof}
Suppose $|\mathcal{R}| \bmod 3 = 1$ after initialising $\mathcal{R}$. Then there exists some constant ${k_1} \geq 0$ such that $|\mathcal{R}| = 3{k_1} + 1$, so the number of agents in $N$ that belong to $4$-cycles is $4|\mathcal{R}| = 12{k_1} + 4$. It follows that the number of agents in $N$ that do not belong to $4$-cycles is $3n - 12{k_1} - 4$. Since $(3n - 12{k_1} - 4) \bmod 3 = 2$ there exists some constant ${k_2} \geq 0$ where the number of agents in $N$ that do not belong to $4$-cycles is $3{k_2} + 2$. By Proposition~\ref{prop:threed_efr_as_wjenvy_maxdeg2_algo_ucupbigcuppi}, $S \cup \bigcup T$ is the set of agents in $N$ that do not belong to $4$-cycles. Since $|S \cup \bigcup T| = 3{k_2} + 2$ and $T$ is a set of disjoint triples it must be that $|S| \geq 2$.
\end{proof}

We now show that Algorithm~\algorithmfont{wjPathsCycles} is bound to terminate and has a linear running time with respect to the number of agents.

\begin{lem}
\label{lem:threed_efr_as_wjenvy_max_degree_2_algo_valid_runningtime}
Algorithm~\algorithmfont{wjPathsCycles} terminates in $O(|N|)$ time.
\end{lem}
\begin{proof}
The pseudocode describes the algorithm at a high level. To analyse the worst-case asymptotic time complexity we describe one possible system of data structures and analyse the algorithm with respect to the number of basic operations on these data structures. We begin the analysis at the start of the pseudocode.

The initialisation of $\pi$, $\hat{T}$, and $\hat{S}$ can be performed in constant time. The set of components $\mathcal{C}$ that are not $4$-cycles can be identified in $O(|N|)$ time using breadth-first search, since the maximum degree of $(N, E)$ is $2$.

It is straightforward to show, using Lemma~\ref{lem:threed_efr_as_wjenvy_max_degree_2_nonC4stimecomplex}, that the call to Subroutine~\algorithmfont{nonC4Components} takes $O(|N|)$ time.

Like $\mathcal{C}$, the set of components $\mathcal{R}$ that are $4$-cycles can be constructed in $O(|N|)$ time. Each nested branch of the ``if/else'' statement involves removing a constant number of elements from $S$, at most two calls to Subroutine~\algorithmfont{oneC4TwoSingles} (which has constant running time), and an assignment to $\hat{T}$ and $\hat{S}$ (which can be performed in $O(|N|)$ time). It follows that the total running time of the ``if/else'' statement is $O(|N|)$.

By Lemma~\ref{lem:threed_efr_as_wjenvy_max_degree_2_subroutine_multipleOfThreeC4s_running_time}, Subroutine~\algorithmfont{multipleOfThreeC4s} has $O(|\mathcal{R}|) = O(|N|)$ running time. By Lemma~\ref{lem:threed_efr_as_wjenvy_max_degree_2_subroutine_configureSurplusAgents_running_time}, Subroutine~\algorithmfont{configureSurplusAgents} has $O(|S|) = O(|N|)$ running time. It follows that the asymptotic worst-case running time of Algorithm~\algorithmfont{wjPathsCycles} is $O(|N|)$.
\end{proof}

Having established that Algorithm~\algorithmfont{wjPathsCycles} is bound to terminate, we prove its correctness using a sequence of lemmas. First we show that if the ASHG belongs to \iwjnomaxdegreetwofamily/ then the algorithm correctly identifies it as such.

\begin{lem}
\label{lem:threed_efr_as_wjenvy_max_degree_2_algo_valid_part1_ino}
If the ASHG represented by $(N, E)$ belongs to \iwjnomaxdegreetwofamily/ then Algorithm~\algorithmfont{wjPathsCycles} returns ``ASHG belongs to \iwjnomaxdegreetwofamily/''.
\end{lem}
\begin{proof}
Suppose the ASHG represented by $(N, E)$ belongs to \iwjnomaxdegreetwofamily/. In the algorithm, the set of components $\mathcal{C}$ that are not $4$-cycles contains exactly one element $C_1$ where $C_1$ contains a single agent $c_1$. By Lemma~\ref{lem:threed_efr_as_max_degree_2_subgraph_nonC4componentsiscorrect}, Subroutine~\algorithmfont{nonC4Components} must return $(\varnothing, \{ c_1 \})$ so $T = \varnothing$ and $S = \{ c_1 \}$. Consider the outermost ``if/else'' statement in the algorithm. By the definition of \iwjnomaxdegreetwofamily/ (Definition~\ref{def:threed_efr_as_wjef_max_deg_2_ino}), it must be that $4|\mathcal{R}| + 1 \bmod 3 = 0$. It follows that $|\mathcal{R}| + 1 \bmod 3 = 0$ and thus that $|\mathcal{R}| \bmod 3 = 2$. It follows that the algorithm enters the first branch of the outermost ``if/else'' statement. Since $|S| = 1 < 4$ and $T = \varnothing$ the algorithm must then return ``ASHG belongs to \iwjnomaxdegreetwofamily/''.
\end{proof}

We now consider the case in which the ASHG represented by $(N, E)$ does not belong to \iwjnomaxdegreetwofamily/.

\begin{lem}
\label{lem:threed_efr_as_wjenvy_max_degree_2_algo_valid_part1pointfive_pit}
If the ASHG represented by $(N, E)$ does not belong to \iwjnomaxdegreetwofamily/ then Algorithm~\algorithmfont{wjPathsCycles} returns a partition into triples $\pi$.
\end{lem}
\begin{proof}
Consider an arbitrary component $C$ in $(N, E)$. Let $k$ be the number of agents in $C$. We show that each agent in $C$ is added to exactly one triple in $\pi$. 

Suppose $C$ is not a $4$-cycle. By the definition of Algorithm~\algorithmfont{wjPathsCycles}, exactly one call is made to Subroutine~\algorithmfont{nonC4Components} with argument $C$. Consider an arbitrary agent $c_i$ in $C$. There are two cases: either $i \leq \lfloor k/3 \rfloor$ or $i > \lfloor k/3 \rfloor$. In the former case, exactly one triple containing $c_i$ is added to $T$ in Subroutine~\algorithmfont{nonC4Components}, which is then added to $\pi$ in the main algorithm. In the latter case, $c_i$ is eventually added to $S$. We can see from Subroutine~\algorithmfont{configureSurplusAgents} that $c_i$ is therefore eventually added to exactly one triple in $\pi$.

Suppose $C$ is a $4$-cycle, meaning $C \in \mathcal{R}$. If $C \in \mathcal{R}'$ then each agent in $C$ is added to exactly one triple in $\pi$ in some call to Subroutine~\algorithmfont{multipleOfThreeC4s}. If $C \notin \mathcal{R}'$ then some call to Subroutine~\algorithmfont{oneC4TwoSingles} is made with the first argument equal to $C$ and the returned set of two triples is then added to $\pi$. It follows that each agent in each $4$-cycle is added to exactly one triple in $\pi$.
\end{proof} 

We now show that if the ASHG represented by $(N, E)$ does not belong to \iwjnomaxdegreetwofamily/ then the algorithm returns a partition into triples $\pi$ that is wj-envy-free. In the next four lemmas we consider certain subsets of $N$ and show that in each subset no agent is wj-envied in $\pi$.

\begin{lem}
\label{lem:threed_efr_as_wjenvy_max_deg_2_no_agent_in_bigcupT_is_wjenvied}
If Algorithm~\algorithmfont{wjPathsCycles} returns a partition into triples $\pi$ then no agent in $\bigcup T$ is wj-envied in $\pi$.
\end{lem}
\begin{proof}
Suppose Algorithm~\algorithmfont{wjPathsCycles} has returned some partition into triples~$\pi$. Consider an arbitrary triple $t$ in $T$. By the definition of Algorithm~\algorithmfont{wjPathsCycles} there are two possibilities: either $t \in \hat{T}$ or $t$ was labelled $\hat{t}$. If $t \in \hat{T}$ then by Lemma~\ref{lem:threed_efr_as_max_degree_2_subgraph_nonC4s} no agent in $t$ is wj-envied in $\pi$. Suppose then that $t$ was labelled $\hat{t}$. By Algorithm~\algorithmfont{wjPathsCycles}, for any agent $c_i$ in $\hat{t}$ it must be that some call was made to Subroutine~\algorithmfont{oneC4TwoSingles} in which the second or third argument was equal to $c_i$ and then the two triples returned by the subroutine were added to $\pi$. By Lemma~\ref{lem:threed_efr_as_max_degree_2_subgraph_oneC4TwoSingles}, it follows that no agent in $\hat{t}$ is wj-envied in $\pi$.
\end{proof}

\begin{lem}
\label{lem:threed_efr_as_wjenvy_max_deg_2_no_agent_in_Uhat_is_wjenvied}
If Algorithm~\algorithmfont{wjPathsCycles} returns a partition into triples $\pi$ then no agent in $\hat{S}$ is wj-envied in $\pi$.
\end{lem}
\begin{proof}
Suppose Algorithm~\algorithmfont{wjPathsCycles} has returned some partition into triples $\pi$ in which some agent $\alpha_i$ has wj-envy for some agent $\hat{s}_{j_1}$ in $\hat{S}$. By the pseudocode, it must be that $\pi(\hat{s}_{j_1})$ contains three agents in $\hat{S}$ so we label $\pi(\hat{s}_{j_1}) = \{ \hat{s}_{j_1}, \hat{s}_{j_2}, \hat{s}_{j_3} \}$. Note that since $|\hat{S}| > 0$ it must be that $\hat{T} = T$.

Since $\alpha_i$ has wj-envy for $\hat{s}_{j_1}$ it must be that $u_{\alpha_i}(\{ \hat{s}_{j_2}, \hat{s}_{j_3} \}) \geq 1$ so without loss of generality assume that $\{ \alpha_i, \hat{s}_{j_2} \} \in E$.  We now consider two possibilities: either $\alpha_i \in S$ or $\alpha_i \notin S$.

First, suppose $\alpha_i \in S$. If $\alpha_i \in \hat{S}$ then Lemma~\ref{lem:threed_efr_as_max_degree_2_subgraph_configureSurplusAgents} is contradicted, so it must be that $\alpha_i \in S \setminus \hat{S}$. By the pseudocode, $\alpha_i$ was labelled either $w_1$, $w_2$, $w_3$ or $w_4$ during algorithm execution and must belong to some set of agents returned by a call to Subroutine~\algorithmfont{pickLowDegree}. Since $\{ \alpha_i, \hat{s}_{j_2}  \} \subset S$ the degree of $\alpha_i$ in the subgraph induced by $S$ in $(N, E)$ is $1$. By the definition of Subroutine~\algorithmfont{pickLowDegree} it must be that the degree of each agent in $\hat{S}$ in $(N, E)$ is also $1$. Now consider the call $\algorithmfont{configureSurplusAgents}(\hat{S})$ and the execution within Subroutine~\algorithmfont{configureSurplusAgents}. Since the degree of each agent in $\hat{S}$ in $(N, E)$ is $1$, it must be that $|\mathcal{Q}| = |\hat{S}|/2 \geq |\hat{S}|/3$ so $\mathcal{X} \subset \mathcal{Q}$. It follows that, by the definition of Subroutine~\algorithmfont{configureSurplusAgents}, each triple in the set of triples returned by this subroutine contains two agents that are adjacent. It follows that either $\{ \hat{s}_{j_1}, \hat{s}_{j_2} \} \in E$, $\{ \hat{s}_{j_2}, \hat{s}_{j_3} \} \in E$, or $\{ \hat{s}_{j_1}, \hat{s}_{j_3} \} \in E$. If $\{ \hat{s}_{j_1}, \hat{s}_{j_2} \} \in E$ or $\{ \hat{s}_{j_2}, \hat{s}_{j_3} \} \in E$ then the degree of $\hat{s}_{j_2}$ in the subgraph induced by $S$ in $(N, E)$ is $2$, which is a contradiction. It remains that $\{ \hat{s}_{j_1}, \hat{s}_{j_3} \} \in E$. Since $\alpha_i$ has wj-envy for $\hat{s}_{j_1}$ it must be that $v_{\hat{s}_{j_3}}(\alpha_i) \geq v_{\hat{s}_{j_3}}(\hat{s}_{j_1}) = 1$. It follows that the degree of $\hat{s}_{j_3}$ in the subgraph induced by $S$ in $(N, E)$ is at least $2$, which is a contradiction.

Second, suppose $\alpha_i \notin S$. Since $\{ \alpha_i, \hat{s}_{j_2} \} \in E$, by the definition of Algorithm~\algorithmfont{wjPathsCycles} it must be that $\alpha_i$ belongs to the same component in $(N, E)$ as $\hat{s}_{j_2}$. Since $\hat{s}_{j_2}\in S$, by the pseudocode of Algorithm~\algorithmfont{wjPathsCycles} it must be that the component that contains $\alpha_i$ and $\hat{s}_{j_2}$ is not a $4$-cycle so belongs to $\mathcal{C}$. Since $\alpha_i \notin S$, by Lemma~\ref{lem:threed_efr_as_max_degree_2_subgraph_nonC4componentsiscorrect} it must be that some triple in $T$ contains $\alpha_i$. Since $T = \hat{T}$ it follows that $T(c_j) = \hat{T}(c_j)$ belongs to $\pi$, so by Lemma~\ref{lem:threed_efr_as_max_degree_2_subgraph_nonC4s_part0} $\alpha_i$ is not wj-envious in $\pi$, which is a contradiction.
\end{proof}

\begin{lem}
\label{lem:threed_efr_as_wjenvy_max_deg_2_no_agent_in_U_is_wjenvied}
If Algorithm~\algorithmfont{wjPathsCycles} returns a partition into triples $\pi$ then no agent in $S$ is wj-envied in $\pi$.
\end{lem}
\begin{proof}
Suppose Algorithm~\algorithmfont{wjPathsCycles} has returned some partition into triples $\pi$. Consider an arbitrary agent $s_i$ in $S$. If $s_i \in \hat{S}$ then by Lemma~\ref{lem:threed_efr_as_wjenvy_max_deg_2_no_agent_in_Uhat_is_wjenvied} it must be that $s_i$ is not wj-envied in $\pi$. It remains that $s_i \notin \hat{S}$. There are three cases: either $|\mathcal{R}|\bmod 3 = 2$, $|S|\geq 4$, and $s_i$ was labelled $w_1$, $w_2$, $w_3$, or $w_4$; $|\mathcal{R}|\bmod 3 = 2$, $|T| \geq 1$, and $s_i$ was labelled $w_1$; or $|\mathcal{R}|\bmod 3 = 1$ and $s_i$ was labelled either $w_1$ or $w_2$. In each of the three cases, some call was then made to Subroutine~\algorithmfont{oneC4TwoSingles} in which the second or third argument was $s_i$ and then two triples returned by the subroutine were added to $\pi$. By Lemma~\ref{lem:threed_efr_as_max_degree_2_subgraph_oneC4TwoSingles} it follows that $s_i$ is not wj-envied in $\pi$.
\end{proof}

\begin{lem}
\label{lem:threed_efr_as_wjenvy_max_deg_2_no_agent_in_bigcupR_is_wjenvied}
If Algorithm~\algorithmfont{wjPathsCycles} returns a partition into triples $\pi$ then no agent in any component in $\mathcal{R}$ is wj-envied in $\pi$.
\end{lem}
\begin{proof}
Consider an arbitrary $R_j$ in $\mathcal{R}$. We show that no agent in $R_j$ is wj-envied in $\pi$. In this case, let $l'$ be the final value assigned to the variable $l$ before the algorithm terminated. There are two possibilities: either $j>l'$ or $j\leq l'$.

Suppose $j>l'$. It must be that $R_j \in \mathcal{R}'$, by the construction of $\mathcal{R}'$ in Algorithm~\algorithmfont{wjPathsCycles}. By Lemma~\ref{lem:threed_efr_as_max_degree_2_subgraph_multipleOfThreeC4s}, it follows that no agent in $R_j$ is wj-envied in $\pi$.

It remains that $j\leq l'$. By the design of Algorithm~\algorithmfont{wjPathsCycles} there are two possibilities: either $|\mathcal{R}| \bmod 3 = 2$ and $l'=2$, or $|\mathcal{R}| \bmod 3 = 1$ and $l'=1$. In either case, by the definition of Algorithm~\algorithmfont{wjPathsCycles} it must be that some call to Subroutine~\algorithmfont{oneC4TwoSingles} was made with the first argument $R_j$, after which the returned set of two triples was added to $\pi$. It follows by Lemma~\ref{lem:threed_efr_as_max_degree_2_subgraph_oneC4TwoSingles} that no agent in $R_j$ is wj-envied in $\pi$.
\end{proof}

\begin{lem}
\label{lem:threed_efr_as_wjenvy_max_degree_2_algo_valid_part2_EF}
If the ASHG represented by $(N, E)$ does not belong to \iwjnomaxdegreetwofamily/ then Algorithm~\algorithmfont{wjPathsCycles} returns a partition into triples $\pi$ that is wj-envy-free.
\end{lem}
\begin{proof}
By Lemma~\ref{lem:threed_efr_as_wjenvy_max_degree_2_algo_valid_part1pointfive_pit}, Algorithm~\algorithmfont{wjPathsCycles} returns a partition into triples $\pi$. By Lemma~\ref{lem:threed_efr_as_max_degree_2_subgraph_nonC4componentsiscorrect}, $S \cup \bigcup T$ contains each agent in a component in $\mathcal{C}$. It follows by Lemmas~\ref{lem:threed_efr_as_wjenvy_max_deg_2_no_agent_in_U_is_wjenvied} and~\ref{lem:threed_efr_as_wjenvy_max_deg_2_no_agent_in_bigcupT_is_wjenvied} that no agent in $\mathcal{C}$ is wj-envied in $\pi$. In addition, by Lemma~\ref{lem:threed_efr_as_wjenvy_max_deg_2_no_agent_in_bigcupR_is_wjenvied}, no agent in any component in $\mathcal{R}$ is wj-envied in $\pi$. Since $\mathcal{C} \cup \mathcal{R}$ is the set of all components in $(N, E)$, it then follows that no agent in $N$ is wj-envied in $\pi$.
\end{proof}

We can now prove the main theorem.

\begin{thm}
\label{thm:threed_efr_as_wjef_algowjpathscycles}
Consider an ASHG with binary and symmetric preferences and maximum degree $2$. There exists an $O(|N|)$-time algorithm that either finds a wj-envy-free partition into triples in the instance or reports that the instance belongs to \iwjnomaxdegreetwofamily/, and thus contains no such partition.
\end{thm}
\begin{proof}
Lemma~\ref{lem:threed_efr_as_wjenvy_max_degree_2_algo_valid_runningtime} shows that Algorithm~\algorithmfont{wjPathsCycles} terminates in $O(|N|)$ time. Lemmas~\ref{lem:threed_efr_as_wjenvy_max_degree_2_algo_valid_part1_ino} and~\ref{lem:threed_efr_as_wjenvy_max_degree_2_algo_valid_part2_EF} establish the correctness of this algorithm and show that the algorithm either returns a wj-envy-free partition into triples or reports ``ASHG belongs to \iwjnomaxdegreetwofamily/''. In the latter case, Lemma~\ref{lem:threed_efr_as_wjef_maxdeg2_ino} shows that this ASHG contains no wj-envy-free partition into triples.
\end{proof}

\subsection{Binary and symmetric preferences with maximum degree \texorpdfstring{$3$}{3}}
\label{sec:threed_efr_as_wjef_maxdeg3}

As before in Section~\ref{sec:threed_efr_as_envyfreeness_maxdeg3}, in this section we consider ASHGs with binary and symmetric preferences and maximum degree $3$. We show that deciding if a given ASHG contains a wj-envy-free partition into triples is $\NP$-complete even when preferences are binary and symmetric and the maximum degree is $3$.

Also as before, we reduce from \porschenxsatvariant/. Recall that by definition $|X(C)|=m$. In this section we assume that the number of clauses $m$ satisfies $m=4l$ for some integer $l\geq 1$. We can show that \porschenxsatvariant/ remains $\NP$-complete under this restriction as follows. Construct four distinct copies of the set of variables $X(C)$ and formula $C$. Construct a new formula $C'$ as the union of the four copies of $C$. It is straightforward to show that $C'$ is exactly satisfiable if and only if each of the four copies is exactly satisfiable, which is true if and only if the original formula $C$ is exactly satisfiable. Note that since $|X(C)|=m=4l$ it must be that $l$ is divisible by $3$.

The overall design of this reduction is similar to the analogous reduction in Section~\ref{sec:threed_efr_as_envyfreeness_maxdeg3}. The main difference between the two is in that here we associate true literals with closed variable gadgets and false literals with open variable gadgets. Another difference is that we construct a number of so-called garbage collector gadgets. 

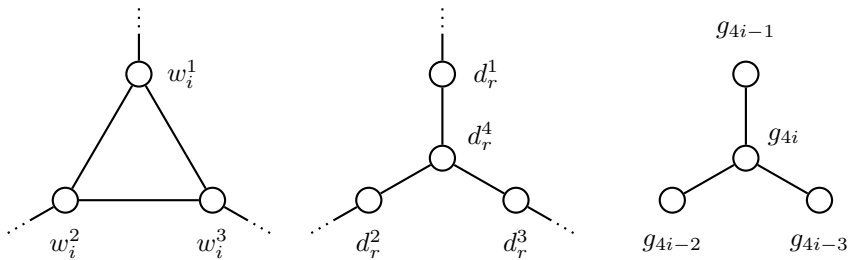
\begin{figure}[ht]
    \centering
    \begin{tikzpicture}
\begin{scope}[every node/.style={circle,draw, minimum size=2.4mm}, scale=0.8]
    \begin{scope}
        \begin{scope}[shift={(-5.0, 0.0)}]
            \node[thick, circle, label={[label distance=0.4cm]0:$w_i^1$}] (v1) at (0,{1.0*1.4}) {};
            \node[thick, circle, label={[label distance=0.4cm]270:$w_i^2$}] (v2) at ({-0.866*1.4},{-0.5*1.4}) {};
            \node[thick, circle, label={[label distance=0.4cm]270:$w_i^3$}] (v3) at ({0.866*1.4},{-0.5*1.4}) {};]
            
            \node[draw=none] (v1a) at (0,{1.0*2.5}) {};
            \node[draw=none] (v2a) at ({-0.866*2.5},{-0.5*2.5}) {};
            \node[draw=none] (v3a) at ({0.866*2.5},{-0.5*2.5}) {};
            
        \end{scope}
        
        \begin{scope}[shift={(0.0, 0.0)}]
            \node[thick, circle, label={[label distance=0.4cm]0:$d_r^1$}] (dr1) at (0,{1.0*1.4}) {};
            \node[thick, circle, label={[label distance=0.4cm]270:$d_r^2$}] (dr2) at ({-0.866*1.4},{-0.5*1.4}) {};
            \node[thick, circle, label={[label distance=0.4cm]270:$d_r^3$}] (dr3) at ({0.866*1.4},{-0.5*1.4}) {};]
            \node[thick, circle, label={[label distance=0.4cm]30:$d_r^4$}] (dr4) at ({0.0},{0.0}) {};]
            
            \node[draw=none] (dr1a) at (0,{1.0*2.5}) {};
            \node[draw=none] (dr2a) at ({-0.866*2.5},{-0.5*2.5}) {};
            \node[draw=none] (dr3a) at ({0.866*2.5},{-0.5*2.5}) {};
            
        \end{scope}
        
        \begin{scope}[shift={(5.0, 0.0)}]
            \node[thick, circle, label={[label distance=0.4cm]90:$g_{4i-1}$}] (gi1) at (0,{1.0*1.4}) {};
            \node[thick, circle, label={[label distance=0.4cm]270:$g_{4i-2}$}] (gi2) at ({-0.866*1.4},{-0.5*1.4}) {};
            \node[thick, circle, label={[label distance=0.4cm]270:$g_{4i-3}$}] (gi3) at ({0.866*1.4},{-0.5*1.4}) {};]
            \node[thick, circle, label={[label distance=0.4cm]30:$g_{4i}$}] (gi4) at ({0.0},{0.0}) {};]
            
            
        \end{scope}
        
    \end{scope}

\end{scope}

\begin{scope}
    \foreach \from/\to in {v1/v2, v2/v3, v3/v1}
        \draw [thick] (\from) -- (\to);
        
    \path[-sss,thick] (v1) edge (v1a);
    \path[-sss,thick] (v2) edge (v2a);
    \path[-sss,thick] (v3) edge (v3a);
        
    \foreach \from/\to in {dr1/dr4, dr2/dr4, dr3/dr4}
        \draw [thick] (\from) -- (\to);

    \path[-sss,thick] (dr1) edge (dr1a);
    \path[-sss,thick] (dr2) edge (dr2a);
    \path[-sss,thick] (dr3) edge (dr3a);
        
    \foreach \from/\to in {gi1/gi4, gi2/gi4, gi3/gi4}
        \draw [thick] (\from) -- (\to);
        
    
    

\end{scope}
\end{tikzpicture}
    \caption[The reduction from \porschenxsatvariant/ to the problem of deciding if a given ASHG contains a wj-envy-free partition into triples]{The reduction from \porschenxsatvariant/ to the problem of deciding if a given ASHG contains a wj-envy-free partition into triples. A variable gadget $W_i$, clause gadget $D_r$, and garbage collector gadget $G_i$ are represented as undirected graphs.}
    \label{fig:threed_efr_as_wj_envy_free_reduction}
\end{figure}

The reduction, illustrated in Figure~\ref{fig:threed_efr_as_wj_envy_free_reduction}, is as follows. Suppose $C$ is an arbitrary instance of \porschenxsatvariant/. We shall construct an ASHG represented by an underlying graph $(N, E)$.

For each variable $x_i$ in $X(C)$, construct a set of three agents $W_i = \{ w_i^1, w_i^2, w_i^3 \}$, which we refer to as the \emph{$i\textsuperscript{th}$ variable gadget}. Add edges $\{ w_i^1, w_i^2 \}$, $\{ w_i^2, w_i^3 \}$, and $\{ w_i^3, w_i^1 \}$. Next, for each clause $c_r$ in $C$ construct a set of four agents $D_r = \{ d_r^1, d_r^2, d_r^3, d_r^4 \}$, which we refer to as the \emph{$r\textsuperscript{th}$ clause gadget}. Add edges $\{ d_r^1, d_r^4 \}$, $\{ d_r^2, d_r^4 \}$, and $\{ d_r^3, d_r^4 \}$. Construct a set of $12l$ agents labelled $g_1, g_2, \dots, g_{12l}$. For any $i$ where $1\leq i\leq 3l$, we shall refer to $G_i = \{ g_{4i-3}, g_{4i-2}, g_{4i-1}, g_{4i} \}$ as the \emph{$i\textsuperscript{th}$ garbage collector gadget}. For each such $i$, Add edges $\{ g_{4i}, g_{4i-1} \}$, $\{ g_{4i}, g_{4i-2} \}$, and $\{ g_{4i}, g_{4i-3} \}$. We remark that there are now $40l$ agents.


We shall connect the variable and clause gadgets in a similar way as in the reduction in Section~\ref{sec:threed_efr_as_envyfreeness_maxdeg3}. Consider each clause $c_r = \{ x_i, x_j, x_k \}$. If $c_r$ contains the $j\textsuperscript{th}$ occurrence of $x_i$ then add the edge $\{ d_r^1, w_i^j \}$. Similarly, add an edge between $d_r^2$ and an agent in $W_j$ depending on the index of the occurrence of $x_j$ in the clause $c_r$ and an edge between $d_r^3$ and an agent in $W_k$ depending on the index of the occurrence of $x_k$ in the clause $c_r$.

The construction of $(N, E)$ is now complete. Note that each agent in a variable gadget has degree $3$. For any $r$ where $1 \leq r \leq m$, $d_r^4$ has degree $3$ and each of $d_r^1$, $d_r^2$, and $d_r^3$ has degree $2$. For any $i$ where $1\leq i\leq 3l$, $g_{4i}$ has degree $3$ and each of $g_{4i-3}$, $g_{4i-2}$, and $g_{4i-1}$ has degree $1$. It follows that the maximum degree of $(N, E)$ is $3$.

It is straightforward to show that this reduction can be performed in polynomial time. To prove that the reduction is correct we show that the ASHG represented by $(N, E)$ contains a wj-envy-free partition into triples if and only if the \porschenxsatvariant/ instance $C$ is exactly satisfiable.

We first prove an ancillary lemma. Recall that for any set of agents $S$, $\sigma(S, \pi)$ is the number of triples in $\pi$ that each contains at least one agent in $S$. Recall also that for any agent $\alpha_i$ we write $\mathcal{N}(\alpha_i)$ to mean the open neighbourhood of $\alpha_i$.

\begin{lem}
\label{lem:threed_efr_as_wjef_util1_neighbourhoodthreetriples}
Suppose $\pi$ is a partition into triples in the ASHG represented by $(N, E)$. For any agent $\alpha_i$, if $u_{\alpha_i}(\pi)=1$ and $\sigma(\mathcal{N}(\alpha_i), \pi)=\deg(\alpha_i)$ then $\alpha_i$ is not wj-envious in $\pi$.
\end{lem}
\begin{proof}
As for Lemma~\ref{lem:threed_efr_as_envyfree_util1_neighbourhoodthreetriples} in Section~\ref{sec:threed_efr_as_envyfreeness_maxdeg3}.
\end{proof}

We now show that if the \porschenxsatvariant/ instance $C$ is exactly satisfiable then the ASHG represented by $(N, E)$ contains a wj-envy-free partition into triples.

\begin{lem}
\label{lem:threed_efr_as_wjenvy_firstdirection}
If $C$ is exactly satisfiable then the ASHG represented by $(N, E)$ contains a wj-envy-free partition into triples.
\end{lem}
\begin{proof}
Suppose $f$ is an exact model of $C$. We shall construct a partition into triples $\pi$ that is wj-envy-free. For each variable $x_i$ in $X(C)$ where $f(x_i)$ is true, add $\{ w_i^1, w_i^2, w_i^3 \}$ to $\pi$. Now consider each clause $c_r = \{ x_i, x_j, x_k \}$ and the corresponding clause gadget $D_r$, labelling $i, j, k$ such that $W_i$ contains an agent adjacent to $d_r^1$, $W_j$ contains an agent adjacent to $d_r^2$, and $W_k$ contains an agent adjacent to $d_r^3$. There are three cases: $f(x_i)$ is true while both $f(x_j)$ and $f(x_k)$ are false, $f(x_j)$ is true while both $f(x_i)$ and $f(x_k)$ are false, and $f(x_k)$ is true while both $f(x_i)$ and $f(x_j)$ are false. In the first case, suppose $c_r$ contains the $a\textsuperscript{th}$ occurrence of $x_j$ and the $b\textsuperscript{th}$ occurrence of $x_k$. Add to $\pi$ the triples $\{ d_r^1, d_r^4, g_{3r} \}$, $\{ d_r^2, w_j^a, g_{3r-1} \}$, and $\{ d_r^3, w_k^b, g_{3r-2} \}$. The constructions in the second and third cases are symmetric: in the second case, suppose $c_r$ contains the $a\textsuperscript{th}$ occurrence of $x_i$ and the $b\textsuperscript{th}$ occurrence of $x_k$. Add to $\pi$ the triples $\{ d_r^2, d_r^4, g_{3r} \}$, $\{ d_r^1, w_i^a, g_{3r-1} \}$, and $\{ d_r^3, w_k^b, g_{3r-2} \}$. In the third case, suppose $c_r$ contains the $a\textsuperscript{th}$ occurrence of $x_i$ and the $b\textsuperscript{th}$ occurrence of $x_j$. Add to $\pi$ the triples $\{ d_r^3, d_r^4, g_{3r} \}$, $\{ d_r^1, w_i^a, g_{3r-1} \}$, and $\{ d_r^2, w_j^b, g_{3r-2} \}$.

The construction of $\pi$ is now complete. Note that there are three kinds of triple in $\pi$. Specifically, for any triple $t$ in $\pi$, either $t = W_i$ for some variable gadget $W_i$; $t = \{ d_r^4, d_r^a, g_{3r} \}$ where $1\leq r\leq m$ and $1\leq a \leq 3$; or $t = \{ d_r^a, w_i^b, g_j \}$ where $1\leq i \leq m$, $1\leq r \leq m$, $1\leq a \leq 3$, $1\leq b \leq 3$, and $1\leq j\leq 12l$. We will show that in each case no agent in $t$ is envious.

First, consider some triple $t$ in $\pi$ of the first kind, where $t = W_i$ for some variable gadget $W_i$. Since each agent in $t$ has utility $2$, clearly no agent in $t$ is envious.

Second, consider some triple $t$ in $\pi$ of the second kind, where $t=\{ d_r^4, d_r^a, g_{3r} \}$, $1\leq r\leq m$, and $1\leq a \leq 3$. By the construction of $\pi$, it must be that $\sigma(\mathcal{N}(d_r^4), \pi) = 3$. Since $u_{d_r^4}(\pi)=1$, it follows by Lemma~\ref{lem:threed_efr_as_wjef_util1_neighbourhoodthreetriples} that $d_r^4$ is not envious. Similarly, since $\pi(d_r^a)=\{ d_r^4, d_r^a, g_{3r} \}$ it follows that $\sigma(\mathcal{N}(d_r^a), \pi) = \deg(d_r^a) = 2$. Since $u_{d_r^a}(\pi)=1$ it follows by Lemma~\ref{lem:threed_efr_as_wjef_util1_neighbourhoodthreetriples} that $d_r^a$ is also not envious. Now suppose for a contradiction that $g_{3r}$ is wj-envious and has wj-envy for some agent $\alpha_j$. It must be that $u_{g_{3r}}(\pi(\alpha_j) \setminus \{ \alpha_j \})\geq 1$ so $\pi(\alpha_j)$ must contain some agent $g_q$ where $\{ g_{3r}, g_q \} \in E$. Label $\pi(\alpha_j) = \{ \alpha_j, g_q, \alpha_k \}$. By the construction of $\pi$, the only possibility is that $\{ \alpha_j, \alpha_k \} \subset D_s$ for some clause gadget $D_s$, where $u_{\alpha_j}(\pi) = u_{\alpha_k}(\pi) = 1$. It follows that $\{ \alpha_k, g_q \} \notin E$ and thus that $u_{\alpha_k}(\pi) = u_{\alpha_k}(\{ \alpha_j, g_q \}) = 1 > 0 = u_{\alpha_k}(\{ g_q, g_{3r} \})$, which contradicts the fact that $g_{3r}$ has wj-envy for $\alpha_j$.

Third, consider some triple $t$ in $\pi$ of the third kind, where $t=\{ d_r^a, w_i^b, g_j \}$,  $1\leq i \leq m$, $1\leq r \leq m$, $1\leq a \leq 3$, $1\leq b \leq 3$, and $1\leq j\leq 12l$. By construction, $\sigma(\mathcal{N}(w_i^b), \pi) = \deg(w_i^b) = 3$ and $u_{w_i^b}(\pi) \geq 1$ so it follows by Lemma~\ref{lem:threed_efr_as_wjef_util1_neighbourhoodthreetriples} that $w_i^b$ is not envious. Similarly, since $\pi(d_r^a)=\{ d_r^a, w_i^b, g_{j} \}$ it follows that $\sigma(\mathcal{N}(d_r^a), \pi) = \deg(d_r^a) = 2$. Since $u_{d_r^a}(\pi)=1$ it follows by Lemma~\ref{lem:threed_efr_as_wjef_util1_neighbourhoodthreetriples} that $d_r^a$ is not envious. As before, suppose for a contradiction that $g_{j}$ is wj-envious and has wj-envy for some agent $\alpha_k$. It must be that $u_{g_{j}}(\pi(\alpha_k) \setminus \{ \alpha_k \})\geq 1$ so $\pi(\alpha_k)$ must contain some agent $g_q$ where $\{ g_{j}, g_q \} \in E$. Label $\pi(\alpha_k) = \{ \alpha_k, g_q, \alpha_h \}$. By construction of $\pi$, the only possibility is that $\{ \alpha_k, \alpha_h \} \subset D_s$ for some clause gadget $D_s$, where $u_{\alpha_k}(\pi) = u_{\alpha_h}(\pi) = 1$. It follows that $\{ \alpha_h, g_q \} \notin E$ and thus that $u_{\alpha_h}(\pi) = u_{\alpha_h}(\{ \alpha_k, g_q \}) = 1 > 0 = u_{\alpha_h}(\{ g_q, g_j \})$, which contradicts the fact that $g_j$ has wj-envy for $\alpha_k$.
\end{proof}

We now show, using a sequence of lemmas, that if the ASHG represented by $(N, E)$ contains a wj-envy-free partition into triples then the \porschenxsatvariant/ instance $C$ is exactly satisfiable.

\begin{lem}
\label{lem:threed_efr_as_wjenvy_giscores0}
If the ASHG represented by $(N, E)$ contains a wj-envy-free partition into triples $\pi$ then $u_{g_i}(\pi)=0$ for any $i$ where $1\leq i\leq 12l$.
\end{lem}
\begin{proof}
Suppose $\pi$ is a wj-envy-free partition into triples. 

By the construction of $(N, E)$, the structure of each garbage collector gadget $G_i$ is identical. Thus, to simplify the proof assume without loss of generality that $G_i = G_1 = \{ g_1, g_2, g_3, g_4 \}$. We shall prove that $\sigma(G_1, \pi)=4$, from which it follows directly that $u_{g_1}(\pi)=u_{g_2}(\pi)=u_{g_3}(\pi)=u_{g_4}(\pi)=0$. 

Since $|G_1|=4$ by definition $\sigma(G_1, \pi) \leq 4$. Suppose for a contradiction that $\sigma(G_1, \pi) \leq 3$. Then there exist two agents $g_a, g_b$ in $G_1$ in the same triple in $\pi$. Label the third agent in that triple as $\alpha_j$. 

By symmetry, we need only consider two cases. In the first case $a=1$ and $b=4$. In the second case $a=1$ and $b=2$. First, suppose $a=1$ and $b=4$. Since $\{ g_1, g_4 \} \subset \pi(g_4)$, by construction it must be that either $u_{g_2}(\pi)=0$ or $u_{g_3}(\pi)=0$. Assume without loss of generality that $u_{g_2}(\pi)=0$. It follows that $g_2$ wj-envies $\alpha_j$, since $u_{g_2}(\pi) = 0 < 1 = u_{g_2}(\{ g_1, g_4 \})$, $u_{g_1}(\{ g_4, \alpha_j \}) = 1 = u_{\alpha_j}(\{ g_4, g_2 \})$, and $u_{g_4}(\{ g_1, \alpha_j \}) = 1 < 2 = u_{g_4}(\{ g_1, g_2 \})$.

Second, suppose $a=1$ and $b=2$. There are two cases: either $g_4 = \alpha_j$ or $g_4\neq \alpha_j$. If $g_4=\alpha_j$ then $g_3$ wj-envies $g_2$, since $u_{g_3}(\pi) = 0 < 1 = u_{g_3}(\{ g_1, g_4 \})$, $u_{g_1}(\{ g_2, g_4 \}) = 1 = u_{g_1}(\{ g_3, g_4 \})$, and $u_{g_4}(\{ g_1, g_2 \}) = 2 = u_{g_4}(\{ g_1, g_3 \})$. On the other hand, if $g_4 \neq \alpha_j$ then it must be that $u_{g_4}(\pi) \leq 1$. Now $g_4$ wj-envies $\alpha_j$, since $u_{g_4}(\pi) \leq 1 < 2 = u_{g_4}(\{ g_1, g_2 \})$, $u_{g_1}(\pi) = 0 < 1 = u_{g_1}(\{ g_2, g_4 \})$, and $u_{g_2}(\pi) = 0 < 1 = u_{g_2}(\{ g_1, g_4 \})$.
\end{proof}

As before in Section~\ref{sec:threed_efr_as_envyfreeness_maxdeg3}, for any partition into triples $\pi$ and any variable gadget $W_i$, if $\sigma(W_i, \pi) = 3$ then let us say that $W_i$ is \emph{open}. If $\sigma(W_i, \pi) = 1$ then let us say that $W_i$ is \emph{closed}.

\begin{lem}
\label{lem:threed_efr_as_wjenvy_seconddirection_triangle_split_stay}
If the ASHG represented by $(N, E)$ contains a wj-envy-free partition into triples $\pi$ then any variable gadget is either open or closed.
\end{lem}
\begin{proof}
The proof is essentially the same as for Lemma~\ref{lem:threed_efr_as_regularenvy_seconddirection_triangle_split_stay}. In short, if some triple in $\pi$ contains exactly two agents in $W_i$ then the third agent in $W_i$ is wj-envious.
\end{proof}

For any set of agents $S$ and any clause gadget $D_r$, let us say that $S$ \emph{intersects} $D_r$, and vice-versa, if $|t \cap D_r| \geq 1$.

\begin{lem}
\label{lem:threed_efr_as_wjenvy_gi011}
If the ASHG represented by $(N, E)$ contains a wj-envy-free partition into triples $\pi$ then for any $i$ where $1\leq i\leq 12l$ it must be that $\pi(g_i)$ intersects some clause gadget and $\pi(g_i) = \{ g_i, \alpha_a, \alpha_b \}$ where $\{ \alpha_a, \alpha_b \} \in E$.
\end{lem}
\begin{proof}
Suppose $\pi$ is a wj-envy-free partition into triples. Consider an arbitrary $g_i$ where $1\leq i \leq 12l$, labelling $\pi(g_i) = \{ g_i, \alpha_a, \alpha_b \}$. There must exist some agent $g_j$ such that $\{ g_i, g_j \} \in E$. By Lemma~\ref{lem:threed_efr_as_wjenvy_giscores0}, $u_{g_i}(\pi) = u_{g_j}(\pi) = 0$. 

Suppose for a contradiction that $\{ \alpha_a, \alpha_b \} \notin E$. It follows that $u_{\alpha_a}(\pi) = u_{\alpha_b}(\pi) = u_{g_i}(\pi) = 0$. Now $g_j$ wj-envies $\alpha_a$, since $u_{g_j}(\pi) = 0 < 1 \leq u_{g_j}(\{ g_i, \alpha_b \})$, $u_{g_i}(\pi) = 0 < 1 \leq u_{g_i}(\{ g_j, \alpha_b \})$, and $u_{\alpha_b}(\pi) = 0 < 1 \leq u_{\alpha_b}(\{ g_i, g_j\})$.

We have now shown that $\{ \alpha_a, \alpha_b \} \in E$. Suppose for a contradiction that $\pi(g_i)$ does not intersect any clause gadget. It must be that $\alpha_a$ and $\alpha_b$ both belong to variable gadgets. In fact, since $\{ \alpha_a, \alpha_b \} \in E$ it must be that both $\alpha_a$ and $\alpha_b$ belong to the same variable gadget. Since $g_i$ belongs to a garbage collector gadget, this contradicts Lemma~\ref{lem:threed_efr_as_wjenvy_seconddirection_triangle_split_stay}.
\end{proof}

\begin{lem}
\label{lem:threed_efr_as_wjenvy_seconddirection}
If the ASHG represented by $(N, E)$ contains a wj-envy-free partition into triples then $C$ is exactly satisfiable.
\end{lem}
\begin{proof}
Suppose $\pi$ is a wj-envy-free partition into triples $(N, E)$. By Lemma~\ref{lem:threed_efr_as_wjenvy_seconddirection_triangle_split_stay}, any variable gadget is either open or closed. Construct a truth assignment $f$ in $C$ by setting $f(x_i)$ to be true if $W_i$ is closed and false otherwise. Each variable $x_i$ corresponds to exactly one variable gadget so it follows that $f$ is a valid truth assignment. By the construction of $(N, E)$, each clause $c_r$ corresponds to exactly one clause gadget $D_r$. Recall that each clause gadget is adjacent to three variable gadgets that correspond to the three variables in that clause. To show that $f$ is an exact model of $C$, it is now sufficient to show that each clause gadget is adjacent to exactly one closed variable gadget.

By Lemma~\ref{lem:threed_efr_as_wjenvy_gi011}, for any $i$ where $1\leq i\leq 12l$ there exists some triple $\pi(g_i) = \{ g_i, \alpha_a, \alpha_b \}$ where $\{ \alpha_a, \alpha_b \} \in E$ and $\pi(g_i)$ intersects some clause gadget. Let $T \subset \pi$ be the set of $12l$ such triples. Since each clause gadget contains four agents, by the definition of $T$ it is impossible for any clause gadget to intersect four or more triples in $T$. It follows that any clause gadget intersects at most three triples in $T$. Since $|T|=12l$ and there are exactly $m=4l$ clause gadgets, it must be that there are on average $12l/4l=3$ triples in $T$ that intersect each clause gadget. 

It follows that each clause gadget intersects exactly three triples in $T$. In fact, by the construction of each clause gadget, the only possibility is that each clause gadget $D_r$ intersects exactly three triples in $T$ and exactly two of these triples each intersect some variable gadget that is adjacent to $D_r$ and must be open. It follows that each clause gadget is adjacent to exactly two open variable gadgets and exactly one closed variable gadget, as desired.
\end{proof}

We have now shown that the ASHG represented by $(N, E)$ contains a wj-envy-free partition into triples if and only if the \porschenxsatvariant/ instance $C$ is exactly satisfiable. This shows that the reduction is correct.

\begin{thm}
\label{thm:threed_efr_as_wjef_npcomplete}
Deciding if a given ASHG contains a wj-envy-free partition into triples is $\NP$-complete even when preferences are binary and symmetric and maximum degree is $3$.
\end{thm}
\begin{proof}
It is straightforward to show that this decision problem belongs to $\NP$, since for any two agents $\alpha_i$ and $\alpha_j$ in $N$ we can test if $\alpha_i$ wj-envies $\alpha_j$ in constant time. 

We have presented a polynomial-time reduction from \porschenxsatvariant/, which is $\NP$-complete~\cite[Lemma~5]{PSSW14}. Given an arbitrary instance $C$ of \porschenxsatvariant/, the reduction constructs an ASHG represented by its underlying graph $(N, E)$ which has binary and symmetric preferences and maximum degree $3$. Together, Lemmas~\ref{lem:threed_efr_as_wjenvy_firstdirection} and~\ref{lem:threed_efr_as_wjenvy_seconddirection} show that this ASHG contains a wj-envy-free partition into triples if and only if $C$ is exactly satisfiable, and thus that this decision problem is $\NP$-hard.
\end{proof}

\section{Justified envy-freeness}
\label{sec:threed_efr_as_jef}
In this section we consider j-envy-freeness.

We begin, in Section~\ref{sec:threed_efr_as_jef_binary}, by noting that any stable partition is j-envy-free. We recall a previous result that if preferences are binary and symmetric then a stable partition into triples must exist and can be found in polynomial time. We observe that a j-envy-free partition must therefore also exist and can also be found in polynomial time (Observation~\ref{obs:threed_efr_as_jef_binary_symmetric_from_stability}). We then strengthen this result to show that a j-envy-free partition into triples must exist and can be found in polynomial time when preferences are binary, but not necessarily symmetric (Theorem~\ref{thm:threed_efr_as_jef_binary_algorithm}). As we shall see, this contrasts with an analogous result for stability.

Next, in Section~\ref{sec:threed_efr_as_jef_ternary}, we consider ASHGs with ternary preferences. We show that, in general, such an ASHG may not contain a j-envy-free partition into triples, and the associated existence problem is $\NP$-complete (Theorem~\ref{thm:threed_efr_as_jef_terasym_npcomplete}).

Finally, in Section~\ref{sec:threed_efr_as_jef_symmetricnonbinrary}, we consider ASHGs with symmetric, non-binary preferences. As before, we show that such an ASHG may not contain a j-envy-free partition into triples, and the associated existence problem is $\NP$-complete (Theorem~\ref{thm:threed_efr_as_jef_symmetric_6_npcomplete}).

\subsection{Binary preferences}
\label{sec:threed_efr_as_jef_binary}
It is straightforward to show that if there exists an agent with j-envy then there exists a blocking triple. It follows that if a partition into triples is stable then it is j-envy-free. If preferences are binary and symmetric then a stable partition into triples must exist and can be constructed in polynomial time~\cite{ManloveMcKay3DSRAS2021}. Observation~\ref{obs:threed_efr_as_jef_binary_symmetric_from_stability} follows directly.

\begin{observation}
\label{obs:threed_efr_as_jef_binary_symmetric_from_stability}
Given an ASHG with binary and symmetric preferences, a j-envy-free partition into triples always exists and can be found in polynomial time.
\end{observation}

It is known that if preferences are binary but not necessarily symmetric then a stable partition into triples may not exist and the associated decision problem is $\NP$-complete~\cite{ManloveMcKay3DSRAS2021}. Interestingly, in contrast we now show that a j-envy-free partition into triples is bound to exist and can be found in polynomial time.

To do this, we consider a simple algorithm that involves iteratively ``satisfying'' any agent with j-envy. We show that this algorithm terminates in polynomial time using a standard proof technique involving a \emph{potential function}~\cite{GairingSavani19}, which strictly increases after each iteration and is polynomial in terms of the problem input. Informally, the potential function that we define here is ``the total number of pairs of agents that belong to the same triple and have a mutual non-zero valuation''.

Formally, the algorithm is as follows. Suppose $(N, V)$ is an ASHG with binary preferences. First let $\pi$ be an arbitrary partition into triples. While there exists some agent $\alpha_i$ that has j-envy in $(N, V)$ for some other agent $\alpha_j$, swap $\alpha_i$ and $\alpha_j$ in the partition. Once there is no such $\alpha_i$, return $\pi$, which must be j-envy-free.  


To prove that this algorithm terminates in polynomial time, we define some new terminology. For any two agents $\alpha_i$ and $\alpha_j$ in $N$, we say that $\{ \alpha_i, \alpha_j \}$ is a \emph{bidirected pair} if $v_{\alpha_i}(\alpha_j) = v_{\alpha_j}(\alpha_i) = 1$. For any set of agents $S$, if both $\alpha_i$ and $\alpha_j$ belong to $S$ then we say that $\{ \alpha_i, \alpha_j \}$ is a \emph{bidirected pair in $S$}. For some partition into triples $\pi$, let the \emph{number of bidirected pairs in $\pi$} be the total of the number of bidirected pairs in each triple in $\pi$.

We show in Lemma~\ref{lem:threed_efr_as_jef_potfunc} that after each swap, the total number of bidirected pairs in $\pi$ strictly increases.

\begin{lem}
\label{lem:threed_efr_as_jef_potfunc}
If $\pi_1$ is the partition before some swap and $\pi_2$ is the partition after that swap then the number of bidirected pairs in $\pi_2$ is strictly greater than the number of bidirected pairs in $\pi_1$.
\end{lem}
\begin{proof}
Without loss of generality, assume that $\alpha_1$ swaps with $\alpha_4$, where $\pi_1(\alpha_1) = \{ \alpha_1, \alpha_2, \alpha_3 \}$ and $\pi_1(\alpha_4) = \{ \alpha_4, \alpha_5, \alpha_6 \}$, so that $\pi_2(\alpha_1) = \{ \alpha_1, \alpha_5 \alpha_6 \}$ and $\pi_2(\alpha_4) = \{ \alpha_4, \alpha_2, \alpha_3 \}$. It is straightforward to show that the number of bidirected pairs in $\pi_2$ is strictly greater than the number of bidirected pairs in $\pi_1$ if the number of bidirected pairs in $\{ \{\alpha_1, \alpha_5\}, \{\alpha_1, \alpha_6\}, \{\alpha_4, \alpha_1\}, \{\alpha_4, \alpha_2\} \}$ is strictly greater than the number of bidirected pairs in $\{ \{\alpha_1, \alpha_2\}, \{\alpha_1, \alpha_3\}, \{\alpha_4, \alpha_5\}, \{\alpha_4, \alpha_6\} \}$.

Since $\alpha_1$ has j-envy for $\alpha_4$ in $\pi_1$ it must be that $2 \geq u_{\alpha_1}(\pi_2) = u_{\alpha_1}(\{ \alpha_5, \alpha_6 \}) > u_{\alpha_1}(\pi_1) = u_{\alpha_1}(\{ \alpha_2, \alpha_3 \}) \geq 0$. Note that since preferences in $(N, V)$ are binary, it must also be that $v_{\alpha_{5}}(\alpha_{1}) = v_{\alpha_{6}}(\alpha_{1}) = 1 > 0 = v_{\alpha_{5}}(\alpha_{4}) = v_{\alpha_{6}}(\alpha_{4})$. It follows that neither $\{ \alpha_4, \alpha_5 \}$ nor $\{ \alpha_4, \alpha_6 \}$ is a bidrected pair. There are now two possibilities: either $u_{\alpha_1}(\{ \alpha_5, \alpha_6 \}) = 1$ or $u_{\alpha_1}(\{ \alpha_5, \alpha_6 \}) = 2$.

First, suppose $u_{\alpha_1}(\{ \alpha_5, \alpha_6 \}) = 1$. It must be that either $v_{\alpha_1}(\alpha_5) = 1$ or $v_{\alpha_1}(\alpha_6) = 1$. Without loss of generality assume that $v_{\alpha_1}(\alpha_5) = 1$. Since $\alpha_1$ has j-envy for $\alpha_4$ it must be that $u_{\alpha_1}(\pi_1) = u_{\alpha_1}(\{ \alpha_2, \alpha_3 \}) = 0 < 1 = u_{\alpha_1}(\{ \alpha_5, \alpha_6 \})$. It follows that neither $\{\alpha_1, \alpha_2 \}$ nor $\{\alpha_1, \alpha_3 \}$ is a bidirected pair. Since $v_{\alpha_1}(\alpha_5) = v_{\alpha_5}(\alpha_1) = 1$ it follows that $\{ \alpha_1, \alpha_5 \}$ is a bidirected pair and thus that the number of bidirected pairs in $\{ \{\alpha_1, \alpha_5\}, \{\alpha_1, \alpha_6\}, \{\alpha_4, \alpha_1\}, \{\alpha_4, \alpha_2\} \}$ is strictly greater than the number of bidirected pairs in $\{ \{\alpha_1, \alpha_2\}, \{\alpha_1, \alpha_3\}, \{\alpha_4, \alpha_5\}, \{\alpha_4, \alpha_6\} \}$, as required.

Second, suppose $u_{\alpha_1}(\{ \alpha_5, \alpha_6 \}) = 2$. Since $v_{\alpha_{5}}(\alpha_{1}) = v_{\alpha_{6}}(\alpha_{1}) = 1$ it must be that both $\{ \alpha_1, \alpha_5 \}$ and $\{ \alpha_1, \alpha_6 \}$ are bidirected pairs. Since $\alpha_1$ has j-envy for $\alpha_4$ it must be that $u_{\alpha_1}(\pi_1) = u_{\alpha_1}(\{ \alpha_2, \alpha_3 \}) < 2 = u_{\alpha_1}(\{ \alpha_5, \alpha_6 \})$. It follows that either $v_{\alpha_1}(\alpha_2) = 0$ or $v_{\alpha_1}(\alpha_3) = 0$. The only possibility is that at most one of $\{ \alpha_1, \alpha_2 \}$ and $\{ \alpha_1, \alpha_3 \}$ is a bidirected pair. The number of bidirected pairs in $\{ \{\alpha_1, \alpha_5\}, \{\alpha_1, \alpha_6\}, \{\alpha_4, \alpha_1\}, \{\alpha_4, \alpha_2\} \}$ is therefore strictly greater than the number of bidirected pairs in $\{ \{\alpha_1, \alpha_2\}, \{\alpha_1, \alpha_3\}, \{\alpha_4, \alpha_5\}, \{\alpha_4, \alpha_6\} \}$, as required.
\end{proof}

\begin{thm}
\label{thm:threed_efr_as_jef_binary_algorithm}
Given an ASHG with binary preferences, a j-envy-free partition into triples must exist and can be found in polynomial time.
\end{thm}
\begin{proof}
By Lemma~\ref{lem:threed_efr_as_jef_potfunc}, after each swap the number of bidirected pairs in $\pi$ strictly increases. Since the number of bidirected pairs in $\pi$ is at most $|N|$ it follows that at most $|N|$ swaps occur.
\end{proof}

\subsection{Ternary preferences}
\label{sec:threed_efr_as_jef_ternary}
A natural question is whether the polynomial-time algorithm described in the proof of Theorem~\ref{thm:threed_efr_as_jef_binary_algorithm} can be extended to the setting in which preferences are ternary, i.e.\ $v_{\alpha_i}(\alpha_j) \in \{ 0, 1, 2 \}$. We show that, assuming $\P \neq \NP$, this is not the case. Specifically, we show that a given ASHG may not contain a j-envy-free partition into triples and the associated decision problem is $\NP$-complete, even when preferences are ternary.

We present a polynomial-time reduction from a special case of \emph{Directed Triangle Packing} (DTC, Problem~\ref{pr:DTC}).

\begin{myproblem}[Directed Triangle Packing (DTC)]
\label{pr:DTC}
\begin{samepage}
\begin{adjustwidth}{8pt}{8pt}
\inp a simple directed graph $G=(W, A)$ where $W=\{ w_1, w_2, \dots, w_{3q} \}$ for some integer $q$\\
\ques Can the vertices of $G$ be partitioned into $q$ disjoint sets $X=\{X_1, X_2, \dots, X_q\}$, each set containing exactly three vertices, such that each $X_p=\{ w_i,w_j,w_k \}$ in $X$ is a directed $3$-cycle, i.e.\ $( w_i,w_j ) \in A$, $( w_j, w_k ) \in A$, and $( w_k, w_i ) \in A$?
\end{adjustwidth}
\end{samepage}
\end{myproblem}

As shown by Cechl\'arov\'a, Fleiner, and Manlove, DTC is $\NP$-complete even when $G$ is antisymmetric~\cite{CFM05} (i.e.\ it contains no bidirectional arcs). We first describe the reduction, from this special case of DTC, and then provide some intuition with respect to its design. 

\begin{figure}
    \centering
    \begin{tikzpicture}
\begin{scope}[scale=0.8]
\begin{scope}[every node/.style={circle,draw, minimum size=2.4mm}]
    \begin{scope}
        \begin{scope}[shift={(0.0, {2.3})}]
            \def\hlabeldist{0.4cm}
            \begin{scope}[every node/.style={circle,thick,draw,minimum size=2.4mm}, scale=1.15]
                \node[label={[label distance=\hlabeldist]90:$h_2$}] (h2) at (0,0) {};
                \node[label={[label distance=\hlabeldist]18:$h_3$}] (h3) at (1.9021,-1.3820) {};
                \node[label={[label distance=\hlabeldist]306:$h_4$}] (h4) at (1.1756,-3.6180) {};
                \node[label={[label distance=\hlabeldist]234:$h_5$}] (h5) at (-1.1756,-3.6180) {};
                \node[label={[label distance=\hlabeldist]162:$h_1$}] (h1) at (-1.9021,-1.3820) {};
            \end{scope}
            \begin{scope}
                \foreach \from/\to in {h1/h2, h2/h3, h3/h4, h4/h5, h5/h1}
                    \draw [thick, darrow] (\from) -- (\to);
                    
                \foreach \from/\to in {h1/h3, h2/h4, h3/h5, h4/h1, h5/h2}
                    \draw [thick, darrow12] (\from) -- (\to);
            \end{scope}
        \end{scope}

        \begin{scope}[every node/.style={circle,draw, minimum size=2.4mm}, shift={(5.5, 0.0)}]
            \def\scalefactorl{2.7}
            \begin{scope}[yscale=1.1]
                \node[thick, circle, label={[label distance=0.4cm]90:$l_1$}] (l1) at (\scalefactorl*-0.5,\scalefactorl*0.5) {};
                \node[thick, circle, label={[label distance=0.4cm]90:$l_3$}] (l3) at (\scalefactorl*0.5,\scalefactorl*0.5) {};
                \node[thick, circle, label={[label distance=0.4cm]-90:$l_2$}] (l2) at (\scalefactorl*-0.5,\scalefactorl*-0.5) {};
                \node[thick, circle, label={[label distance=0.4cm]-90:$l_4$}] (l4) at (\scalefactorl*0.5,\scalefactorl*-0.5) {};
            \end{scope}
            
            \foreach \from/\to in {l1/l4, l2/l3, l1/l3, l2/l4}
                    \draw [thick, darrow] (\from) -- (\to);
            \foreach \from/\to in {l1/l2, l3/l4}
                    \draw [thick, darrow22] (\from) -- (\to);
        \end{scope}
        
        \begin{scope}[shift={(9.6, 0.0)}]
            \draw[darrow, thick] (-1.5, {1.1 * sin(150)}) -- (l3);
            \draw[darrow12, thick] (-1.5, {1.1 * sin(210)}) -- (l4);
            
            \filldraw[color=enclosure_color, fill=white, rounded corners=0.5cm] (-1.7, -0.7) rectangle (1.7, 0.7) {};
            \node[draw=none] (clabel) at (0.0, -0.05) {$c_1, c_2, \dots, c_{3q}$};
        \end{scope}
        
    \end{scope}
\end{scope}
\end{scope}
\end{tikzpicture}
    \caption{The reduction from DTC to the problem of deciding if a given ASHG with ternary preferences contains a j-envy-free partition into triples}
    \label{fig:threed_efr_as_jef_terasym_reduction}
\end{figure}
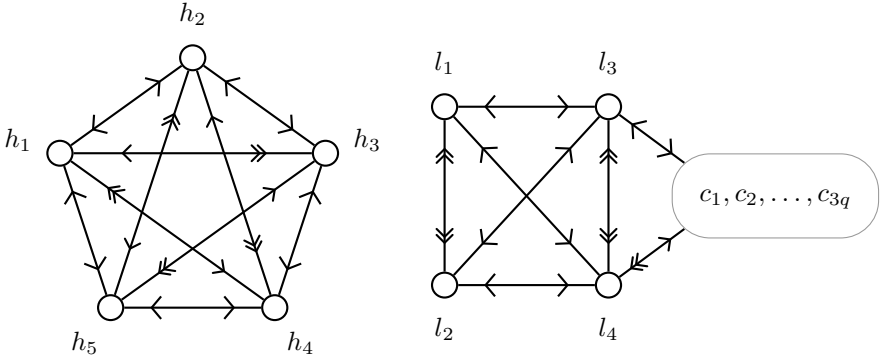

The reduction, illustrated in Figure~\ref{fig:threed_efr_as_jef_terasym_reduction}, is as follows. Suppose $G=(W, A)$ is an arbitrary instance of DTC. We shall construct an ASHG $(N, V)$. Unless otherwise specified, assume that $v_{\alpha_i}(\alpha_j)=0$ for any $\alpha_i$ and $\alpha_j$ in $N$. To simplify the description of the reduction, in this section we write $i \myoplus y$ meaning $((i + y - 1) \bmod 5) + 1$.

First construct a set of five agents $H = \{ h_1, h_2, h_3, h_4, h_5 \}$. For each $i$ where $1\leq i \leq 5$ let $v_{h_i}(h_{i \myoplus 1}) = v_{h_{i \myoplus 1}}(h_i) = 1$, $v_{h_i}(h_{i \myoplus 3}) = 1$, and $v_{h_i}(h_{i \myoplus 2}) = 2$. Next, construct a set $L = \{ l_1, l_2, l_3, l_4 \}$ of four agents. Let $v_{l_1}(l_2) = v_{l_2}(l_1) = v_{l_3}(l_4) = v_{l_4}(l_3) = 2$ and $v_{l_1}(l_3) = v_{l_1}(l_4) = v_{l_2}(l_3) = v_{l_2}(l_4) = v_{l_3}(l_1) = v_{l_3}(l_2) = v_{l_4}(l_1) = v_{l_4}(l_2) = 1$. Next, construct a set $C = \{ c_1, c_2, \dots, c_{3q} \}$ of $3q$ agents. For each $i$ where $1\leq i \leq 3q$, let $v_{c_i}(l_3) = v_{l_3}(c_i) = v_{l_4}(c_i) = 1$ and $v_{c_i}(l_4) = 2$. For each $i$ and $j$ where $1\leq i, j \leq 3q$, let $v_{c_i}(c_j) = 2$ if $( w_i, w_j ) \in A$ otherwise $1$. The construction of $(N, V)$ is now complete. Note that the structure of the valuations among the agents in $C$ now reflects the directed graph $G$. 

We remark that the design of $H$ is derived from a particular instance that contains no j-envy-free partition into triples. To construct this instance, delete every agent in $N$ other than the agents in $H$ and $l_1$. The accompanying proof, which shows that this instance contains no j-envy-free partition into triples, can be derived straightforwardly from the proof of Lemma~\ref{lem:threed_efr_as_jef_terasym_atleasttwotriplescontainoneagentinH}, which appears later in this section.

It is straightforward to show that the reduction runs in polynomial time. To prove that the reduction is correct we show that the ASHG $(N, V)$ contains a j-envy-free partition into triples if and only if the DTC instance $G$ contains a directed triangle cover.

We first show that if the DTC instance $G$ contains a directed triangle cover then the ASHG $(N, V)$ contains a j-envy-free partition into triples. 

\begin{lem}
\label{lem:threed_efr_as_jef_terasym_first_direction}
If $G$ contains a directed triangle cover then $(N, V)$ contains a j-envy-free partition into triples.
\end{lem}
\begin{proof}
Suppose $G$ contains a directed triangle cover $X = \{ X_1, X_2, \dots, X_q \}$. We shall construct a partition into triples $\pi$ that is j-envy-free. First, add $\{ h_1, h_2, h_3 \}$, $\{ h_4, l_1, l_2 \}$, and $\{ h_5, l_3, l_4 \}$ to $\pi$. Next, for each directed $3$-cycle $X_p = \{ w_i, w_j, w_k \}$ in $X$, add $\{ c_i, c_j, c_k \}$ to $\pi$.

Suppose for a contradiction that some agent $\alpha_j$ exists where $\alpha_j$ has j-envy for some other agent $\alpha_{k_1}$ where $\pi(\alpha_{k_1}) = \{ \alpha_{k_1}, \alpha_{k_2}, \alpha_{k_3} \}$. Since $N = H \cup L \cup C$ it must be that either $\alpha_{k_1} \in H$, $\alpha_{k_1} \in L$, or $\alpha_{k_1} \in C$. We show that each case leads to a contradiction. It then follows that no such $\alpha_j$ exists and thus that $\pi$ is j-envy-free.
\begin{itemize}
    \item Suppose $\alpha_{k_1} \in H$. By the construction of $\pi$ there are two possibilities: either $\alpha_{k_1} \in \{ h_1, h_2, h_3 \}$ or $\alpha_{k_1} \in \{ h_4, h_5 \}$. 
    \begin{itemize}
    \item Suppose firstly that $\alpha_{k_1} \in \{ h_4, h_5 \}$. By the construction of $\pi$ either $\{ \alpha_{k_2}, \alpha_{k_3} \} = \{ l_1, l_2 \}$ or $\{ \alpha_{k_2}, \alpha_{k_3} \} = \{ l_3, l_4 \}$. Note that $u_{l_1}(\pi) = u_{l_2}(\pi) = u_{l_3}(\pi) = u_{l_4}(\pi) = 2$. Since $u_{l_1}(\{ l_3, l_4 \}) = 2$ and $u_{l_2}(\{ l_3, l_4 \}) = 2$, neither $l_1$ nor $l_2$ has j-envy for $\alpha_{k_1}$, so $\alpha_j \notin \{ l_1, l_2 \}$. Similarly, since $u_{l_3}(\{ l_1, l_2 \}) = 2$ and $u_{l_4}(\{ l_1, l_2 \}) = 2$ neither $l_3$ nor $l_4$ has j-envy for $\alpha_{k_1}$, so $\alpha_j \notin \{ l_3, l_4 \}$. Since $u_{c_i}(\pi) = 3$, $u_{c_i}(\{ l_1, l_2 \}) = 0$, and $u_{c_i}(\{ l_1, l_2 \}) = 2$ for any $i$ where $1\leq i \leq 3q$, it must be no agent in $C$ has j-envy for $\alpha_{k_1}$, so $\alpha_j \notin C$. It remains that $\alpha_{j} \in H$. Since this implies $v_{l_1}(\alpha_j) = v_{l_2}(\alpha_j) = v_{l_3}(\alpha_j) = v_{l_4}(\alpha_j) = 0$ it follows that $\alpha_j$ does not have j-envy for $\alpha_{k_1}$ and thus that $\alpha_{k_1} \notin \{ h_4, h_5 \}$.
    
    \item Suppose then that $\alpha_{k_1} \in \{ h_1, h_2, h_3 \}$. Since $\alpha_j$ has j-envy for $\alpha_{k_1}$ it must be that $v_{\alpha_j}(\alpha_{k_2}) \geq 1$ so it follows that $\alpha_j \in \{ h_4, h_5 \}$. If $\alpha_{k_1} = h_1$ and $\alpha_j = h_4$ then we reach a contradiction since $h_4$ has j-envy for $h_1$ but $v_{h_3}(h_1) = 1 = v_{h_3}(h_4)$. Similarly, if $\alpha_{k_1} = h_1$ and $\alpha_j = h_5$ then we reach a contradiction since $v_{h_2}(h_1) = 1 = v_{h_2}(h_5)$. If $\alpha_{k_1} = h_2$ or $\alpha_{k_1} = h_3$ then we also reach a contradiction since $v_{h_1}(h_4) = v_{h_1}(h_5) = 1 = v_{h_1}(h_2) < 2 = v_{h_1}(h_3)$.
    \end{itemize}
    
    \item Suppose $\alpha_{k_1} \in C$. By the construction of $\pi$ it must be that $\alpha_{k_2} \in C$ and $\alpha_{k_3} \in C$ so we label $\alpha_{k_1} = c_{i_1}$, $\alpha_{k_2} = c_{i_2}$, and $\alpha_{k_3} = c_{i_3}$. By the construction of $(N, V)$ in the reduction it follows that $v_{c_{i_2}}(c_{i_1}) \geq 1$ and $v_{c_{i_3}}(c_{i_1}) \geq 1$. Since $\alpha_j$ has j-envy for $c_{i_1}$ it follows then that $v_{c_{i_2}}(\alpha_j) = 2$ and $v_{c_{i_3}}(\alpha_j) = 2$. By the construction of the instance there are two possibilities: either $\alpha_j = l_4$ or $\alpha_j \in C$. If $\alpha_j = l_4$ then $u_{l_4}(\{ c_{i_2}, c_{i_3} \}) = 2$ which is a contradiction since by assumption $l_4$ has j-envy for $c_{i_1}$ but $u_{l_4}(\pi) = 2$. If $\alpha_j \in C$ then label $\alpha_j = c_{i_4}$. Since $v_{c_{i_2}}(c_{i_4}) = 2$ and $v_{c_{i_3}}(c_{i_4}) = 2$, by the construction of $C$ it must be that $( w_{i_2}, w_{i_4} ) \in A$ and $( w_{i_3}, w_{i_4} ) \in A$, where the vertices $w_{i_2}, w_{i_3}, w_{i_4}$ are the vertices in $W$ that correspond respectively to the agents $c_{i_2}, c_{i_3}, c_{i_4}$ in $C$. Since $G$ is antisymmetric, it follows that $( w_{i_4}, w_{i_2} ) \notin A$ and $( w_{i_4}, w_{i_3} ) \notin A$ so it must be that $v_{c_{i_4}}(c_{i_2}) = v_{c_{i_4}}(c_{i_3}) = 1$. This is also a contradiction since by assumption $c_{i_4}$ has j-envy for $c_{i_1}$ but $u_{c_{i_4}}(\{ c_{i_2}, c_{i_3} \}) = 2$ and by the construction of $\pi$ it must be that $u_{c_{i_4}}(\pi) = 3$.
    
    \item Suppose $\alpha_{k_1} \in L$. It must be that $\alpha_{k_1} = l_{i_1}$, $\alpha_{k_2} = l_{i_2}$, and $\alpha_{k_3} = h_{i_3}$, where $1\leq i_1, i_2 \leq 4$ and $i_3 \in \{ 4, 5 \}$. If $\alpha_j \in H$ then it must be that $v_{l_{i_2}}(\alpha_j) = 0$ which contradicts the supposition that $\alpha_j$ has j-envy for $l_{i_1}$. Otherwise, if $\alpha_j \notin H$ then $v_{h_{i_3}}(\alpha_j) = 0$, which also contradicts the supposition that $\alpha_j$ has j-envy for $l_{i_1}$.\qedhere
\end{itemize}
\end{proof}

We now show that if the ASHG $(N, V)$ contains a j-envy-free partition into triples then the DTC instance $G$ contains a directed triangle cover. Recall that for any set of agents $S$, $\sigma(S, \pi)$ is the number of triples in $\pi$ that each contains at least one agent in $S$.

\begin{lem}
\label{lem:threed_efr_as_jef_terasym_hbelongsto3}
If $(N, V)$ contains a j-envy-free partition into triples $\pi$ then $\sigma(H, \pi) \geq 3$.
\end{lem}
\begin{proof}
Since $|H|=5$ it must be that $\sigma(H, \pi) \geq 2$. Suppose for a contradiction that $\sigma(H, \pi) = 2$. It must be that one triple in $\pi$ contains three agents in $H$ and one triple in $\pi$ contains two agents in $H$. Suppose the former triple is $\{ h_{i_1}, h_{i_2}, h_{i_3} \}$ and the latter triple is $\{ h_{i_4}, h_{i_5}, \alpha_j \}$, where $1\leq i_1, i_2, \dots, i_5 \leq 5$ and $\alpha_j \in N \setminus H$. There are five symmetries in $H$ and $\binom{5}{2}=10$ possible assignments of $\{ h_{i_4}, h_{i_5} \}$ to two agents in $H$, so we need only consider the two assignments $i_4 = 1$, $i_5 = 2$ and $i_4 = 1$, $i_5 = 3$, which are not symmetric. If $i_4 = 1$ and $i_5 = 2$ then it remains that $\{ i_1, i_2, i_3 \} = \{ 3, 4, 5 \}$. In this case, $h_5$ has j-envy for $\alpha_j$ since $u_{h_5}(\pi) = 2 < 3 \leq u_{h_5}(\{ h_1, h_2 \})$, $v_{h_1}(\alpha_j) = 0 < 1 = v_{h_1}(h_5)$, and $v_{h_2}(\alpha_j) = 0 < 1 = v_{h_2}(h_5)$. If $i_4 = 1$ and $i_5 = 3$ then it remains that $\{ i_1, i_2, i_3 \} = \{ 2, 4, 5 \}$. In this case, $h_4$ has j-envy for $\alpha_j$ since $u_{h_4}(\pi) = 2 < 3 \leq u_{h_4}(\{ h_1, h_3 \})$, $v_{h_1}(\alpha_j) = 0 < 1 = v_{h_1}(h_4)$, and $v_{h_3}(\alpha_j) = 0 < 1 = v_{h_3}(h_4)$. 
\end{proof}

\begin{lem}
\label{lem:threed_efr_as_jef_terasym_atleasttwotriplescontainoneagentinH}
If $(N, V)$ contains a j-envy-free partition into triples $\pi$  then at least two triples in $\pi$ each contains exactly one agent in $H$.
\end{lem}
\begin{proof}
By Lemma~\ref{lem:threed_efr_as_jef_terasym_hbelongsto3}, $\sigma(H, \pi) \geq 3$. If, contrary to the lemma statement, at most one triple in $\pi$ contains exactly one agent in $H$ then it must be that two triples in $\pi$ each contains two agents in $H$ and one triple in $\pi$ contains exactly one agent in $H$. Suppose one of the two former triples is $\{ h_{i_1}, h_{i_2}, \alpha_{j_1} \}$ and the latter triple is $\{ h_{i_3}, \alpha_{j_2}, \alpha_{j_3} \}$, where $1\leq i_1, i_2, i_3 \leq 5$ and $\alpha_{j_1}, \alpha_{j_2}, \alpha_{j_3} \in N \setminus H$. By the construction of the instance it must be that $v_{h_{i_1}}(\alpha_{j_1}) = v_{h_{i_2}}(\alpha_{j_1}) = 0$,  $v_{h_{i_1}}(h_{i_3}) \geq 1$ and $v_{h_{i_2}}(h_{i_3}) \geq 1$. It follows that $h_{i_3}$ has j-envy for $\alpha_{j_1}$ since $u_{h_{i_3}}(\pi) = 0 < 2 \leq u_{h_{i_3}}(\{ h_{i_1}, h_{i_2} \})$, $v_{h_{i_1}}(\alpha_j) = 0 < 1 \leq v_{h_{i_1}}(h_{i_3})$, and $v_{h_{i_2}}(\alpha_j) = 0 < 1 \leq v_{h_{i_2}}(h_{i_3})$. This contradicts the supposition that $\pi$ is j-envy-free.
\end{proof}

We have shown in Lemma~\ref{lem:threed_efr_as_jef_terasym_atleasttwotriplescontainoneagentinH} that if $(N, V)$ contains a j-envy-free partition into triples $\pi$ then at least two triples in $\pi$ each contains exactly one agent in $H$. Suppose $t_{\beta}$ and $t_{\gamma}$ are two such triples where $t_{\beta} = \{ h_{a_1}, \alpha_{b_1}, \alpha_{b_2} \}$ and $t_{\gamma} = \{ h_{a_2}, \alpha_{b_3}, \alpha_{b_4} \}$.

\begin{lem}
\label{lem:threed_efr_as_jef_terasym_lequalsthefourisolated}
If $(N, V)$ contains a j-envy-free partition into triples then $\{ \alpha_{b_1},\allowbreak\alpha_{b_2}, \alpha_{b_3},\allowbreak\alpha_{b_4} \} = L$.
\end{lem}
\begin{proof}
Suppose for a contradiction that $\{ \alpha_{b_1}, \alpha_{b_2}, \alpha_{b_3}, \alpha_{b_4} \} \neq L$.

By definition, $\{ \alpha_{b_1}, \alpha_{b_2}, \alpha_{b_3}, \alpha_{b_4} \} \cap H = \varnothing$ and $\{ \alpha_{b_1}, \alpha_{b_2}, \alpha_{b_3}, \alpha_{b_4} \} \neq L$ it must be that at least one agent in $\{ \alpha_{b_1}, \alpha_{b_2}, \alpha_{b_3}, \alpha_{b_4} \}$ belongs to $C$. Assume without loss of generality that $\alpha_{b_1} \in C$.

Note that by construction of the instance, the valuation of any agent not in $H$ for any other agent not in $H$ is at least $1$. 

Since $\alpha_{b_2} \notin H$, it must be that $u_{\alpha_{b_2}}(\pi) = v_{\alpha_{b_2}}(\alpha_{b_1})$. By the design of the instance, since $\alpha_{b_2} \notin H$ and $\alpha_{b_1} \notin H$ it must be that $v_{\alpha_{b_2}}(\alpha_{b_1}) \in \{ 1, 2 \}$. We consider each possibility of $u_{\alpha_{b_2}}(\pi) = v_{\alpha_{b_2}}(\alpha_{b_1})$.

Firstly, suppose $u_{\alpha_{b_2}}(\pi) = 1$. As noted earlier in this proof, since $\alpha_{b_2}, \alpha_{b_3}, \alpha_{b_4} \in N \setminus H$ it must be that $v_{\alpha_{b_2}}(\alpha_{b_3}) \geq 1$ and $v_{\alpha_{b_2}}(\alpha_{b_4}) \geq 1$. It follows that $\alpha_{b_2}$ has j-envy for $h_{a_2}$, since $u_{\alpha_{b_2}}(\pi) = 1 < 2 \leq u_{\alpha_{b_2}}(\{ \alpha_{b_3}, \alpha_{b_4} \})$, $v_{\alpha_{b_3}}(h_{a_2}) = 0 < 1 \leq v_{\alpha_{b_3}}(\alpha_{b_2})$, and $v_{\alpha_{b_4}}(h_{a_2}) = 0 < 1 \leq v_{\alpha_{b_4}}(\alpha_{b_2})$. This contradicts the supposition that $\pi$ is j-envy-free.

Suppose then that $u_{\alpha_{b_2}}(\pi) = 2$, so $v_{\alpha_{b_2}}(\alpha_{b_1}) = 2$. Since $\alpha_{b_1} \in C$ by assumption, by the design of the instance it must be that $\alpha_{b_2} \in C$. For the remainder of this lemma only, label $\alpha_{b_1} = c_{i_1}$ and $\alpha_{b_2} = c_{i_2}$. Since $v_{c_{i_2}}(c_{i_1}) = 2$ it follows that $( w_{i_2}, w_{i_1} ) \in A$. Since $G$ is antisymmetric it must be that $( w_{i_1}, w_{i_2} ) \notin A$ and thus that $v_{c_{i_1}}(c_{i_2}) = 1$. Since $\pi(c_{i_1}) = \{ c_{i_1}, c_{i_2}, h_{a_1} \}$ it follows that $u_{c_{i_1}}(\pi) = v_{c_{i_1}}(c_{i_2}) = 1$. Now $\alpha_{b_1}$ has j-envy for $h_{a_2}$, since $u_{\alpha_{b_1}}(\pi) = 1 < 2 \leq u_{\alpha_{b_1}}(\{ \alpha_{b_3}, \alpha_{b_4} \})$, $v_{\alpha_{b_3}}(h_{a_2}) = 0 < 1 \leq v_{\alpha_{b_3}}(\alpha_{b_1})$, and $v_{\alpha_{b_4}}(h_{a_2}) = 0 < 1 \leq v_{\alpha_{b_4}}(\alpha_{b_1})$. This contradicts the supposition that $\pi$ is j-envy-free.
\end{proof}

\begin{lem}
\label{lem:threed_efr_as_jef_terasym_structureofL}
If $(N, V)$ contains a j-envy-free partition into triples then $\{ \{ \alpha_{b_1},\allowbreak\alpha_{b_2} \}, \{ \alpha_{b_3},\allowbreak\alpha_{b_4} \} \} = \{ \{ l_1,\allowbreak l_2 \}, \{ l_3,\allowbreak l_4 \} \}$.
\end{lem}
\begin{proof}
By Lemma~\ref{lem:threed_efr_as_jef_terasym_lequalsthefourisolated}, $\{ \alpha_{b_1}, \alpha_{b_2}, \alpha_{b_3}, \alpha_{b_4} \} = L$. There are now three possibilities: first that $\{ \{ \alpha_{b_1}, \alpha_{b_2} \}, \{ \alpha_{b_3}, \alpha_{b_4} \} \} = \{ \{ l_1, l_3 \}, \{ l_2, l_4 \} \}$, second that $\{ \{ \alpha_{b_1}, \alpha_{b_2} \}, \{ \alpha_{b_3}, \alpha_{b_4} \} \} = \{ \{ l_1, l_4 \}, \{ l_2, l_3 \} \}$, and third that $\{ \{ \alpha_{b_1}, \alpha_{b_2} \}, \{ \alpha_{b_3}, \alpha_{b_4} \} \} = \{ \{ l_1, l_2 \}, \{ l_3, l_4 \} \}$.

First suppose $\{ \{ \alpha_{b_1}, \alpha_{b_2} \}, \{ \alpha_{b_3}, \alpha_{b_4} \} \} = \{ \{ l_1, l_3 \}, \{ l_2, l_4 \} \}$. Without loss of generality assume that $\alpha_{b_1} = l_1$. Now $l_1$ has j-envy for $h_{a_2}$ since $u_{l_1}(\{ h_{a_1}, l_3 \}) = 1 < 3 = u_{l_1}(\{ l_2, l_4 \})$, $v_{l_2}(h_{a_2}) = 0 < 2 = v_{l_2}(l_1)$, and $v_{l_4}(h_{a_2}) = 0 < 1 = v_{l_4}(l_1)$.

Second suppose $\{ \{ \alpha_{b_1}, \alpha_{b_2} \}, \{ \alpha_{b_3}, \alpha_{b_4} \} \} = \{ \{ l_1, l_4 \}, \{ l_2, l_3 \} \}$. Without loss of generality assume that $\alpha_{b_1} = l_1$. As before, $l_1$ has j-envy for $h_{a_2}$ since $u_{l_1}(\{ h_{a_1}, l_4 \}) = 1 < 3 = u_{l_1}(\{ l_2, l_3 \})$, $v_{l_2}(h_{a_2}) = 0 < 2 = v_{l_2}(l_1)$, and $v_{l_3}(h_{a_2}) = 0 < 1 = v_{l_3}(l_1)$.

It remains that $\{ \{ \alpha_{b_1}, \alpha_{b_2} \}, \{ \alpha_{b_3}, \alpha_{b_4} \} \} = \{ \{ l_1, l_2 \}, \{ l_3, l_4 \} \}$.
\end{proof}

By Lemma~\ref{lem:threed_efr_as_jef_terasym_structureofL}, either $\{ \alpha_{b_1}, \alpha_{b_2} \} = \{ l_1, l_2 \}$ or $\{ \alpha_{b_1}, \alpha_{b_2} \} = \{ l_3, l_4 \}$. Without loss of generality assume that $\{ \alpha_{b_1}, \alpha_{b_2} \} = \{ l_3, l_4 \}$.

\begin{lem}
\label{lem:threed_efr_as_jef_terasym_eachcigets3}
If $(N, V)$ contains a j-envy-free partition into triples $\pi$ then $u_{c_i}(\pi) \geq 3$ for each agent $c_i$ in $C$.
\end{lem}
\begin{proof}
Suppose to the contrary that some $1\leq i \leq 3q$ exists where $u_{c_i}(\pi) < 3$. Then $c_i$ has j-envy for $h_{a_1}$ since $u_{c_i}(\pi) \leq 2 < 3 = u_{c_i}(\{ l_3, l_4 \})$, $v_{l_3}(h_{a_1}) = 0 < 1 = v_{l_3}(c_i)$, and $v_{l_4}(h_{a_1}) = 0 < 1 = v_{l_4}(c_i)$.
\end{proof}

\begin{lem}
\label{lem:threed_efr_as_jef_terasym_second_direction}
If $(N, V)$ contains a j-envy-free partition into triples $\pi$ then $G$ contains a directed triangle cover.
\end{lem}
\begin{proof}
Suppose $(N, V)$ contains a j-envy-free partition into triples $\pi$. Lemma~\ref{lem:threed_efr_as_jef_terasym_eachcigets3} shows that $u_{c_i}(\pi) \geq 3$ for each agent $c_i$ in $C$. By construction, it follows that $\pi(c_i)$ contains two agents $c_j, c_k$ such that $v_{c_i}(c_j) \geq 1$ and $v_{c_i}(c_k) = 2$. Hence $c_k$ corresponds to a vertex $w_k$ in $W$ where $( w_i, w_k ) \in A$ and, since $G$ is antisymmetric, $( w_k, w_i ) \notin A$. Since $c_i$ was chosen arbitrarily it follows that $\{ w_i, w_j, w_k \}$ is a directed $3$-cycle in $G$. It follows thus that there are exactly $q$ triples in $\pi$ each containing three agents $\{ c_i, c_j, c_k \}$ where the three corresponding vertices $\{ w_i, w_j, w_k \}$ form a directed $3$-cycle in $G$. From these triples a directed triangle cover $X$ can be easily constructed.
\end{proof}


We have now shown that the ASHG $(N, V)$ contains a j-envy-free partition into triples if and only if the DTC instance $G$ contains a directed triangle cover. This shows that the reduction is correct.

\begin{thm}
\label{thm:threed_efr_as_jef_terasym_npcomplete}
Deciding if a given ASHG contains a j-envy-free partition into triples is $\NP$-complete even when preferences are ternary.
\end{thm}
\begin{proof}
It is straightforward to show that this decision problem belongs to $\NP$, since for any two agents $\alpha_i, \alpha_j \in N$ we can test if $\alpha_i$ j-envies $\alpha_j$ in constant time. 

We have presented a polynomial-time reduction from a special case of DTC, which is $\NP$-complete~\cite{CFM05}. Given a directed antisymmetric graph $G$, the reduction constructs an ASHG with ternary preferences $(N, V)$. Together, Lemmas~\ref{lem:threed_efr_as_jef_terasym_first_direction} and~\ref{lem:threed_efr_as_jef_terasym_second_direction} show that $(N, V)$ contains a j-envy-free partition into triples if and only if $G$ contains a directed triangle cover, and thus that this decision problem is $\NP$-hard.
\end{proof}

\subsection{Non-binary and symmetric preferences}
\label{sec:threed_efr_as_jef_symmetricnonbinrary}
From Theorems~\ref{thm:threed_efr_as_jef_binary_algorithm} and~\ref{thm:threed_efr_as_jef_terasym_npcomplete}, a natural question arises: is it the symmetry of agents' preferences that guarantees the existence of a j-envy-free partition into triples? 

In this section we show that this is not the case, and a j-envy-free partition into triples may not exist even when agents' preferences are symmetric, and the associated existence problem is $\NP$-complete. In fact, we show that this problem is $\NP$-complete even when agents' valuations are symmetric and between $0$ and $6$ inclusive. It remains open whether this result also applies to further restricted cases in which agents' valuations are strictly less than $6$. In particular, whether this result can be strengthened to the restricted case in which preferences are ternary and symmetric.

It seems tricky to design an ASHG in which preferences are non-binary and symmetric that does not contain a j-envy-free partition into triples. In fact, the authors were unable to do this by hand and instead relied on random search, in part inspired by a strategy used by Bullinger~\cite{bullinger21} in a similar endeavour. The search involved repeatedly generating candidate instances and testing, using an integer programming-based solution, whether any candidate did not contain a j-envy-free partition into triples. Certain assumptions were made in order to reduce the search space. For example, it was assumed that any such instance would contain at most $12$ agents. It was also assumed that there would be a single agent for which any other agent has a valuation of $0$. This assumption was inspired by the idea of an ``undesired guest''~\cite{BJ02}\cite[Example~3]{GS62}. It was also assumed that the core part of the instance would have a number of structural symmetries.

After the search identified an initial ``no'' instance, the authors were then able to simplify it further. In order to avoid repetition, we do not describe the final instance explicitly. Instead, we first describe the associated polynomial-time reduction and then how to construct the final instance from a gadget used in the reduction. The accompanying proof, which shows that this instance contains no j-envy-free partition into triples, can be derived straightforwardly from the proof of Lemma~\ref{lem:threed_efr_as_jef_second_direction}, which appears later in this section.

The polynomial-time reduction that we present is from \emph{Partition into Triangles} (PIT, Problem~\ref{prob:pit}). It is similar to the reduction we presented in Section~\ref{sec:threed_efr_as_jef_ternary} for the analogous problem involving ternary preferences that are not (necessarily) symmetric. In that section we reduced from Directed Triangle Packing (DTC) but here we reduce from PIT.

\begin{myproblem}[Partition Into Triangles (PIT)]
\label{prob:pit}
\begin{samepage}
\begin{adjustwidth}{8pt}{8pt}
\inp a simple undirected graph $G=(W, E)$ where $|W|=3q$ for some integer $q$\\
\ques Can the vertices of $G$ be partitioned into $q$ disjoint sets $X=\{ X_1, X_2, \dots, X_q \}$, each set containing exactly three vertices, such that each $X_p = \{ w_i, w_j, w_k \}$ where $1\leq p\leq q$ is a triangle?
\end{adjustwidth}
\end{samepage}
\end{myproblem}

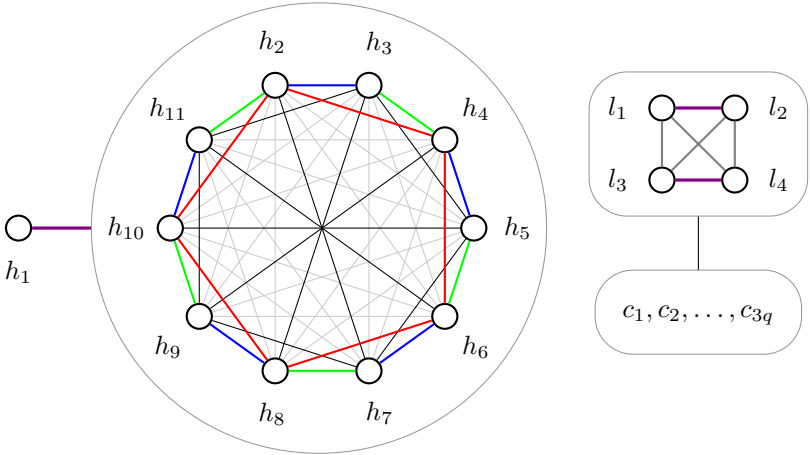
\begin{figure}
    \centering
    \vspace*{4mm}
    \definecolor{figurecolourschemewt1_adjusted}{rgb}{0.8,0.8,0.8}
\definecolor{figurecolourschemewt1}{rgb}{0.5,0.5,0.5}
\definecolor{figurecolourschemewt2}{rgb}{0.5,0.0,0.5}
\definecolor{figurecolourschemewt3}{rgb}{0,0,0}
\definecolor{figurecolourschemewt4}{rgb}{0,0,1}
\definecolor{figurecolourschemewt5}{rgb}{0,1,0}
\definecolor{figurecolourschemewt6}{rgb}{1,0,0}

\begin{tikzpicture}

\begin{scope}[every node/.style={circle,draw, minimum size=2.4mm}, scale=0.8]
    \begin{scope}
        \def\honedist{4.6}
        \def\honeangle{90} 
        \node[thick, circle, label={[label distance=0.4cm]270:$h_1$}] (h1) at ({\honedist*sin(\honeangle)*-1.0 - 1.6}, {\honedist*cos(\honeangle)}) {};
        
        
        \draw[color=figurecolourschemewt2, very thick] (h1) -- (0.0, 0.0);
    
        \begin{scope}[shift={(-1.2, 0.0)}]
            \def\scalefactor{0.25}
            \def\hlabeldist{0.4cm}
            
            \filldraw[color=enclosure_color, fill=white](-0.05, 0) circle (\scalefactor*15);
            
            \node[thick, circle, label={[label distance=\hlabeldist]76:$h_3$}] (h3) at (\scalefactor*3.09,\scalefactor*9.51) {};
            \node[thick, circle, label={[label distance=\hlabeldist]94:$h_2$}] (h2) at (\scalefactor*-3.09,\scalefactor*9.51) {};
            \node[thick, circle, label={[label distance=\hlabeldist]-225:$h_{11}$}] (h11) at (\scalefactor*-8.09,\scalefactor*5.88) {};
            \node[thick, circle, label={[label distance=\hlabeldist]-180:$h_{10}$}] (h10) at (\scalefactor*-10.0,\scalefactor*0.0) {};
            \node[thick, circle, label={[label distance=\hlabeldist]-135:$h_9$}] (h9) at (\scalefactor*-8.09,\scalefactor*-5.88) {};
            \node[thick, circle, label={[label distance=\hlabeldist]-94:$h_8$}] (h8) at (\scalefactor*-3.09,\scalefactor*-9.51) {};
            \node[thick, circle, label={[label distance=\hlabeldist]-76:$h_7$}] (h7) at (\scalefactor*3.09,\scalefactor*-9.51) {};
            \node[thick, circle, label={[label distance=\hlabeldist]-45:$h_6$}] (h6) at (\scalefactor*8.09,\scalefactor*-5.88) {};
            \node[thick, circle, label={[label distance=\hlabeldist]0:$h_5$}] (h5) at (\scalefactor*10.0,\scalefactor*0.0) {};
            \node[thick, circle, label={[label distance=\hlabeldist]45:$h_4$}] (h4) at (\scalefactor*8.09,\scalefactor*5.88) {};
        \end{scope}

        \begin{scope}[shift={(5.0, 0.0)}]
            \draw[color=figurecolourschemewt3] (0.0,-2.0) -- (0.0, 2.0);
        
        \end{scope}
        
        \begin{scope}[shift={(5.0, -1.4)}]
            \filldraw[color=enclosure_color, fill=white, rounded corners=0.5cm] (-1.7, -0.7) rectangle (1.7, 0.7) {};
            \node[draw=none] (clabel) at (0.0, -0.05) {$c_1, c_2, \dots, c_{3q}$};
            
        \end{scope}
        
        \begin{scope}[shift={(5.0, 1.4)}]
            \def\scalefactorl{1.2}
            \filldraw[color=enclosure_color, fill=white, rounded corners=0.5cm] (\scalefactorl*-1.5,\scalefactorl*-1.0) rectangle (\scalefactorl*1.5,\scalefactorl*1.0) {};
            \node[thick, circle, label={[label distance=0.4cm]180:$l_1$}] (l1) at (\scalefactorl*-0.5,\scalefactorl*0.5) {};
            \node[thick, circle, label={[label distance=0.4cm]0:$l_2$}] (l2) at (\scalefactorl*0.5,\scalefactorl*0.5) {};
            \node[thick, circle, label={[label distance=0.4cm]180:$l_3$}] (l3) at (\scalefactorl*-0.5,\scalefactorl*-0.5) {};
            \node[thick, circle, label={[label distance=0.4cm]0:$l_4$}] (l4) at (\scalefactorl*0.5,\scalefactorl*-0.5) {};
        \end{scope}
        
        \begin{scope}
            \draw[color=figurecolourschemewt1, thick] (l1) -- (l3) -- (l2) -- (l4) -- (l1);
            \draw[color=figurecolourschemewt2, very thick] (l1) -- (l2);
            \draw[color=figurecolourschemewt2, very thick] (l3) -- (l4);
        \end{scope}
        
        \begin{scope}
            \foreach \i in {2,3,4,5,6,7,8,9,10,11} \draw[color=figurecolourschemewt1_adjusted] let \n1={int(mod(\i+1,10)+2)} in (h\i) -- (h\n1);
            \foreach \i in {2,3,4,5,6,7,8,9,10,11} \draw[color=figurecolourschemewt1_adjusted] let \n1={int(mod(\i+2,10)+2)} in (h\i) -- (h\n1);
            
            \foreach \i in {3,5,7,9,11} \draw[color=figurecolourschemewt3] let \n1={int(mod(\i,10) + 2)} in (h\i) -- (h\n1);
            \foreach \i in {2,3,4,5,6} \draw[color=figurecolourschemewt3] let \n1={int(mod(\i,10) + 5)} in (h\i) -- (h\n1);
            
            \foreach \i in {2,4,6,8,10} \draw[color=figurecolourschemewt4, thick] let \n1={int(mod(\i,11) + 1)} in (h\i) -- (h\n1);
            
            \foreach \i in {3,5,7,9,11} \draw[color=figurecolourschemewt5, thick] let \n1={int(mod(\i,10) + 1)} in (h\i) -- (h\n1);
            
            \foreach \i in {2,4,6,8,10} \draw[color=figurecolourschemewt6, thick] let \n1={int(mod(\i,10) + 2)} in (h\i) -- (h\n1);
        \end{scope}
    \end{scope}
\end{scope}
\begin{scope}
\end{scope}
\end{tikzpicture}
    \vspace*{4mm}
    \caption[The reduction from PIT to the problem of deciding if an ASHG with symmetric preferences contains a j-envy-free partition into triples]{The reduction from PIT to the problem of deciding if an ASHG with symmetric preferences contains a j-envy-free partition into triples. Valuation colour key: \textcolor{figurecolourschemewt6}{\figurecolorschemewtsixname} - 6, \textcolor{figurecolourschemewt5}{\figurecolorschemewtfivename} - 5, \textcolor{figurecolourschemewt4}{\figurecolorschemewtfourname} - 4, \textcolor{figurecolourschemewt3}{\figurecolorschemewtthreename} - 3, \textcolor{figurecolourschemewt2}{\figurecolorschemewttwoname} - 2, \textcolor{figurecolourschemewt1}{\figurecolorschemewtonename} - 1.}
    \label{fig:threed_efr_as_jef_symmetric_reduction}
\end{figure}

PIT is $\NP$-complete~\cite[Theorem~3.7]{GJ79}. The reduction, illustrated in Figure~\ref{fig:threed_efr_as_jef_symmetric_reduction}, is as follows. Suppose $G$ is an arbitrary instance of PIT. We shall construct an ASHG $(N, V)$ that has symmetric preferences and maximum valuation $6$. Since the valuations in $(N, V)$ will be symmetric, we shall usually specify valuations in one direction only. For example, instead of writing ``let $v_{\alpha_i}(\alpha_j)=v_{\alpha_j}(\alpha_i)=1$'' we write ``let $v_{\alpha_i}(\alpha_j)=1$''. Unless otherwise specified, assume that $v_{\alpha_i}(\alpha_j)=0$ for any $\alpha_i, \alpha_j \in N$. To simplify the description of the reduction, in this section we write $i \myoplus y$ meaning $((i + y - 2) \bmod 10) + 2$. 

First, construct a set of eleven agents $H = \{ h_1, h_2, \dots, h_{11} \}$. For each $i$ where $2\leq i \leq 11$ let $v_{h_1}(h_i) = 2$. For each $i$ where $2\leq i \leq 11$, let:
\begin{itemize}
    \item $v_{h_i}(h_{i \myoplus 1}) = 4$ if $i$ is even otherwise $5$
    \item $v_{h_i}(h_{i \myoplus 2}) = 6$ if $i$ is even otherwise $3$
    \item $v_{h_i}(h_{i \myoplus 3}) = 1$
    \item $v_{h_i}(h_{i \myoplus 4}) = 1$
    \item $v_{h_i}(h_{i \myoplus 5}) = 3$.
\end{itemize}
Next, construct a set of four agents $L = \{ l_1, l_2, l_3, l_4 \}$. Let $v_{l_1}(l_2) = v_{l_3}(l_4) = 2$ and $v_{l_1}(l_3) = v_{l_1}(l_4) = v_{l_2}(l_3) = v_{l_2}(l_4) = 1$.

Next, construct a set of $3q$ agents $C = \{ c_1, c_2, \dots, c_{3q} \}$. Let $v_{c_i}(l_r) = 3$ for each $i$ and $r$ where $1\leq i \leq 3q$ and $1\leq r \leq 4$. For each $i$ and $j$ where $1\leq i, j \leq 3q$ let $v_{c_i}(c_j) = 3$ if $\{ w_i, w_j \} \in E$ otherwise $2$. The construction of $(N, V)$ is now complete. Note that the structure of the valuations among the agents in $C$ reflects the graph $G$.

We remark that the design of $H$ is derived from a particular instance that contains no j-envy-free partition into triples. To construct this instance, delete every agent in $N$ other than the agents in $H$ and $l_1$. The resulting instance thus contains valuations between $0$ and $6$ inclusive.

It is straightforward to show that the reduction runs in polynomial time. To prove that the reduction is correct we show that the ASHG $(N, V)$ contains a j-envy-free partition into triples if and only if the PIT instance $G$ contains a partition into triangles.

We first show that if the PIT instance $G$ contains a partition into triangles then the ASHG $(N, V)$ contains a j-envy-free partition into triples.

\begin{lem}
\label{lem:threed_efr_as_jef_first_direction}
If $G$ contains a partition into triangles then $(N, V)$ contains a j-envy-free partition into triples.
\end{lem}
\begin{proof}
Suppose $G$ contains a partition into triangles $X = \{ X_1, X_2, \dots, X_q \}$. We shall construct a partition into triples $\pi$ that is j-envy-free. First, add $\{ h_2, h_{10}, h_{11} \}$, $\{ h_5, h_6, h_8 \}$, $\{ h_1, h_9, h_4 \}$, $\{ h_3, l_1, l_2 \}$ and $\{ h_7, l_3, l_4 \}$ to $\pi$. Next, for each triangle $X_p = \{ w_i, w_j, w_k \}$ in $X$, add $\{ c_i, c_j, c_k \}$ to $\pi$.

Suppose for a contradiction that some agent $\alpha_j$ exists where $\alpha_j$ has j-envy for some other agent $\alpha_{k_1}$ where $\pi(\alpha_{k_1}) = \{ \alpha_{k_1}, \alpha_{k_2}, \alpha_{k_3} \}$. Since $N = H \cup L \cup C$ it must be that either $\alpha_{k_1} \in H$, $\alpha_{k_1} \in L$, or $\alpha_{k_1} \in C$. We show that each case leads to a contradiction. It follows that no such $\alpha_j$ exists and thus that $\pi$ is j-envy-free.
\begin{itemize}
    \item Suppose $\alpha_{k_1} \in H$. Either $\alpha_{k_1} \in \{ h_3, h_7 \}$ or $\alpha_{k_1} \in H \setminus \{ h_3, h_7 \}$.
    \begin{itemize}
        \item Suppose $\alpha_{k_1} \in \{ h_3, h_7 \}$. Then it must be that either $\{ \alpha_{k_2}, \alpha_{k_3} \} = \{ l_1, l_2 \}$ or $\{ \alpha_{k_2}, \alpha_{k_3} \} = \{ l_3, l_4 \}$. Suppose firstly that $\{ \alpha_{k_2}, \alpha_{k_3} \} = \{ l_1, l_2 \}$. We can see immediately that $\alpha_j \notin H$ since otherwise $u_{\alpha_j}(\{ l_1, l_2 \}) = 0$. It must also be that $\alpha_j \notin C$, since  $u_{c_{a}}(\{ l_1, l_2 \}) = 6 = u_{c_a}(\pi)$ for any $c_a$ in $C$. Similarly, $u_{l_3}(\{ l_1, l_2 \}) = 2 = u_{l_3}(\pi)$ and $u_{l_4}(\{ l_1, l_2 \}) = 2 = u_{l_4}(\pi)$ so $\alpha_j \neq l_3$ and $\alpha_j \neq l_4$. This shows $\alpha_j \notin L$. We have shown that $\alpha_j \notin H$, $\alpha_j \notin C$, and $\alpha_j \notin L$, which is a contradiction. The proof for the case in which $\{ \alpha_{k_2}, \alpha_{k_3} \} = \{ l_3, l_4 \}$ is symmetric and also leads to a contradiction.
        \item Suppose $\alpha_{k_1} \in H\setminus \{ h_3, h_7 \}$. If $\alpha_{k_1} = h_1$ then $\{ \alpha_{k_2}, \alpha_{k_3} \} = \{ h_4, h_9 \}$ so it must be that $v_{h_4}(\alpha_j) > 2 = v_{h_4}(h_1)$ and $v_{h_9}(\alpha_j) > 2 = v_{h_9}(h_1)$, which is impossible by the design of $H$. The proof for every other assignment of $\alpha_{k_1}$ is similar: if $\alpha_{k_1} = h_2$ then $v_{h_{11}}(\alpha_j) > 5$, which is impossible. If $\alpha_{k_1} = h_4$ then $v_{h_1}(\alpha_j) > 2$, which is impossible. If $\alpha_{k_1} = h_5$ then $v_{h_6}(\alpha_j) > 5$ and $v_{h_8}(\alpha_j) > 1$, which is impossible. If $\alpha_{k_1} = h_6$ then $v_{h_8}(\alpha_j) > 6$, which is impossible. If $\alpha_{k_1} = h_8$ then $v_{h_6}(\alpha_j) > 6$, which is impossible. If $\alpha_{k_1} = h_9$ then $v_{h_1}(\alpha_j) > 2$, which is impossible. If $\alpha_{k_1} = h_{10}$ then $v_{h_2}(\alpha_j) > 6$, which is impossible. If $\alpha_{k_1} = h_{11}$ then $v_{h_2}(\alpha_j) > 5$ and $v_{h_{10}}(\alpha_j) > 4$, which is impossible.
    \end{itemize}
    \item Suppose $\alpha_{k_1} \in C$. By construction, it must be that $\alpha_{k_1} = c_{i_1}$, $\alpha_{k_2} = c_{i_2}$, and $\alpha_{k_3} = c_{i_3}$ where $\{ c_{i_1}, c_{i_2}, c_{i_3} \} \subseteq C$, where the corresponding $(vert)$ices $\{ w_{i_1}, w_{i_2}, w_{i_3} \}$ in $G$ are a triangle. It follows that  $v_{c_{i_2}}(c_{i_1}) = 3$. By assumption, $\alpha_j$ has j-envy for $c_{i_1}$ so it must be that $v_{c_{i_2}}(\alpha_j) > v_{c_{i_2}}(c_{i_1}) = 3$, which is impossible by the design of $C$.
    \item Suppose $\alpha_{k_1} \in L$. It must be that $\alpha_{k_1} = l_{i_1}$ for some integer $i_1$ where $1\leq i_1 \leq 4$, $\alpha_{k_2} = l_{i_2}$ for some integer $i_2$ where $1\leq i_2 \leq 4$ and $\alpha_{k_3} = h_{i_3}$ where $i_3 \in \{ 3, 7 \}$. If $\alpha_j \in H$ then $v_{l_{i_2}}(\alpha_j) = 0$ which contradicts the supposition that $\alpha_j$ has j-envy for $l_{i_1}$. Otherwise, if $\alpha_j \notin H$ then $v_{h_{i_3}}(\alpha_j) = 0$, which also contradicts the supposition that $\alpha_j$ has j-envy for $l_{i_1}$.\qedhere
\end{itemize}
\end{proof}


The next step is to show that if the ASHG $(N, V)$ contains a j-envy-free partition into triples then the PIT instance $G$ contains a partition into triangles. The full proof involves an extensive case analysis so is deferred to Appendix~\ref{sec:appendix_jef_sym_non_binary_proof}.

\begin{restatable}{lem}{lemthreedefrasjefseconddirection}
\label{lem:threed_efr_as_jef_second_direction}
If $(N, V)$ contains a j-envy-free partition into triples then $G$ contains a partition into triangles.
\end{restatable}
\begin{proof}[Proof sketch]
First, we assume that $(N, V)$ contains a j-envy-free partition into triples $\pi$. Next, we consider $H$ and show that there is essentially only one possible configuration of the triples in $\pi$ that contain agents in $H$. Specifically, we show that some triple in $\pi$ comprises $l_1$, $l_2$, and some agent in $H$, which we call $h_{a_1}$. It follows that $u_{c_i}(\pi) = 6$ for each $i$ where $1\leq i \leq 3q$, since otherwise there exists some agent $c_i$ in $C$ with j-envy for $h_{a_1}$. It is then straightforward to show that there are exactly $q$ triples in $\pi$, each of which contains three agents in $C$, that correspond to a triangle in $G$. These $q$ triples reveal a partition into triangles in $G$.
\end{proof}



\begin{thm}
\label{thm:threed_efr_as_jef_symmetric_6_npcomplete}
Deciding if a given ASHG contains a j-envy-free partition into triples is $\NP$-complete even when preferences are symmetric and each agent's valuations are between $0$ and $6$.
\end{thm}
\begin{proof}
It is straightforward to show that this decision problem belongs to $\NP$, since for any two agents $\alpha_i$ and $\alpha_j$ in $N$ we can test if $\alpha_i$ j-envies $\alpha_j$ in constant time. 

We have presented a polynomial-time reduction from PIT, which is $\NP$-complete~\cite{GJ79}. Given an undirected graph $G$, the reduction constructs an ASHG with symmetric preferences $(N, V)$ in which each agent's valuations are between $0$ and $6$ inclusive. Together, Lemmas~\ref{lem:threed_efr_as_jef_first_direction} and~\ref{lem:threed_efr_as_jef_second_direction} show that $(N, V)$ contains a j-envy-free partition into triples if and only if $G$ contains a partition into triangles, and thus that this decision problem is $\NP$-hard.
\end{proof}
\section{Conclusion}
\label{sec:threed_efr_as_conclusion}

In this paper we considered three successively weaker solution concepts: envy-freeness, weakly justified envy-freeness, and justified envy-freeness, and studied the existence of partitions into triples that satisfy each concept together with associated search problems. We imposed various restrictions on the agents' preferences and presented a complete complexity classification in terms of these restrictions.

Our polynomial-time algorithms may have practical applications. For example, our algorithm for justified envy-freeness could be applied to a real-life situation involving ``friends'' and ``neutrals''~\cite{ijcai2017-51}. Our algorithms for envy-free\-ness and wj-envy-freeness can only be applied to settings in which the agents' preferences are heavily restricted. While such a restriction may not necessarily be realistic in practice, in theory characterising the frontier between solvability and $\NP$-completeness is a first step.

The scope for future work is wide. An immediate open question is whether Theorem~\ref{thm:threed_efr_as_jef_symmetric_6_npcomplete} can be strengthened to ASHGs with ternary and symmetric preferences. As we remarked in Section~\ref{sec:threed_efr_as_jef_symmetricnonbinrary}, it seems tricky to design an ASHG with symmetric preferences in which there is no j-envy-free partition into triples. We discussed the random search technique used to discover such an instance, which was partly inspired by previous work~\cite{bullinger21}. Although this technique was effective, it seems hard to provide intuition as to why the discovered instance contains no j-envy-free partition into triples. It is also still unknown whether any simpler instance exists, for example with fewer than $12$ agents. 

Another immediate question is the extent to which our results generalise to other restrictions on coalition size, for example if coalitions must have size $k$, for some fixed $k$ where $k\geq 3$. We conjecture that all of our $\NP$-hardness reductions can be generalised to this setting.

Studying more general solution concepts, such as envy-free up to $r$ (EF-$r$)~\cite{Li2023journal}, may lead to more interesting efficient algorithms. It is also possible to define a new type of envy between j-envy and wj-envy such that $\alpha_i$ has \emph{$r$-j-envy} for $\alpha_j$ if $\alpha_i$ has weakly justified envy for $\alpha_j$ and at least $r$ agents in $\pi(\alpha_j)$ prefer $\alpha_i$ to $\alpha_j$. In a model in which coalitions must have size $k$, $k$-j-envy coalesces with j-envy.

Similarly, it could also be interesting to investigate ordinal preferences, for example considering complete, $\mathscr{B}$-, $\mathscr{W}$-, or lexicographic preferences~\cite{CH02,CH04,Haj06}. We conjecture that in a hedonic game model with $\mathscr{B}$-preferences~\cite{CH02}, an envy-free partition may not exist, and the associated existence problem is solvable in polynomial time.

Another possibility is to identify other restrictions on the agents' valuations in which an envy-free, wj-envy-free or j-envy-free partition into triples can be found in polynomial time. The gadgets used in our reductions are highly regular and it might be that there exist interesting classes of instances that must contain an envy-free, wj-envy-free, or j-envy-free partition into triples. Alternatively, we could study these problems from the perspective of parameterised complexity. For example, in the case of binary and symmetric preferences, one could consider the tree-width~\cite{Robertson84} of the underlying graph. 

It could also be interesting to estimate the probability that a randomly chosen ASHG contains an envy-free, wj-envy-free, or j-envy-free partition into triples, or to estimate the same probability in a random ASHG with binary or ternary preferences. Our complexity results indicate that, among instances with binary and symmetric preferences and maximum degree $2$, the set of ASHGs that contain a j-envy-free partition into triples (i.e.\ all instances) is larger than the set of ASHGs that contain a wj-envy-free partition into triples, which is in turn larger than the set of ASHGs that contain an envy-free partition into triples. We conjecture that, in general, the probability that a given ASHG contains an envy-free partition into triples is smaller than the probability that it contains a wj-envy-free partition into triples, which is in turn smaller than the probability that it contains a j-envy-free partition into triples. 
In this direction, it might be possible to apply probabilistic techniques from graph theory, such as the Erd\H{o}s-R\'enyi model of a random graph. Of course, the probabilistic events in which agents have envy for other agents are not independent, which complicates the analysis. Alternatively, an empirical approach might be informative.

\backmatter





\section*{Acknowledgements}
Michael McKay was supported by Doctoral Training Partnership grant number EP/R513222/1 from the the Engineering and Physical Sciences Research Council. \'Agnes Cseh was supported by the J\'anos Bolyai Research Fellowship. David Manlove was supported by grant number EP/X013618/1 from the Engineering and Physical Sciences Research Council.  The authors are very grateful to the anonymous reviewers who have given detailed comments on versions of this paper. Their suggestions have helped us to significantly improve the presentation and style of the paper.










\begin{appendices}
\section{Proof of Theorem~\ref{thm:threed_efr_as_ef_algorithm}}
\label{sec:appendix_efalgo}

Recall that Lemma~\ref{lem:threed_efr_as_ef_if_and_only_if} shows a necessary and sufficient condition for the existence of an envy-free partition into triples. It is relatively straightforward to adapt the proof of this lemma to show that there exists an $O(|N|)$-time algorithm that either constructs an envy-free partition into triples or reports that no such partition exists. We state this formally as Theorem~\ref{thm:threed_efr_as_ef_algorithm}.

\thmthreedefrasefalgorithm*
\begin{proof}
We shall describe an algorithm that either outputs $\bot$, if no envy-free partition into triples exists, or a labelling $\tau$ of each agent $\alpha_i$ such that $1\leq \tau(\alpha_i) \leq n$ that represents the index of $\pi(\alpha_i)$ in an arbitrary ordering of the triples in an envy-free partition into triples $\pi$. Define $\mathcal{Q}$ and $\mathcal{R}$ as in Lemma~\ref{lem:threed_efr_as_ef_if_and_only_if}.

The algorithm has three phases. In the first phase, the algorithm constructs a stack $P$ that contains all isolated agents in $(N, E)$. It also constructs a stack $T$, which contains exactly one agent in each component of two or more agents, such that if a component is a path then $T$ contains one of its endpoints. The construction of $P$ and $T$ can be completed in $O(|N|)$ time.

The algorithm now enters the second phase. In this phase, the algorithm maintains a counter $r$ to track the label of the agent last labelled. Initially, $r=1$. The algorithm pops an unmarked agent $m_i$ from the stack $T$ and marks $\tau(m_i)=1$. It sets a new counter $c$ to $1$, which will track the number of agents in the ``current'' component, which contains $m_i$. It then identifies successive adjacent agents and labels each one with $r$, incrementing $r$ by one every third agent, following the path or cycle in the underlying graph. The successive agents are therefore marked $1,1,1,2,2,2,3,3,3\dots$ The counter $c$ is updated to ensure that $c$ is the number of agents in this component. Eventually, either some agent with degree $1$ or some previously labelled agent is discovered. In this case, there are three possibilities. 

The first possibility is that $c=3{k_3}$ for some integer $k_3$ where ${k_3}\geq 1$. In this case the algorithm pops some yet unlabelled $m_i$ from the stack $T$ and repeats the above process. 

The second possibility is that $c=3{k_2}-1$ for some integer $k_2$ where ${k_2}\geq 1$. In this case the algorithm pops some isolated agent $p_i$ from the stack $P$, labels $\tau(p_i)=r$, pops some yet-unlabelled $m_i$ from the stack $T$, and repeats the above process. If the stack $P$ is empty, then it must be that $2|\mathcal{Q}| + |\mathcal{R}| > |P|$. The algorithm then returns $\bot$, since by Lemma~\ref{lem:threed_efr_as_ef_if_and_only_if} no envy-free partition into triples exists. 

The third possibility is that $c=3{k_1}-2$ for some integer $k_1$ where ${k_1}>1$. In this case it must be that exactly one agent has been labelled with the current value of $r$. We call this agent $\alpha_j$. Next, the algorithm identifies the last agent $\alpha_l$ that was labelled with $r - 1$, which must be adjacent to $\alpha_j$. It relabels $\alpha_l$ so that $\tau(\alpha_l) = r$. It now follows that exactly two adjacent agents are labelled with $r - 1$ and exactly two adjacent agents are labelled with $r$. The algorithm then pops two agents $p_g$ and $p_h$ from the stack $P$, labels $\tau(p_g) = r$ and $\tau(p_h) = r-1$, and then pops some yet-unlabelled $m_i$ from the stack $T$ and repeats the above process. If $|P| < 2$ then it must be that $2|\mathcal{Q}| + |\mathcal{R}| > |P|$. The algorithm then returns $\bot$, since by Lemma~\ref{lem:threed_efr_as_ef_if_and_only_if} no envy-free partition into triples exists. 

In the third phase, since the algorithm has not yet returned $\bot$, it must be that each agent with degree $1$ or more has been labelled and therefore assigned to some triple in $\pi$. The algorithm arbitrarily assigns the remaining agents in $P$ to triples in $\pi$ by popping the next agent $p_i$ from the stack $P$, labelling $\tau(p_i) = r$, and incrementing $r$ every third agent. The argument from Lemma~\ref{lem:threed_efr_as_ef_if_and_only_if} shows that the partition into triples represented by $\tau$ is envy-free.
\end{proof}

\section{Proof of Lemma~\ref{lem:threed_efr_as_jef_second_direction}}
\label{sec:appendix_jef_sym_non_binary_proof}

Recall that $G$ is an arbitrary instance of PIT and and $(N, V)$ is the ASHG constructed by the reduction. Our goal is to show that if $(N, V)$ contains a j-envy-free partition into triples then $G$ contains a partition into triangles. To this end we prove a sequence of intermediary results in Lemmas~\ref{lem:threed_efr_as_jef_two_and_one_in_h}--\ref{lem:threed_efr_as_jef_eachpigets6}.

To begin, we define two possible configurations of $H$ in an arbitrary j-envy-free partition into triples $\pi$. 

If some triple $t$ in $\pi$ contains exactly one agent in $H$ then let us say that $H$ has an \emph{open configuration in $\pi$}. Otherwise, let us say that $H$ has a \emph{closed configuration in $\pi$}. We will eventually show, in Lemma~\ref{lem:threed_efr_as_jef_hopen}, that the only possible configuration of $H$ in $\pi$ is an open configuration. 


\begin{lem}
\label{lem:threed_efr_as_jef_two_and_one_in_h}
If $(N, V)$ contains a j-envy-free partition into triples $\pi$ then no triples $t_1, t_2$ in $\pi$ exist such that $t_1$ contains exactly two agents in $H$ and $t_2$ contains exactly one agent in $H$.
\end{lem}
\begin{proof}
Suppose for a contradiction that there exists some such $t_1$ and $t_2$ in $\pi$. Suppose $t_1 = \{ h_{i_1}, h_{i_2}, \alpha_{j_1} \}$ and $t_2 = \{ h_{i_3}, \alpha_{j_2}, \alpha_{j_3} \}$ where $1\leq i_1, i_2, i_3 \leq 11$ and $\alpha_{j_1}, \alpha_{j_2}, \alpha_{j_3} \in N \setminus H$. Now $h_{i_3}$ has j-envy for $\alpha_{j_1}$ since $u_{h_{i_3}}(\pi) = 0 < 2 \leq u_{h_{i_3}}(\{ h_{i_1}, h_{i_2} \})$, $v_{h_{i_1}}(\alpha_{j_1}) = 0 < 1 \leq v_{h_{i_1}}(h_{i_3})$ and $v_{h_{i_2}}(\alpha_{j_1}) = 0 < 1 \leq v_{h_{i_2}}(h_{i_3})$. This contradicts our supposition that $\pi$ is j-envy-free.
\end{proof}

\begin{lem}
\label{lem:threed_efr_as_jef_3332_case_part1}
If $(N, V)$ contains a j-envy-free partition into triples $\pi$, $\sigma(H, \pi) = 4$, and $u_{h_1}(\pi) < 4$ then $H$ has an open configuration in $\pi$.
\end{lem}
\begin{proof}
Suppose to the contrary that $\sigma(H, \pi) = 4$, $u_{h_1}(\pi) < 4$, and $H$ has a closed configuration in $\pi$. Since $\sigma(H, \pi) = 4$ it must be that three triples in $\pi$ each contains exactly three agents in $H$ and one triple in $\pi$ contains exactly two agents in $H$. Suppose then that the triples $t_1$, $t_2$, and $t_3$ in $\pi$ each contains exactly three agents in $H$ and some triple $t_4$ in $\pi$ contains exactly two agents in $H$. Since $u_{h_1}(\pi) < 4$ by assumption, by the design of $H$ it follows that $\pi(h_1)$ contains at most one agent in $H \setminus \{ h_1 \}$ and therefore $h_1 \in t_4$. It follows that $t_4 = \{ h_1, h_{i_1}, \alpha_{j} \}$ where $2 \leq i_1 \leq 11$ and $\alpha_{j} \in N \setminus H$. We use a case analysis to prove a contradiction occurs for each possible assignment of $i_1$. 

\begin{itemize}
    \item Suppose first $i_1 = 2$. If $u_{h_4}(\pi) \leq 7$ then $h_4$ has j-envy for $\alpha_{j}$, since $u_{h_4}(\pi) \leq 7 < 8 = u_{h_4}(\{ h_1, h_2 \})$, $v_{h_1}(\alpha_{j}) = 0 < 2 = v_{h_1}(h_4)$, and $v_{h_2}(\alpha_{j}) = 0 < 6 = v_{h_2}(h_4)$. It follows that $u_{h_4}(\pi) \geq 8$. By the assumptions regarding the structure of $\pi$, it must be that $\pi(h_4)$ contains three agents in $H$. Since $t_4 = \{ h_1, h_2, \alpha_{j} \}$ it must be that $\pi(h_4) = \{ h_4, h_{i_2}, h_{i_3} \}$ where $\{ i_2, i_3 \} \subset \{ 3, 5, 6, 7, 8, 9, 10, 11 \}$. Recall that $v_{h_4}(h_3)=5$, $v_{h_4}(h_5)=4$, $v_{h_4}(h_6)=6$, $v_{h_4}(h_7)=1$, $v_{h_4}(h_8)=1$, $v_{h_4}(h_9)=3$, $v_{h_4}(h_{10})=1$, and $v_{h_4}(h_{11})=1$. Since we established $u_{h_4}(\pi) \geq 8$ it follows that there are 5 possibilities: $\{ i_2, i_3 \} = \{ 3, 9 \}$, $\{ i_2, i_3 \} = \{ 3, 5 \}$, $\{ i_2, i_3 \} = \{ 3, 6 \}$, $\{ i_2, i_3 \} = \{ 5, 6 \}$, and $\{ i_2, i_3 \} = \{ 6, 9 \}$, which we shall now consider.
\begin{itemize}
    \item Suppose $\{ i_2, i_3 \} = \{ 3, 9 \}$. It follows that $h_2$ has j-envy for $h_9$, since $u_{h_2}(\pi) = 2 < 10 = u_{h_2}(\{ h_3, h_4 \})$, $v_{h_3}(h_9) = 1 < 4 = v_{h_3}(h_2)$, and $v_{h_4}(h_9) = 3 < 6 = v_{h_4}(h_2)$. This contradicts the supposition that $\pi$ is j-envy-free.
    \item Suppose $\{ i_2, i_3 \} = \{ 3, 5 \}$. It follows that $h_2$ has j-envy for $h_5$, since $u_{h_2}(\pi) = 2 < 10 = u_{h_2}(\{ h_3, h_4 \})$, $v_{h_3}(h_5) = 3 < 4 = v_{h_3}(h_2)$, and $v_{h_4}(h_5) = 4 < 6 = v_{h_4}(h_2)$. 
    \item Suppose $\{ i_2, i_3 \} = \{ 3, 6 \}$. Consider $h_{11}$. If $u_{h_{11}}(\pi) \leq 6$ then $h_{11}$ has j-envy for $\alpha_{j}$, since $u_{h_{11}}(\pi) \leq 6 < 7 = u_{h_{11}}(\{ h_1, h_2 \})$, $v_{h_1}(\alpha_{j}) = 0 < 2 = v_{h_1}(h_{11})$, and $v_{h_2}(\alpha_{j}) = 0 < 5 = v_{h_2}(h_{11})$. It follows that $u_{h_{11}}(\pi) \geq 7$. We have established that $h_2 \notin \pi(h_{11})$, $h_3 \notin \pi(h_{11})$, and $h_6 \notin \pi(h_{11})$ so, by the design of $H$, it must be that $\pi(h_{11}) = \{ h_9, h_{10}, h_{11} \}$. Now $h_2$ has j-envy for $h_9$, since $u_{h_2}(\pi) = 2 < 11 = u_{h_2}(\{ h_{10}, h_{11} \})$, $v_{h_{10}}(h_9) = 5 < 6 = v_{h_{10}}(h_2)$, and $v_{h_{11}}(h_9) = 3 < 5 = v_{h_{11}}(h_2)$.
    \item Suppose $\{ i_2, i_3 \} = \{ 5, 6 \}$. Consider $h_3$. If $u_{h_3}(\pi) \leq 5$ then $h_3$ has j-envy for $\alpha_{j}$, since $u_{h_3}(\pi) \leq 5 < 6 = u_{h_3}(\{ h_1, h_2 \})$, $v_{h_1}(\alpha_{j}) = 0 < 2 = v_{h_1}(h_3)$, and $v_{h_2}(\alpha_{j}) = 0 < 4 = v_{h_2}(h_3)$. It follows that $u_{h_3}(\pi) \geq 6$. We have established that $h_2 \notin \pi(h_3)$, $h_4 \notin \pi(h_3)$, and $h_5 \notin \pi(h_3)$ so, by the design of $H$, it must be that $\pi(h_3) = \{ h_3, h_8, h_{11} \}$. Now $h_{11}$ has j-envy for $\alpha_{j}$, since $u_{h_{11}}(\pi) = 4 < 7 = u_{h_{11}}(\{ h_1, h_2 \})$, $v_{h_1}(\alpha_{j}) = 0 < 2 = v_{h_1}(h_{11})$, and $v_{h_2}(\alpha_{j}) = 0 < 5 = v_{h_2}(h_{11})$.
    \item Suppose $\{ i_2, i_3 \} = \{ 6, 9 \}$. Consider $h_{11}$. If $u_{h_{11}}(\pi) \leq 6$ then $h_{11}$ has j-envy for $\alpha_{j}$, since $u_{h_{11}}(\pi) \leq 6 < 7 = u_{h_{11}}(\{ h_1, h_2 \})$, $v_{h_1}(\alpha_{j}) = 0 < 2 = v_{h_1}(h_{11})$, and $v_{h_2}(\alpha_{j}) = 0 < 5 = v_{h_2}(h_{11})$. It follows that $u_{h_{11}}(\pi) \geq 7$. We have established that $h_2 \notin \pi(h_{11})$, $h_6 \notin \pi(h_{11})$, and $h_9 \notin \pi(h_{11})$ so, by the design of $H$, it must be that $\pi(h_{11}) = \{ h_3, h_{10}, h_{11} \}$. Now $h_2$ has j-envy for $h_{10}$, since $u_{h_2}(\pi) = 2 < 9 = u_{h_2}(\{ h_3, h_{11} \})$, $v_{h_3}(h_{10}) = 1 < 4 = v_{h_3}(h_2)$, and $v_{h_{11}}(h_{10}) = 4 < 5 = v_{h_{11}}(h_2)$.
\end{itemize}
\item Suppose next $i_1 = 3$. If $u_{h_4}(\pi) \leq 6$ then $h_4$ has j-envy for $\alpha_{j}$, since $u_{h_4}(\pi) \leq 6 < 7 = u_{h_4}(\{ h_1, h_3 \})$, $v_{h_1}(\alpha_{j}) = 0 < 2 = v_{h_1}(h_4)$, and $v_{h_3}(\alpha_{j}) = 0 < 5 = v_{h_3}(h_4)$. It follows that $u_{h_4}(\pi) \geq 7$. Since $\pi(h_4) \neq t_4$ it must be that $\pi(h_4) = \{ h_4, h_{i_2}, h_{i_3} \}$ where $\{ i_2, i_3 \} \subset \{ 2, 5, 6, 7, 8, 9, 10, 11 \}$. Recall that $v_{h_4}(h_2)=6$, $v_{h_4}(h_5)=4$, $v_{h_4}(h_6)=6$, $v_{h_4}(h_7)=1$, $v_{h_4}(h_8)=1$, $v_{h_4}(h_9)=3$, $v_{h_4}(h_{10})=1$, and $v_{h_4}(h_{11})=1$. Since we established $u_{h_4}(\pi) \geq 7$ it follows that there are $14$ possibilities: $\{ i_2, i_3 \} = \{ 2, 5 \}$, $\{ i_2, i_3 \} = \{ 2, 6 \}$, $\{ i_2, i_3 \} = \{ 2, 7 \}$, $\{ i_2, i_3 \} = \{ 2, 8 \}$, $\{ i_2, i_3 \} = \{ 2, 9 \}$, $\{ i_2, i_3 \} = \{ 2, 10 \}$, $\{ i_2, i_3 \} = \{ 2, 11 \}$, $\{ i_2, i_3 \} = \{ 5, 6 \}$, $\{ i_2, i_3 \} = \{ 5, 9 \}$, $\{ i_2, i_3 \} = \{ 6, 7 \}$, $\{ i_2, i_3 \} = \{ 6, 8 \}$, $\{ i_2, i_3 \} = \{ 6, 9 \}$, $\{ i_2, i_3 \} = \{ 6, 10 \}$, and $\{ i_2, i_3 \} = \{ 6, 11 \}$, which we shall now consider.
\begin{itemize}
    \item Suppose $\{ i_2, i_3 \} = \{ 2, 5 \}$. It follows that $h_3$ has j-envy for $h_5$, since $u_{h_3}(\pi) = 2 < 9 = u_{h_3}(\{ h_2, h_4 \})$, $v_{h_2}(h_5) = 1 < 4 = v_{h_2}(h_3)$, and $v_{h_4}(h_5) = 4 < 5 = v_{h_4}(h_3)$.
    \item Suppose $\{ i_2, i_3 \} = \{ 2, 6 \}$. If $u_{h_5}(\pi) \leq 4$ then $h_5$ has j-envy for $\alpha_{j}$, since $u_{h_5}(\pi) \leq 4 < 5 = u_{h_5}(\{ h_1, h_3 \})$, $v_{h_1}(\alpha_{j}) = 0 < 2 = v_{h_1}(h_5)$, and $v_{h_3}(\alpha_{j}) = 0 < 3 = v_{h_3}(h_5)$. It follows that $u_{h_5}(\pi) \geq 5$. We have established that $h_3 \notin \pi(h_5)$, $h_4 \notin \pi(h_5)$, and $h_6 \notin \pi(h_5)$ so, by the design of $H$, it must be that $\pi(h_5) = \{ h_5, h_7, h_{10} \}$. It remains that $\pi(h_{11}) = \{ h_8, h_9, h_{11} \}$. Now $h_{11}$ has j-envy for $\alpha_{j}$, since $u_{h_{11}}(\pi) = 4 < 5 = u_{h_{11}}(\{ h_1, h_3 \})$, $v_{h_1}(\alpha_{j}) = 0 < 2 = v_{h_1}(h_{11})$, and $v_{h_3}(\alpha_{j}) = 0 < 3 = v_{h_3}(h_{11})$.
    \item Suppose $\{ i_2, i_3 \} = \{ 2, 7 \}$. It follows that $h_3$ has j-envy for $h_7$, since $u_{h_3}(\pi) = 2 < 9 = u_{h_3}(\{ h_2, h_4 \})$, $v_{h_2}(h_7) = 3 < 4 = v_{h_2}(h_3)$, and $v_{h_4}(h_7) = 1 < 5 = v_{h_4}(h_3)$.
    \item Suppose $\{ i_2, i_3 \} = \{ 2, 8 \}$. It follows that $h_3$ has j-envy for $h_8$, since $u_{h_3}(\pi) = 2 < 9 = u_{h_3}(\{ h_2, h_4 \})$, $v_{h_2}(h_8) = 1 < 4 = v_{h_2}(h_3)$, and $v_{h_4}(h_8) = 1 < 5 = v_{h_4}(h_3)$.
    \item Suppose $\{ i_2, i_3 \} = \{ 2, 9 \}$. It follows that $h_3$ has j-envy for $h_9$, since $u_{h_3}(\pi) = 2 < 9 = u_{h_3}(\{ h_2, h_4 \})$, $v_{h_2}(h_9) = 1 < 4 = v_{h_2}(h_3)$, and $v_{h_4}(h_9) = 3 < 5 = v_{h_4}(h_3)$.
    \item Suppose $\{ i_2, i_3 \} = \{ 2, 10 \}$. If $u_{h_{11}}(\pi) \leq 4$ then $h_{11}$ has j-envy for $\alpha_{j}$, since $u_{h_{11}}(\pi) \leq 4 < 5 = u_{h_5}(\{ h_1, h_3 \})$, $v_{h_1}(\alpha_{j}) = 0 < 2 = v_{h_1}(h_{11})$, and $v_{h_3}(\alpha_{j}) = 0 < 3 = v_{h_3}(h_{11})$. It follows that $u_{h_{11}}(\pi) \geq 5$. We have established that $h_2 \notin \pi(h_{11})$, $h_3 \notin \pi(h_{11})$, and $h_{10} \notin \pi(h_{11})$ so, by the design of $H$, it must be that $\pi(h_{11}) = \{ h_6, h_9, h_{11} \}$. Since $h_5 \notin t_4$, it must be that $\pi(h_5)$ contains three agents in $H$ and thus $\pi(h_5) = \{ h_5, h_7, h_8 \}$. Now $h_6$ has j-envy for $h_5$, since $u_{h_6}(\pi) = 4 < 10 = u_{h_6}(\{ h_7, h_8 \})$, $v_{h_7}(h_5) = 3 < 4 = v_{h_7}(h_6)$, and $v_{h_8}(h_5) = 1 < 6 = v_{h_8}(h_6)$.
    \item Suppose $\{ i_2, i_3 \} = \{ 2, 11 \}$. If $u_{h_5}(\pi) \leq 4$ then $h_5$ has j-envy for $\alpha_{j}$, since $u_{h_5}(\pi) \leq 4 < 5 = u_{h_5}(\{ h_1, h_3 \})$, $v_{h_1}(\alpha_{j}) = 0 < 2 = v_{h_1}(h_5)$, and $v_{h_3}(\alpha_{j}) = 0 < 3 = v_{h_3}(h_5)$. It follows that $u_{h_5}(\pi) \geq 5$. We have established that $h_3 \notin \pi(h_5)$ and $h_4 \notin \pi(h_5)$ so, by the design of $H$, there are three possibilities: either $\pi(h_5) = \{ h_5, h_6, h_7 \}$, $\pi(h_5) = \{ h_5, h_6, h_{10} \}$, or $\pi(h_5) = \{ h_5, h_7, h_{10} \}$.
    \begin{itemize}
        \item If $\pi(h_5) = \{ h_5, h_6, h_7 \}$ then $h_4$ has j-envy for $h_7$, since $u_{h_4}(\pi) = 7 < 10 = u_{h_4}(\{ h_5, h_6 \})$, $v_{h_5}(h_7) = 3 < 4 = v_{h_5}(h_4)$, and $v_{h_6}(h_7) = 4 < 6 = v_{h_6}(h_4)$.
        \item If $\pi(h_5) = \{ h_5, h_6, h_{10} \}$ then $h_4$ has j-envy for $h_{10}$, since $u_{h_4}(\pi) = 7 < 10 = u_{h_4}(\{ h_5, h_6 \})$, $v_{h_5}(h_{10}) = 3 < 4 = v_{h_5}(h_4)$, and $v_{h_6}(h_{10}) = 1 < 6 = v_{h_6}(h_4)$.
        \item If $\pi(h_5) = \{ h_5, h_7, h_{10} \}$ then it remains that $\pi(h_6) = \{ h_6, h_8, h_9 \}$. Now $h_6$ has j-envy for $h_{10}$, since $u_{h_6}(\pi) = 7 < 9 = u_{h_6}(\{ h_5, h_7 \})$, $v_{h_5}(h_{10}) = 3 < 5 = v_{h_5}(h_6)$, and $v_{h_7}(h_{10}) = 1 < 4 = v_{h_7}(h_6)$.
    \end{itemize}
    \item Suppose $\{ i_2, i_3 \} = \{ 5, 6 \}$. If $u_{h_2}(\pi) \leq 5$ then $h_2$ has j-envy for $\alpha_{j}$, since $u_{h_2}(\pi) \leq 5 < 6 = u_{h_2}(\{ h_1, h_3 \})$, $v_{h_1}(\alpha_{j}) = 0 < 2 = v_{h_1}(h_2)$, and $v_{h_3}(\alpha_{j}) = 0 < 4 = v_{h_3}(h_2)$. It follows that $u_{h_2}(\pi) \geq 6$. Since $h_2 \notin t_4$ it must be that $\pi(h_2)$ contains three agents in $H$. Since $t_4 = \{ h_1, h_3, \alpha_{j} \}$ and $\pi(h_4) = \{ h_4, h_5, h_6 \}$ it follows that $\pi(h_2) = \{ h_2, h_{i_4}, h_{i_5} \}$ where $\{ i_4, i_5 \} \subset \{ 7, 8, 9, 10, 11 \}$. Recall that $v_{h_2}(h_7)=3$, $v_{h_2}(h_8)=1$, $v_{h_2}(h_9)=1$, $v_{h_2}(h_{10})=6$, and $v_{h_2}(h_{11})=5$. Since we established $u_{h_2}(\pi) \geq 6$ it follows that there are seven possibilities: $\{ h_{i_4}, h_{i_5} \} = \{ 7, 10 \}$, $\{ h_{i_4}, h_{i_5} \} = \{ 7, 11 \}$, $\{ h_{i_4}, h_{i_5} \} = \{ 8, 10 \}$, $\{ h_{i_4}, h_{i_5} \} = \{ 8, 11 \}$, $\{ h_{i_4}, h_{i_5} \} = \{ 9, 10 \}$, $\{ h_{i_4}, h_{i_5} \} = \{ 9, 11 \}$, and $\{ h_{i_4}, h_{i_5} \} = \{ 10, 11 \}$, which we shall now consider.
    \begin{itemize}
        \item If $\{ h_{i_4}, h_{i_5} \} = \{ 7, 10 \}$ then it remains that $\pi(h_{11}) = \{ h_8, h_9, h_{11} \}$. Now $h_{11}$ has j-envy for $h_7$, since $u_{h_{11}}(\pi) = 4 < 9 = u_{h_{11}}(\{ h_2, h_{10} \})$, $v_{h_2}(h_7) = 3 < 5 = v_{h_2}(h_{11})$, and $v_{h_{10}}(h_7) = 1 < 4 = v_{h_{10}}(h_{11})$.
        \item If $\{ h_{i_4}, h_{i_5} \} = \{ 7, 11 \}$ then $h_3$ has j-envy for $h_7$, since $u_{h_3}(\pi) = 2 < 7 = u_{h_3}(\{ h_2, h_{11} \})$, $v_{h_2}(h_7) = 3 < 4 = v_{h_2}(h_3)$, and $v_{h_{11}}(h_7) = 1 < 3 = v_{h_{11}}(h_3)$.
        \item If $\{ h_{i_4}, h_{i_5} \} = \{ 8, 10 \}$ then it remains that $\pi(h_7) = \{ h_7, h_9, h_{11} \}$. Now $h_8$ has j-envy for $h_{11}$, since $u_{h_8}(\pi) = 7 < 9 = u_{h_8}(\{ h_7, h_9 \})$, $v_{h_7}(h_{11}) = 1 < 5 = v_{h_7}(h_8)$, and $v_{h_9}(h_{11}) = 3 < 4 = v_{h_9}(h_8)$.
        \item If $\{ h_{i_4}, h_{i_5} \} = \{ 8, 11 \}$ then $h_3$ has j-envy for $h_8$, since $u_{h_3}(\pi) = 2 < 7 = u_{h_3}(\{ h_2, h_{11} \})$, $v_{h_2}(h_8) = 1 < 4 = v_{h_2}(h_3)$, and $v_{h_{11}}(h_8) = 1 < 3 = v_{h_{11}}(h_3)$.
        \item If $\{ h_{i_4}, h_{i_5} \} = \{ 9, 10 \}$ then it remains that $\pi(h_7) = \{ h_7, h_8, h_{11} \}$. Now $h_9$ has j-envy for $h_{11}$, since $u_{h_9}(\pi) = 6 < 7 = u_{h_9}(\{ h_7, h_8 \})$, $v_{h_7}(h_{11}) = 1 < 3 = v_{h_7}(h_9)$, and $v_{h_8}(h_{11}) = 1 < 4 = v_{h_8}(h_9)$.
        \item If $\{ h_{i_4}, h_{i_5} \} = \{ 9, 11 \}$ then it remains that $\pi(h_{10}) = \{ h_7, h_8, h_{10} \}$. Now $h_{10}$ has j-envy for $h_9$, since $u_{h_{10}}(\pi) = 7 < 10 = u_{h_{10}}(\{ h_2, h_{11} \})$, $v_{h_2}(h_9) = 1 < 6 = v_{h_2}(h_{10})$, and $v_{h_{11}}(h_9) = 3 < 4 = v_{h_{11}}(h_{10})$.
        \item If $\{ h_{i_4}, h_{i_5} \} = \{ 10, 11 \}$ then it remains that $\pi(h_7) = \{ h_7, h_8, h_9 \}$. Now $h_{10}$ has j-envy for $h_7$, since $u_{h_{10}}(\pi) = 10 < 11 = u_{h_{10}}(\{ h_8, h_9 \})$, $v_{h_8}(h_7) = 5 < 6 = v_{h_8}(h_{10})$, and $v_{h_9}(h_7) = 3 < 5 = v_{h_9}(h_{10})$. 
    \end{itemize}
    \item Suppose $\{ i_2, i_3 \} = \{ 5, 9 \}$. It follows that $h_3$ has j-envy for $h_9$, since $u_{h_3}(\pi) = 2 < 8 = u_{h_3}(\{ h_4, h_5 \})$, $v_{h_4}(h_9) = 3 < 5 = v_{h_4}(h_3)$, and $v_{h_5}(h_9) = 1 < 3 = v_{h_5}(h_3)$.
    \item Suppose $\{ i_2, i_3 \} = \{ 6, 7 \}$. Consider $h_5$. Since $h_3 \notin \pi(h_5)$, $h_4 \notin \pi(h_5)$, $h_6 \notin \pi(h_5)$, and $h_7 \notin \pi(h_5)$ by the design of $H$ it must be that $u_{h_5}(\pi) \leq 4$. It follows that $h_5$ has j-envy for $h_7$, since $u_{h_5}(\pi) \leq 4 < 9 = u_{h_5}(\{ h_4, h_6 \})$, $v_{h_4}(h_7) = 1 < 4 = v_{h_4}(h_5)$, and $v_{h_6}(h_7) = 4 < 5 = v_{h_6}(h_5)$.
    \item Suppose $\{ i_2, i_3 \} = \{ 6, 8 \}$. Consider $h_5$. If $u_{h_5}(\pi) \leq 4$ then $h_5$ has j-envy for $\alpha_{j}$, since $u_{h_5}(\pi) \leq 4 < 5 = u_{h_2}(\{ h_1, h_3 \})$, $v_{h_1}(\alpha_{j}) = 0 < 2 = v_{h_1}(h_5)$, and $v_{h_3}(\alpha_{j}) = 0 < 3 = v_{h_3}(h_5)$. It follows that $u_{h_5}(\pi) \geq 5$. Since $h_3 \notin \pi(h_5)$, $h_4 \notin \pi(h_5)$, and $h_6 \notin \pi(h_5)$, the only possibility is that $\pi(h_5) = \{ h_5, h_7, h_{10} \}$. It remains that $\pi(h_2) = \{ h_2, h_9, h_{11} \}$. Now $h_{10}$ has j-envy for $h_9$, since $u_{h_{10}}(\pi) = 4 < 10 = u_{h_{10}}(\{ h_2, h_{11} \})$, $v_{h_2}(h_9) = 1 < 6 = v_{h_2}(h_{10})$, and $v_{h_{11}}(h_9) = 3 < 4 = v_{h_{11}}(h_{10})$. 
    \item Suppose $\{ i_2, i_3 \} = \{ 6, 9 \}$. Consider $h_5$. Since $h_3 \notin \pi(h_5)$, $h_4 \notin \pi(h_5)$, and $h_6 \notin \pi(h_5)$ by the design of $H$ it must be that $u_{h_5}(\pi) \leq 6$. It follows that $h_5$ has j-envy for $h_9$, since $u_{h_5}(\pi) \leq 6 < 9 = u_{h_5}(\{ h_4, h_6 \})$, $v_{h_4}(h_9) = 3 < 4 = v_{h_4}(h_5)$, and $v_{h_6}(h_9) = 1 < 5 = v_{h_6}(h_5)$.
    \item Suppose $\{ i_2, i_3 \} = \{ 6, 10 \}$. Consider $h_5$. Since $h_3 \notin \pi(h_5)$, $h_4 \notin \pi(h_5)$, $h_6 \notin \pi(h_5)$, and $h_{10} \notin \pi(h_5)$ by the design of $H$ it must be that $u_{h_5}(\pi) \leq 4$. It follows that $h_5$ has j-envy for $h_{10}$, since $u_{h_5}(\pi) \leq 4 < 9 = u_{h_5}(\{ h_4, h_6 \})$, $v_{h_4}(h_{10}) = 1 < 4 = v_{h_4}(h_5)$, and $v_{h_6}(h_{10}) = 1 < 5 = v_{h_6}(h_5)$.
    \item Suppose $\{ i_2, i_3 \} = \{ 6, 11 \}$. Consider $h_5$. Since $h_3 \notin \pi(h_5)$, $h_4 \notin \pi(h_5)$, and $h_6 \notin \pi(h_5)$ by the design of $H$ it must be that $u_{h_5}(\pi) \leq 6$. It follows that $h_5$ has j-envy for $h_{11}$, since $u_{h_5}(\pi) \leq 6 < 9 = u_{h_5}(\{ h_4, h_6 \})$, $v_{h_4}(h_{11}) = 1 < 4 = v_{h_4}(h_5)$, and $v_{h_6}(h_{11}) = 3 < 5 = v_{h_6}(h_5)$.\qedhere
\end{itemize}
\end{itemize}

\end{proof}

\begin{lem}
\label{lem:threed_efr_as_jef_3332_case_part2}
If $(N, V)$ contains a j-envy-free partition into triples $\pi$, $\sigma(H, \pi) = 4$, and $u_{h_1}(\pi) = 4$ then $H$ has an open configuration in $\pi$.
\end{lem}
\begin{proof}
Suppose to the contrary that $\sigma(H, \pi) = 4$, $u_{h_1}(\pi) = 4$, and $H$ has an open configuration in $\pi$. Since $\sigma(H, \pi) = 4$ it must be that three triples in $\pi$ each contains exactly three agents in $H$ and one triple in $\pi$ contains exactly two agents in $H$. Suppose three triples $t_1$, $t_2$, and $t_3$ in $\pi$ each contains exactly three agents in $H$ and some triple $t_4$ in $\pi$ contains exactly two agents in $H$. Since $u_{h_1}(\pi) = 4$, by the design of $H$ it follows that $\pi(h_1)$ contains two agents in $H \setminus \{ h_1 \}$ and therefore $h_1 \notin t_4$. It follows that $t_4 = \{ h_{i_1}, h_{i_2}, \alpha_{j} \}$ where $2 \leq i_1, i_2 \leq 11$ and $\alpha_{j} \in (N \setminus H)$.  We use a case analysis to prove a contradiction occurs for each possible assignment of $\{ h_{i_1}, h_{i_2} \}$ where $2\leq i_1, i_2 \leq 11$.

As before, in the proof of Lemma~\ref{lem:threed_efr_as_jef_3332_case_part1}, the symmetries of $H$ allow us to shorten the case analysis. Recall that the structure of the valuations between agents $H \setminus \{ h_1 \}$ has five symmetries. It follows that for any possible assignment of $\{ {i_1}, {i_2} \}$ there exist in total five symmetric assignments, where the case analysis for one assignment is symmetric to the case analysis for the other four assignments. Since there are $\binom{10}{2}=45$ possible assignments of $\{ {i_1}, {i_2} \}$ where $2\leq i_1, i_2 \leq 11$ and five symmetries we need only consider the nine assignments $\{ 2, 3 \}$, $\{ 2, 4 \}$, $\{ 2, 5 \}$, $\{ 2, 6 \}$, $\{ 2, 7 \}$, $\{ 3, 4 \}$, $\{ 3, 5 \}$, $\{ 3, 6 \}$, $\{ 3, 7 \}$, of which no two are symmetric.
\begin{itemize}
    \item Suppose $\{ {i_1}, {i_2} \} = \{ 2, 3 \}$. Since $h_2 \notin \pi(h_4)$ and $h_3 \notin \pi(h_4)$ it must be that $u_{h_4}(\pi) \leq 10$. It follows that $h_4$ has j-envy for $\alpha_{j}$ since $u_{h_4}(\pi) \leq 10 < 12 = u_{h_4}(\{ h_2, h_3 \})$, $v_{h_2}(\alpha_{j}) = 0 < 6 = v_{h_2}(h_4)$, and $v_{h_3}(\alpha_{j}) = 0 < 5 = v_{h_3}(h_4)$. This contradicts the supposition that $\pi$ is j-envy-free. The following proofs concerning other possible assignments of $\{ {i_1}, {i_2} \}$ are similar in technique although in some cases we must make further deductions about the utilities of agents in $H$.
    \item Suppose $\{ {i_1}, {i_2} \} = \{ 2, 4 \}$. Since $h_2 \notin \pi(h_3)$ and $h_4 \notin \pi(h_3)$ it follows that $u_{h_3}(\pi) \leq 6$. It then follows that $h_3$ has j-envy for $\alpha_{j}$ since $u_{h_3}(\pi) \leq 6 < 9 = u_{h_3}(\{ h_2, h_4 \})$, $v_{h_2}(\alpha_{j}) = 0 < 4 = v_{h_2}(h_3)$, and $v_{h_4}(\alpha_{j}) = 0 < 5 = v_{h_4}(h_3)$.
    \item Suppose $\{ {i_1}, {i_2} \} = \{ 2, 5 \}$. If $u_{h_4}(\pi) \leq 9$ then $h_4$ has j-envy for $\alpha_{j}$, since $u_{h_4}(\pi) \leq 9 < 10 = u_{h_4}(\{ h_2, h_5 \})$, $v_{h_2}(\alpha_{j}) = 0 < 6 = v_{h_2}(h_4)$, and $v_{h_5}(\alpha_{j}) = 0 < 4 = v_{h_5}(h_4)$. It follows that $u_{h_4}(\pi) \geq 10$. The only possibility is that $\pi(h_4) = \{ h_3, h_4, h_6 \}$. Now $h_3$ has j-envy for $\alpha_{j}$ since $u_{h_3}(\pi) = 6 < 7 = u_{h_3}(\{ h_2, h_5 \})$, $v_{h_2}(\alpha_{j}) = 0 < 4 = v_{h_2}(h_3)$, and $v_{h_5}(\alpha_{j}) = 0 < 3 = v_{h_5}(h_3)$.
    \item Suppose $\{ {i_1}, {i_2} \} = \{ 2, 6 \}$. Since $h_2 \notin \pi(h_4)$ and $h_6 \notin \pi(h_4)$ it follows that $u_{h_4}(\pi) \leq 9$. Now $h_4$ has j-envy for $\alpha_{j}$ since $u_{h_4}(\pi) \leq 9 < 12 = u_{h_4}(\{ h_2, h_6 \})$, $v_{h_2}(\alpha_{j}) = 0 < 6 = v_{h_2}(h_4)$, and $v_{h_6}(\alpha_{j}) = 0 < 6 = v_{h_6}(h_4)$.
    \item Suppose $\{ {i_1}, {i_2} \} = \{ 2, 7 \}$. If $u_{h_4}(\pi) \leq 6$ then $h_4$ has j-envy for $\alpha_{j}$, since $u_{h_4}(\pi) \leq 6 < 7 = u_{h_4}(\{ h_2, h_7 \})$, $v_{h_2}(\alpha_{j}) = 0 < 6 = v_{h_2}(h_4)$, and $v_{h_7}(\alpha_{j}) = 0 < 1 = v_{h_7}(h_4)$. It follows that $u_{h_4}(\pi) \geq 7$. By the design of $H$, it follows that $\pi(h_4)$ contains three agents in $H$. Since $t_4 = \{ h_2, h_7, \alpha_{j} \}$ it follows that $\pi(h_4) = \{ h_4, h_{i_3}, h_{i_4} \}$ where $\{ i_3, i_4 \} \subset \{ 1, 3, 5, 6, 8, 9, 10, 11 \}$. Recall that $v_{h_4}(h_1) = 2$, $v_{h_4}(h_3) = 5$, $v_{h_4}(h_5) = 4$, $v_{h_4}(h_6) = 6$, $v_{h_4}(h_8) = 1$, $v_{h_4}(h_9) = 3$, $v_{h_4}(h_{10}) = 1$, and $v_{h_4}(h_{11}) = 1$. Since we established $u_{h_4}(\pi) \geq 7$ it follows that there are $11$ possibilities: $\{ {i_3}, {i_4} \} = \{ 1, 3 \}$, $\{ {i_3}, {i_4} \} = \{ 1, 6 \}$, $\{ {i_3}, {i_4} \} = \{ 3, 5 \}$, $\{ {i_3}, {i_4} \} = \{ 3, 6 \}$, $\{ {i_3}, {i_4} \} = \{ 3, 9 \}$, $\{ {i_3}, {i_4} \} = \{ 5, 6 \}$, $\{ {i_3}, {i_4} \} = \{ 5, 9 \}$, $\{ {i_3}, {i_4} \} = \{ 6, 8 \}$, $\{ {i_3}, {i_4} \} = \{ 6, 9 \}$, $\{ {i_3}, {i_4} \} = \{ 6, 10 \}$, and $\{ {i_3}, {i_4} \} = \{ 6, 11 \}$, which we shall now consider.
    \begin{itemize}
        \item Suppose $\{ {i_3}, {i_4} \} = \{ 1, 3 \}$. It follows that $h_2$ has j-envy for $h_1$ since $u_{h_2}(\pi) = 3 < 10 = u_{h_2}(\{ h_3, h_4 \})$, $v_{h_3}(h_1) = 2 < 4 = v_{h_3}(h_2)$, and $v_{h_4}(h_1) = 2 < 6 = v_{h_4}(h_2)$.
        \item Suppose $\{ {i_3}, {i_4} \} = \{ 1, 6 \}$. Consider $h_5$. It must be that $u_{h_5}(\pi) \leq 6$. Now $h_5$ has j-envy for $h_1$ since $u_{h_5}(\pi) \leq 6 < 9 = u_{h_5}(\{ h_4, h_6 \})$, $v_{h_4}(h_1) = 2 < 4 = v_{h_4}(h_5)$, and $v_{h_6}(h_1) = 2 < 5 = v_{h_6}(h_5)$.
        \item Suppose $\{ {i_3}, {i_4} \} = \{ 3, 5 \}$. It follows that $h_2$ has j-envy for $h_5$ since $u_{h_2}(\pi) = 3 < 10 = u_{h_2}(\{ h_3, h_4 \})$, $v_{h_3}(h_5) = 3 < 4 = v_{h_3}(h_2)$, and $v_{h_4}(h_5) = 4 < 6 = v_{h_4}(h_2)$.
        \item Suppose $\{ {i_3}, {i_4} \} = \{ 3, 6 \}$. In this case, consider $\pi(h_1)$. Since $h_1 \notin t_4$ it follows that $\pi(h_1)$ contains three agents in $H$. Suppose $\pi(h_1) = \{ h_1, h_{i_5}, h_{i_6} \}$ where $2 \leq {i_5}, {i_6} \leq 11$. Since we have established $\pi(h_2) = \{ h_2, h_7, \alpha_{j} \}$ and $\pi(h_4) = \{ h_3, h_4, h_6 \}$ it follows that $\{ h_{i_5}, h_{i_6} \} \subset \{ 5, 8, 9, 10, 11 \}$. Thus there are $\binom{5}{2}=10$ possible assignments of $\{ h_{i_5}, h_{i_6} \}$, which we shall now consider.
        \begin{itemize}
            \item If $\{ h_{i_5}, h_{i_6} \} = \{ 5, 8 \}$ then $h_7$ has j-envy for $h_1$ since $u_{h_7}(\pi) = 3 < 8 = u_{h_7}(\{ h_5, h_8 \})$, $v_{h_5}(h_1) = 2 < 3 = v_{h_5}(h_7)$, and $v_{h_8}(h_1) = 2 < 5 = v_{h_8}(h_7)$.
            \item If $\{ h_{i_5}, h_{i_6} \} = \{ 5, 9 \}$ then $h_7$ has j-envy for $h_1$ since $u_{h_7}(\pi) = 3 < 6 = u_{h_7}(\{ h_5, h_9 \})$, $v_{h_5}(h_1) = 2 < 3 = v_{h_5}(h_7)$, and $v_{h_9}(h_1) = 2 < 3 = v_{h_9}(h_7)$.
            \item If $\{ h_{i_5}, h_{i_6} \} = \{ 5, 10 \}$ then $u_{h_{10}}(\pi) = 5$. It follows that $h_{10}$ has j-envy for $\alpha_{j}$ since $u_{h_{10}}(\pi) = 5 < 7 = u_{h_{10}}(\{ h_2, h_7 \})$, $v_{h_2}(\alpha_{j}) = 0 < 6 = v_{h_2}(h_{10})$, and $v_{h_7}(\alpha_{j}) = 0 < 1 = v_{h_7}(h_{10})$.
            \item If $\{ h_{i_5}, h_{i_6} \} = \{ 5, 11 \}$ then $u_{h_{11}}(\pi) = 3$. It follows that $h_{11}$ has j-envy for $\alpha_{j}$ since $u_{h_{11}}(\pi) = 3 < 6 = u_{h_{11}}(\{ h_2, h_7 \})$, $v_{h_2}(\alpha_{j}) = 0 < 5 = v_{h_2}(h_{11})$, and $v_{h_7}(\alpha_{j}) = 0 < 1 = v_{h_7}(h_{11})$.
            \item If $\{ h_{i_5}, h_{i_6} \} = \{ 8, 9 \}$ then $h_7$ has j-envy for $h_1$ since $u_{h_7}(\pi) = 3 < 8 = u_{h_7}(\{ h_8, h_9 \})$, $v_{h_8}(h_1) = 2 < 5 = v_{h_8}(h_7)$, and $v_{h_9}(h_1) = 2 < 3 = v_{h_9}(h_7)$.
            \item If $\{ h_{i_5}, h_{i_6} \} = \{ 8, 10 \}$ then it remains that $\pi(h_9) = \{ h_5, h_9, h_{11} \}$ and thus $u_{h_9}(\pi)=u_{\{ h_{5}, h_{11} \}}=4$. It follows that $h_9$ has j-envy for $h_1$ since $u_{h_9}(\pi) = 4 < 9 = u_{h_9}(\{ h_8, h_{10} \})$, $v_{h_8}(h_1) = 2 < 4 = v_{h_8}(h_9)$, and $v_{h_{10}}(h_1) = 2 < 5 = v_{h_{10}}(h_9)$.
            \item If $\{ h_{i_5}, h_{i_6} \} = \{ 8, 11 \}$ then $u_{h_{11}}(\pi) = 3$. It follows that $h_{11}$ has j-envy for $\alpha_{j}$ since $u_{h_{11}}(\pi) = 3 < 6 = u_{h_{11}}(\{ h_2, h_7 \})$, $v_{h_2}(\alpha_{j}) = 0 < 5 = v_{h_2}(h_{11})$, and $v_{h_7}(\alpha_{j}) = 0 < 1 = v_{h_7}(h_{11})$.
            \item If $\{ h_{i_5}, h_{i_6} \} = \{ 9, 10 \}$ then it remains that $\pi(h_9) = \{ h_5, h_8, h_{11} \}$ and thus $u_{h_{11}}(\pi)=u_{\{ h_{5}, h_{8} \}}=2$. It follows that $h_{11}$ has j-envy for $\alpha_{j}$ since $u_{h_{11}}(\pi) = 2 < 6 = u_{h_{11}}(\{ h_2, h_7 \})$, $v_{h_2}(\alpha_{j}) = 0 < 5 = v_{h_2}(h_{11})$, and $v_{h_7}(\alpha_{j}) = 0 < 1 = v_{h_7}(h_{11})$.
            \item If $\{ h_{i_5}, h_{i_6} \} = \{ 9, 11 \}$ then $u_{h_{11}}(\pi) = 5$. It follows that $h_{11}$ has j-envy for $\alpha_{j}$ since $u_{h_{11}}(\pi) = 5 < 6 = u_{h_{11}}(\{ h_2, h_7 \})$, $v_{h_2}(\alpha_{j}) = 0 < 5 = v_{h_2}(h_{11})$, and $v_{h_7}(\alpha_{j}) = 0 < 1 = v_{h_7}(h_{11})$.
            \item If $\{ h_{i_5}, h_{i_6} \} = \{ 10, 11 \}$ then $h_2$ has j-envy for $h_1$ since $u_{h_2}(\pi) = 3 < 11 = u_{h_2}(\{ h_{10}, h_{11} \})$, $v_{h_{10}}(h_1) = 2 < 6 = v_{h_{10}}(h_2)$, and $v_{h_{11}}(h_1) = 2 < 5 = v_{h_{11}}(h_2)$.
        \end{itemize}
        \item Suppose $\{ {i_3}, {i_4} \} = \{ 3, 9 \}$. It follows that $h_2$ has j-envy for $h_9$ since $u_{h_2}(\pi) = 3 < 10 = u_{h_2}(\{ h_3, h_4 \})$, $v_{h_3}(h_9) = 1 < 4 = v_{h_3}(h_2)$, and $v_{h_4}(h_9) = 3 < 6 = v_{h_4}(h_2)$.
        \item Suppose $\{ {i_3}, {i_4} \} = \{ 5, 6 \}$. Consider $h_3$. If $u_{h_3}(\pi) \leq 4$ then $h_3$ has j-envy for $\alpha_{j}$, since $u_{h_3}(\pi) \leq 4 < 5 = u_{h_3}(\{ h_2, h_7 \})$, $v_{h_2}(\alpha_{j}) = 0 < 4 = v_{h_2}(h_3)$, and $v_{h_7}(\alpha_{j}) = 0 < 1 = v_{h_7}(h_3)$. It follows that $u_{h_3}(\pi) \geq 5$. We have established that $h_4 \notin \pi(h_3)$, $h_5 \notin \pi(h_3)$ and $h_2 \notin \pi(h_3)$ so, by the design of $H$, there are three possibilities: either $\pi(h_3) = \{ h_1, h_3, h_8 \}$, $\pi(h_3) = \{ h_1, h_3, h_{11} \}$, or $\pi(h_3) = \{ h_3, h_8, h_{11} \}$.
        \begin{itemize}
            \item If $\pi(h_3) = \{ h_1, h_3, h_8 \}$ then $u_{h_8}(\pi) = 5$. It follows that $h_8$ has j-envy for $\alpha_{j}$, since $u_{h_8}(\pi) = 5 < 6 = u_{h_8}(\{ h_2, h_7 \})$, $v_{h_2}(\alpha_{j}) = 0 < 1 = v_{h_2}(h_8)$, and $v_{h_7}(\alpha_{j}) = 0 < 5 = v_{h_7}(h_8)$.
            \item If $\pi(h_3) = \{ h_1, h_3, h_{11} \}$ then $u_{h_{11}}(\pi) = 5$. It follows that $h_{11}$ has j-envy for $\alpha_{j}$, since $u_{h_{11}}(\pi) = 5 < 6 = u_{h_{11}}(\{ h_2, h_7 \})$, $v_{h_2}(\alpha_{j}) = 0 < 5 = v_{h_2}(h_{11})$, and $v_{h_7}(\alpha_{j}) = 0 < 1 = v_{h_7}(h_{11})$.
            \item If $\pi(h_3) = \{ h_3, h_8, h_{11} \}$ then $u_{h_8}(\pi) = 4$. It follows that $h_8$ has j-envy for $\alpha_{j}$, since $u_{h_8}(\pi) = 4 < 6 = u_{h_8}(\{ h_2, h_7 \})$, $v_{h_2}(\alpha_{j}) = 0 < 1 = v_{h_2}(h_8)$, and $v_{h_7}(\alpha_{j}) = 0 < 5 = v_{h_7}(h_8)$.
        \end{itemize}
        \item Suppose $\{ {i_3}, {i_4} \} = \{ 5, 9 \}$. Consider $h_6$. It must be that $u_{h_6}(\pi) \leq 9$. Now $h_6$ has j-envy for $h_9$ since $u_{h_6}(\pi) \leq 9 < 11 = u_{h_6}(\{ h_4, h_5 \})$, $v_{h_4}(h_9) = 3 < 6 = v_{h_4}(h_6)$, and $v_{h_5}(h_9) = 1 < 5 = v_{h_5}(h_6)$.
        \item Suppose $\{ {i_3}, {i_4} \} = \{ 6, 8 \}$. Consider $h_{10}$. If $u_{h_{10}}(\pi) \leq 6$ then $h_{10}$ has j-envy for $\alpha_{j}$, since $u_{h_{10}}(\pi) \leq 6 < 7 = u_{h_{10}}(\{ h_2, h_7 \})$, $v_{h_2}(\alpha_{j}) = 0 < 6 = v_{h_2}(h_{10})$, and $v_{h_7}(\alpha_{j}) = 0 < 1 = v_{h_7}(h_{10})$. It follows that $u_{h_{10}}(\pi) \geq 7$. We have established that $h_2 \notin \pi(h_{10})$ and $h_8 \notin \pi(h_{10})$ so, by the design of $H$, there are four possibilities: either $\pi(h_{10}) = \{ h_1, h_9, h_{10} \}$, $\pi(h_{10}) = \{ h_5, h_9, h_{10} \}$, $\pi(h_{10}) = \{ h_5, h_{10}, h_{11} \}$, or $\pi(h_3) = \{ h_9, h_{10}, h_{11} \}$.
        \begin{itemize}
            \item If $\pi(h_{10}) = \{ h_1, h_9, h_{10} \}$ then $h_8$ has j-envy for $h_1$, since $u_{h_8}(\pi) = 7 < 10 = u_{h_8}(\{ h_9, h_{10} \})$, $v_{h_9}(h_1) = 2 < 4 = v_{h_9}(h_8)$, and $v_{h_{10}}(h_1) = 2 < 6 = v_{h_{10}}(h_8)$.
            \item If $\pi(h_{10}) = \{ h_5, h_9, h_{10} \}$ then $h_8$ has j-envy for $h_5$, since $u_{h_8}(\pi) = 7 < 10 = u_{h_8}(\{ h_9, h_{10} \})$, $v_{h_9}(h_5) = 1 < 4 = v_{h_9}(h_8)$, and $v_{h_{10}}(h_5) = 3 < 6 = v_{h_{10}}(h_8)$.
            \item If $\pi(h_{10}) = \{ h_5, h_{10}, h_{11} \}$ then $u_{h_{11}}(\pi) = 5$. It follows that $h_{11}$ has j-envy for $\alpha_{j}$, since $u_{h_{11}}(\pi) = 5 < 6 = u_{h_{11}}(\{ h_2, h_7 \})$, $v_{h_2}(\alpha_{j}) = 0 < 5 = v_{h_2}(h_{11})$, and $v_{h_7}(\alpha_{j}) = 0 < 1 = v_{h_7}(h_{11})$.
            \item If $\pi(h_3) = \{ h_9, h_{10}, h_{11} \}$ then $h_8$ has j-envy for $h_{11}$, since $u_{h_8}(\pi) = 7 < 10 = u_{h_8}(\{ h_9, h_{10} \})$, $v_{h_9}(h_{11}) = 3 < 4 = v_{h_9}(h_8)$, and $v_{h_{10}}(h_{11}) = 4 < 6 = v_{h_{10}}(h_8)$.
        \end{itemize}
        \item Suppose $\{ {i_3}, {i_4} \} = \{ 6, 9 \}$. It must be that $u_{h_5}(\pi) \leq 6$. Now $h_5$ has j-envy for $h_9$ since $u_{h_5}(\pi) \leq 6 < 9 = u_{h_5}(\{ h_4, h_6 \})$, $v_{h_4}(h_9) = 3 < 4 = v_{h_4}(h_5)$, and $v_{h_6}(h_9) = 1 < 5 = v_{h_6}(h_5)$.
        \item Suppose $\{ {i_3}, {i_4} \} = \{ 6, 10 \}$. It must be that $u_{h_5}(\pi) \leq 5$. Now $h_5$ has j-envy for $h_{10}$ since $u_{h_5}(\pi) \leq 5 < 9 = u_{h_5}(\{ h_4, h_6 \})$, $v_{h_4}(h_{10}) = 1 < 4 = v_{h_4}(h_5)$, and $v_{h_6}(h_{10}) = 1 < 5 = v_{h_6}(h_5)$.
        \item Suppose $\{ {i_3}, {i_4} \} = \{ 6, 11 \}$. It must be that $u_{h_5}(\pi) \leq 6$. Now $h_5$ has j-envy for $h_{11}$ since $u_{h_5}(\pi) \leq 6 < 9 = u_{h_5}(\{ h_4, h_6 \})$, $v_{h_4}(h_{11}) = 1 < 4 = v_{h_4}(h_5)$, and $v_{h_6}(h_{11}) = 3 < 5 = v_{h_6}(h_5)$.
    \end{itemize}
\item Suppose $\{ {i_1}, {i_2} \} = \{ 3, 4 \}$. If $u_{h_2}(\pi) \leq 9$ then $h_2$ has j-envy for $\alpha_{j}$, since $u_{h_2}(\pi) \leq 9 < 10 = u_{h_2}(\{ h_3, h_4 \})$, $v_{h_3}(\alpha_{j}) = 0 < 4 = v_{h_3}(h_2)$, and $v_{h_4}(\alpha_{j}) = 0 < 6 = v_{h_4}(h_2)$. It follows that $u_{h_2}(\pi) \geq 10$. The only possibility is that $\pi(h_2) = \{ h_2, h_{10}, h_{11} \}$. Now consider $h_5$. If $u_{h_5}(\pi) \leq 6$ then $h_5$ has j-envy for $\alpha_{j}$, since $u_{h_5}(\pi) \leq 6 < 7 = u_{h_5}(\{ h_3, h_4 \})$, $v_{h_3}(\alpha_{j}) = 0 < 3 = v_{h_3}(h_5)$, and $v_{h_4}(\alpha_{j}) = 0 < 4 = v_{h_4}(h_5)$. It follows that $u_{h_5}(\pi) \geq 7$. Since we have established $\pi(h_2) = \{ h_2, h_{10}, h_{11} \}$ and $t_4 = \{ h_3, h_4, \alpha_{j} \}$, there are just two possibilities: either $\pi(h_5) = \{ h_1, h_5, h_6 \}$ or $\pi(h_5) = \{ h_5, h_6, h_7 \}$.
\begin{itemize}
    \item If $\pi(h_5) = \{ h_1, h_5, h_6 \}$ then $h_4$ has j-envy for $h_1$ since $u_{h_4}(\pi) = 5 < 10 = u_{h_4}(\{ h_5, h_6 \})$, $v_{h_5}(h_1) = 2 < 4 = v_{h_5}(h_4)$, and $v_{h_6}(h_1) = 2 < 6 = v_{h_6}(h_4)$.
    \item If $\pi(h_5) = \{ h_5, h_6, h_7 \}$ then $h_4$ has j-envy for $h_7$ since $u_{h_4}(\pi) = 5 < 10 = u_{h_4}(\{ h_5, h_6 \})$, $v_{h_5}(h_7) = 3 < 4 = v_{h_5}(h_4)$, and $v_{h_6}(h_7) = 4 < 6 = v_{h_6}(h_4)$.
\end{itemize}
\item Suppose $\{ {i_1}, {i_2} \} = \{ 3, 5 \}$. If $u_{h_4}(\pi) \leq 8$ then $h_4$ has j-envy for $\alpha_{j}$, since $u_{h_4}(\pi) \leq 8 < 9 = u_{h_4}(\{ h_3, h_5 \})$, $v_{h_3}(\alpha_{j}) = 0 < 5 = v_{h_3}(h_4)$, and $v_{h_5}(\alpha_{j}) = 0 < 4 = v_{h_5}(h_4)$. It follows that $u_{h_4}(\pi) \geq 9$. There are three possibilities: either $\pi(h_4) = \{ h_2, h_4, h_6 \}$, $\pi(h_4) = \{ h_2, h_4, h_9 \}$, or $\pi(h_4) = \{ h_4, h_6, h_9 \}$.
\begin{itemize}
    \item Suppose $\pi(h_4) = \{ h_2, h_4, h_6 \}$. In this case, consider $\pi(h_1)$. Since $h_1 \notin t_4$ it follows that $\pi(h_1)$ contains three agents in $H$. Suppose $\pi(h_1) = \{ h_1, h_{i_5}, h_{i_6} \}$ where $1 \leq {i_5}, {i_6} \leq 11$. Since we have established $\pi(h_3) = \{ h_3, h_5, \alpha_{j} \}$ and $\pi(h_2) = \{ h_2, h_4, h_6 \}$ it follows that $\{ h_{i_5}, h_{i_6} \} \subset \{ 7, 8, 9, 10, 11 \}$. Thus there are $\binom{5}{2}=10$ possible assignments of $\{ h_{i_5}, h_{i_6} \}$, which we shall now consider.
    \begin{itemize}
        \item If $\{ h_{i_5}, h_{i_6} \} = \{ 7, 8 \}$ then $h_6$ has j-envy for $h_1$, since $u_{h_6}(\pi) = 7 < 10 = u_{h_6}(\{ h_7, h_8 \})$, $v_{h_7}(h_1) = 2 < 4 = v_{h_7}(h_6)$, and $v_{h_8}(h_1) = 2 < 6 = v_{h_8}(h_6)$.
        \item If $\{ h_{i_5}, h_{i_6} \} = \{ 7, 9 \}$ then it remains that $\pi(h_8) = \{ h_8, h_{10}, h_{11} \}$. It follows that $h_8$ has j-envy for $h_1$, since $u_{h_8}(\pi) = 7 < 9 = u_{h_8}(\{ h_7, h_9 \})$, $v_{h_7}(h_1) = 2 < 5 = v_{h_7}(h_8)$, and $v_{h_9}(h_1) = 2 < 4 = v_{h_9}(h_8)$.
        \item If $\{ h_{i_5}, h_{i_6} \} = \{ 7, 10 \}$ then $h_7$ has j-envy for $\alpha_{j}$, since $u_{h_7}(\pi) = 3 < 4 = u_{h_7}(\{ h_3, h_5 \})$, $v_{h_3}(\alpha_{j}) = 0 < 1 = v_{h_3}(h_7)$, and $v_{h_5}(\alpha_{j}) = 0 < 3 = v_{h_5}(h_7)$.
        \item If $\{ h_{i_5}, h_{i_6} \} = \{ 7, 11 \}$ then $h_7$ has j-envy for $\alpha_{j}$, since $u_{h_7}(\pi) = 3 < 4 = u_{h_7}(\{ h_3, h_5 \})$, $v_{h_3}(\alpha_{j}) = 0 < 1 = v_{h_3}(h_7)$, and $v_{h_5}(\alpha_{j}) = 0 < 3 = v_{h_5}(h_7)$.
        \item If $\{ h_{i_5}, h_{i_6} \} = \{ 8, 9 \}$ then it remains that $\pi(h_7) = \{ h_7, h_{10}, h_{11} \}$. It follows that $h_7$ has j-envy for $h_1$, since $u_{h_7}(\pi) = 2 < 8 = u_{h_7}(\{ h_8, h_9 \})$, $v_{h_8}(h_1) = 2 < 5 = v_{h_8}(h_7)$, and $v_{h_9}(h_1) = 2 < 3 = v_{h_9}(h_7)$.
        \item If $\{ h_{i_5}, h_{i_6} \} = \{ 8, 10 \}$ then it remains that $\pi(h_9) = \{ h_7, h_9, h_{11} \}$. It follows that $h_9$ has j-envy for $h_1$, since $u_{h_9}(\pi) = 6 < 9 = u_{h_9}(\{ h_8, h_{10} \})$, $v_{h_8}(h_1) = 2 < 4 = v_{h_8}(h_9)$, and $v_{h_{10}}(h_1) = 2 < 5 = v_{h_{10}}(h_9)$.
        \item If $\{ h_{i_5}, h_{i_6} \} = \{ 8, 11 \}$ then $h_{11}$ has j-envy for $\alpha_{j}$, since $u_{h_{11}}(\pi) = 3 < 4 = u_{h_{11}}(\{ h_3, h_5 \})$, $v_{h_3}(\alpha_{j}) = 0 < 3 = v_{h_3}(h_{11})$, and $v_{h_5}(\alpha_{j}) = 0 < 1 = v_{h_5}(h_{11})$.
        \item If $\{ h_{i_5}, h_{i_6} \} = \{ 9, 10 \}$ then it remains that $\pi(h_{11}) = \{ h_7, h_8, h_{11} \}$. It follows that $h_{11}$ has j-envy for $h_1$, since $u_{h_{11}}(\pi) = 2 < 7 = u_{h_{11}}(\{ h_9, h_{10} \})$, $v_{h_9}(h_1) = 2 < 3 = v_{h_9}(h_{11})$, and $v_{h_{10}}(h_1) = 2 < 4 = v_{h_{10}}(h_{11})$.
        \item If $\{ h_{i_5}, h_{i_6} \} = \{ 9, 11 \}$ then it remains that $\pi(h_{10}) = \{ h_7, h_8, h_{10} \}$. It follows that $h_{10}$ has j-envy for $h_1$, since $u_{h_{10}}(\pi) = 7 < 9 = u_{h_{10}}(\{ h_9, h_{11} \})$, $v_{h_9}(h_1) = 2 < 5 = v_{h_9}(h_{10})$, and $v_{h_{11}}(h_1) = 2 < 4 = v_{h_{11}}(h_{10})$.
        \item If $\{ h_{i_5}, h_{i_6} \} = \{ 10, 11 \}$ then $h_2$ has j-envy for $h_1$, since $u_{h_2}(\pi) = 7 < 11 = u_{h_2}(\{ h_{10}, h_{11} \})$, $v_{h_{10}}(h_1) = 2 < 6 = v_{h_{10}}(h_2)$, and $v_{h_{11}}(h_1) = 2 < 5 = v_{h_{11}}(h_2)$.
    \end{itemize}
    \item Suppose $\pi(h_4) = \{ h_2, h_4, h_9 \}$. It follows that $h_3$ has j-envy for $h_9$, since $u_{h_3}(\pi) = 3 < 9 = u_{h_3}(\{ h_2, h_4 \})$, $v_{h_2}(h_9) = 1 < 4 = v_{h_2}(h_3)$, and $v_{h_4}(h_9) = 3 < 5 = v_{h_4}(h_3)$.
    \item Suppose $\pi(h_4) = \{ h_4, h_6, h_9 \}$. It follows that $h_5$ has j-envy for $h_9$, since $u_{h_5}(\pi) = 3 < 9 = u_{h_5}(\{ h_4, h_6 \})$, $v_{h_4}(h_9) = 3 < 4 = v_{h_4}(h_5)$, and $v_{h_6}(h_9) = 1 < 5 = v_{h_6}(h_5)$.
\end{itemize}
\item Suppose $\{ {i_1}, {i_2} \} = \{ 3, 6 \}$. It must be that $u_{h_4}(\pi) \leq 10$. It follows that $h_4$ has j-envy for $\alpha_{j}$ since $u_{h_4}(\pi) \leq 10 < 11 = u_{h_4}(\{ h_3, h_6 \})$, $v_{h_3}(\alpha_{j}) = 0 < 5 = v_{h_3}(h_4)$, and $v_{h_6}(\alpha_{j}) = 0 < 6 = v_{h_6}(h_4)$.
\item Suppose $\{ {i_1}, {i_2} \} = \{ 3, 7 \}$. If $u_{h_8}(\pi) \leq 7$ then $h_8$ has j-envy for $\alpha_{j}$, since $u_{h_8}(\pi) \leq 7 < 8 = u_{h_8}(\{ h_3, h_7 \})$, $v_{h_3}(\alpha_{j}) = 0 < 3 = v_{h_3}(h_8)$, and $v_{h_7}(\alpha_{j}) = 0 < 5 = v_{h_7}(h_8)$. It follows that $u_{h_8}(\pi) \geq 8$. By the design of $H$, it follows that $\pi(h_8)$ contains three agents in $H$. Since $t_4 = \{ h_3, h_7, \alpha_{j} \}$ it must be that $\pi(h_8) = \{ h_8, h_{i_3}, h_{i_4} \}$ where $\{ i_3, i_4 \} \subset \{ 1, 2, 4, 5, 6, 9, 10, 11 \}$. Recall that $v_{h_8}(h_1) = 2$, $v_{h_8}(h_2) = 1$, $v_{h_8}(h_4) = 1$, $v_{h_8}(h_5) = 1$, $v_{h_8}(h_6) = 6$, $v_{h_8}(h_9) = 4$, $v_{h_8}(h_{10}) = 6$, and $v_{h_8}(h_{11}) = 1$. Since we established $u_{h_8}(\pi) \geq 8$ it follows that there are five possibilities: $\{ {i_3}, {i_4} \} = \{ 1, 6 \}$, $\{ {i_3}, {i_4} \} = \{ 1, 10 \}$, $\{ {i_3}, {i_4} \} = \{ 6, 9 \}$, $\{ {i_3}, {i_4} \} = \{ 6, 10 \}$, and $\{ {i_3}, {i_4} \} = \{ 9, 10 \}$, which we shall now consider.
\begin{itemize}
    \item Suppose $\{ {i_3}, {i_4} \} = \{ 1, 6 \}$. It follows that $h_7$ has j-envy for $h_1$, since $u_{h_7}(\pi) = 1 < 9 = u_{h_7}(\{ h_6, h_8 \})$, $v_{h_6}(h_1) = 2 < 4 = v_{h_6}(h_7)$, and $v_{h_8}(h_1) = 2 < 5 = v_{h_8}(h_7)$.
    \item Suppose $\{ {i_3}, {i_4} \} = \{ 1, 10 \}$. Consider $h_9$. It follows that $u_{h_9}(\pi) \leq 6$. Now $h_9$ has j-envy for $h_1$, since $u_{h_9}(\pi) \leq 6 < 9 = u_{h_9}(\{ h_8, h_{10} \})$, $v_{h_8}(h_1) = 2 < 4 = v_{h_8}(h_9)$, and $v_{h_{10}}(h_1) = 2 < 5 = v_{h_{10}}(h_9)$.
    \item Suppose $\{ {i_3}, {i_4} \} = \{ 6, 9 \}$. It follows that $h_7$ has j-envy for $h_9$, since $u_{h_7}(\pi) = 1 < 9 = u_{h_7}(\{ h_6, h_8 \})$, $v_{h_6}(h_9) = 1 < 4 = v_{h_6}(h_7)$, and $v_{h_8}(h_9) = 4 < 5 = v_{h_8}(h_7)$.
    \item Suppose $\{ {i_3}, {i_4} \} = \{ 6, 10 \}$. In this case, consider $\pi(h_1)$. Since $h_1 \notin t_4$ it follows that $\pi(h_1)$ contains three agents in $H$. Suppose $\pi(h_1) = \{ h_1, h_{i_5}, h_{i_6} \}$ where $2 \leq i_5, i_6 \leq 11$. Since we have established $\pi(h_3) = \{ h_3, h_7, \alpha_{j} \}$ and $\pi(h_6) = \{ h_6, h_8, h_{10} \}$ it follows that $\{ h_{i_5}, h_{i_6} \} \subset \{ 2, 4, 5, 9, 11 \}$. Thus there are $\binom{5}{2}=10$ possible assignments of $\{ h_{i_5}, h_{i_6} \}$, which we shall now consider.
    \begin{itemize}
        \item If $\{ h_{i_5}, h_{i_6} \} = \{ 2, 4 \}$ then $h_3$ has j-envy for $h_1$, since $u_{h_3}(\pi) = 1 < 9 = u_{h_3}(\{ h_2, h_4 \})$, $v_{h_2}(h_1) = 2 < 4 = v_{h_2}(h_3)$, and $v_{h_4}(h_1) = 2 < 5 = v_{h_4}(h_3)$.
        \item If $\{ h_{i_5}, h_{i_6} \} = \{ 2, 5 \}$ then $h_3$ has j-envy for $h_1$, since $u_{h_3}(\pi) = 1 < 7 = u_{h_3}(\{ h_2, h_5 \})$, $v_{h_2}(h_1) = 2 < 4 = v_{h_2}(h_3)$, and $v_{h_5}(h_1) = 2 < 3 = v_{h_5}(h_3)$.
        \item If $\{ h_{i_5}, h_{i_6} \} = \{ 2, 9 \}$ then it remains that $\pi(h_{11}) = \{ h_4, h_5, h_{11} \}$. Now $h_3$ has j-envy for $h_{11}$, since $u_{h_3}(\pi) = 1 < 8 = u_{h_3}(\{ h_4, h_5 \})$, $v_{h_4}(h_{11}) = 1 < 5 = v_{h_4}(h_3)$, and $v_{h_5}(h_{11}) = 1 < 3 = v_{h_5}(h_3)$.
        \item If $\{ h_{i_5}, h_{i_6} \} = \{ 2, 11 \}$ then $h_3$ has j-envy for $h_1$, since $u_{h_3}(\pi) = 1 < 7 = u_{h_3}(\{ h_2, h_{11} \})$, $v_{h_2}(h_1) = 2 < 4 = v_{h_2}(h_3)$, and $v_{h_{11}}(h_1) = 2 < 3 = v_{h_{11}}(h_3)$.
        \item If $\{ h_{i_5}, h_{i_6} \} = \{ 4, 5 \}$ then $h_3$ has j-envy for $h_1$, since $u_{h_3}(\pi) = 1 < 8 = u_{h_3}(\{ h_5, h_5 \})$, $v_{h_4}(h_1) = 2 < 5 = v_{h_4}(h_3)$, and $v_{h_5}(h_1) = 2 < 3 = v_{h_5}(h_3)$.
        \item If $\{ h_{i_5}, h_{i_6} \} = \{ 4, 9 \}$ then $h_4$ has j-envy for $\alpha_{j}$, since $u_{h_4}(\pi) = 5 < 6 = u_{h_4}(\{ h_3, h_7 \})$, $v_{h_3}(\alpha_{j}) = 0 < 5 = v_{h_3}(h_4)$, and $v_{h_7}(\alpha_{j}) = 0 < 1 = v_{h_7}(h_4)$.
        \item If $\{ h_{i_5}, h_{i_6} \} = \{ 4, 11 \}$ then $h_3$ has j-envy for $h_1$, since $u_{h_3}(\pi) = 1 < 8 = u_{h_3}(\{ h_4, h_{11} \})$, $v_{h_4}(h_1) = 2 < 5 = v_{h_4}(h_3)$, and $v_{h_{11}}(h_1) = 2 < 3 = v_{h_{11}}(h_3)$.
        \item If $\{ h_{i_5}, h_{i_6} \} = \{ 5, 9 \}$ then $h_5$ has j-envy for $\alpha_{j}$, since $u_{h_5}(\pi) = 3 < 6 = u_{h_5}(\{ h_3, h_7 \})$, $v_{h_3}(\alpha_{j}) = 0 < 3 = v_{h_3}(h_5)$, and $v_{h_7}(\alpha_{j}) = 0 < 3 = v_{h_7}(h_5)$.
        \item If $\{ h_{i_5}, h_{i_6} \} = \{ 5, 11 \}$ then $h_3$ has j-envy for $h_1$, since $u_{h_3}(\pi) = 1 < 6 = u_{h_3}(\{ h_5, h_{11} \})$, $v_{h_5}(h_1) = 2 < 3 = v_{h_5}(h_3)$, and $v_{h_{11}}(h_1) = 2 < 3 = v_{h_{11}}(h_3)$.
        \item If $\{ h_{i_5}, h_{i_6} \} = \{ 9, 11 \}$ then $h_{10}$ has j-envy for $h_1$, since $u_{h_{10}}(\pi) = 7 < 9 = u_{h_{10}}(\{ h_9, h_{11} \})$, $v_{h_9}(h_1) = 2 < 5 = v_{h_9}(h_{10})$, and $v_{h_{11}}(h_1) = 2 < 4 = v_{h_{11}}(h_{10})$.
    \end{itemize}
    \item Suppose $\{ {i_3}, {i_4} \} = \{ 9, 10 \}$. Consider $h_2$. If $u_{h_2}(\pi) \leq 6$ then $h_2$ has j-envy for $\alpha_{j}$, since $u_{h_2}(\pi) \leq 6 < 7 = u_{h_2}(\{ h_3, h_7 \})$, $v_{h_3}(\alpha_{j}) = 0 < 4 = v_{h_3}(h_2)$, and $v_{h_7}(\alpha_{j}) = 0 < 3 = v_{h_7}(h_2)$. It follows that $u_{h_2}(\pi) \geq 7$. We have established that $h_3 \notin \pi(h_2)$, $h_7 \notin \pi(h_2)$, and $h_{10} \notin \pi(h_2)$ so, by the design of $H$, there are three possibilities: either $\pi(h_2) = \{ h_1, h_2, h_4 \}$, $\pi(h_2) = \{ h_1, h_2, h_{11} \}$, or $\pi(h_2) = \{ h_2, h_4, h_{11} \}$.
    \begin{itemize}
        \item If $\pi(h_2) = \{ h_1, h_2, h_4 \}$ then $h_3$ has j-envy for $h_1$, since $u_{h_3}(\pi) = 1 < 9 = u_{h_3}(\{ h_2, h_4 \})$, $v_{h_2}(h_1) = 2 < 4 = v_{h_2}(h_3)$, and $v_{h_4}(h_1) = 2 < 5 = v_{h_4}(h_3)$.
        \item If $\pi(h_2) = \{ h_1, h_2, h_{11} \}$ then $h_3$ has j-envy for $h_1$, since $u_{h_3}(\pi) = 1 < 7 = u_{h_3}(\{ h_2, h_{11} \})$, $v_{h_2}(h_1) = 2 < 4 = v_{h_2}(h_3)$, and $v_{h_{11}}(h_1) = 2 < 3 = v_{h_{11}}(h_3)$.
        \item If $\pi(h_2) = \{ h_2, h_4, h_{11} \}$ then it remains that $\pi(h_1) = \{ h_1, h_5, h_6 \}$. Now $h_7$ has j-envy for $h_1$, since $u_{h_7}(\pi) = 1 < 7 = u_{h_7}(\{ h_5, h_6 \})$, $v_{h_5}(h_1) = 2 < 3 = v_{h_5}(h_7)$, and $v_{h_6}(h_1) = 2 < 4 = v_{h_6}(h_7)$.\qedhere
    \end{itemize}
\end{itemize}
\end{itemize}
\end{proof}

\begin{lem}
\label{lem:threed_efr_as_jef_3332_case}
If $(N, V)$ contains a j-envy-free partition into triples $\pi$ and $\sigma(H, \pi) = 4$ then $H$ has an open configuration in $\pi$.
\end{lem}
\begin{proof}
Suppose $\sigma(H, \pi) = 4$. Consider $u_{h_1}(\pi)$. By the design of $H$, it must be that $2 \leq u_{h_1}(\pi) \leq 4$. If $u_{h_1}(\pi) < 4$ then Lemma~\ref{lem:threed_efr_as_jef_3332_case_part1} shows that $H$ has an open configuration in $\pi$. If $u_{h_1}(\pi) = 4$ then Lemma~\ref{lem:threed_efr_as_jef_3332_case_part2} shows that $H$ has an open configuration in $\pi$.
\end{proof}

\begin{lem}
\label{lem:threed_efr_as_jef_32222_case}
If $(N, V)$ contains a j-envy-free partition into triples $\pi$ and $\sigma(H, \pi) = 5$ then $H$ has an open configuration in $\pi$.
\end{lem}
\begin{proof}
Suppose, to the contrary, that $\sigma(H, \pi) = 5$ and $H$ has a closed configuration in $\pi$. Then it must be that four triples in $\pi$ each contains exactly two agents in $H$ and one triple in $\pi$ contains exactly three agents in $H$. Suppose four triples $t_1$, $t_2$, $t_3$, and $t_4$ in $\pi$ each contains exactly two agents in $H$ and some triple $t_5$ in $\pi$ contains exactly three agents in $H$, where $t_1 = \{ h_{i_1}, h_{i_2}, \alpha_{j_1} \}$, $1\leq i_1, i_2 \leq 11$, and $\alpha_{j_1} \in N \setminus H$. We use a case analysis on $v_{h_{i_1}}(h_{i_2})$ to prove a contradiction. Note that by the design of $H$ it must be that $1 \leq v_{h_{i_1}}(h_{i_2}) \leq 6$.
\begin{itemize}
    \item Suppose $v_{h_{i_1}}(h_{i_2}) = 6$. By the symmetry of $H$, assume without loss of generality that $\{ i_1, i_2 \} = \{ 2, 4 \}$. It follows that $u_{h_3}(\pi) \leq 6$. Now $h_3$ has j-envy for $\alpha_{j_1}$ since $u_{h_3}(\pi) \leq 6 < 9 = u_{h_3}(\{ h_2, h_4 \})$, $v_{h_2}(\alpha_{j_1}) = 0 < 4 = v_{h_2}(h_3)$, and $v_{h_4}(\alpha_{j_1}) = 0 < 5 = v_{h_4}(h_3)$. This contradicts the supposition that $\pi$ is j-envy-free. We shall use a similar technique to prove a contradiction when considering the other cases of $v_{h_{i_1}}(h_{i_2})$.
    \item Suppose $v_{h_{i_1}}(h_{i_2}) = 5$. Assume without loss of generality that $\{ i_1, i_2 \} = \{ 3, 4 \}$. If $u_{h_2}(\pi) \leq 9$ then $h_2$ has j-envy for $\alpha_{j_1}$ since $u_{h_2}(\pi) \leq 9 < 10 = u_{h_2}(\{ h_3, h_4 \})$, $v_{h_3}(\alpha_{j_1}) = 0 < 4 = v_{h_3}(h_2)$, and $v_{h_4}(\alpha_{j_1}) = 0 < 6 = v_{h_4}(h_2)$. It follows that $u_{h_2}(\pi) \geq 10$. The only possibility is that $\pi(h_2) = \{ h_2, h_{10}, h_{11} \}$. It must be that $\pi(h_2) = t_5$. We now consider $h_6$. Since $h_6 \notin t_5$ it must be that either $h_6 \in t_2$, $h_6 \in t_3$, or $h_6 \in t_4$. Assume without loss of generality that $h_6 \in t_2$ and that $t_2 = \{ h_6, h_{i_3}, \alpha_{j_2} \}$ where $1\leq i_3 \leq 11$ and $\alpha_{j_2} \in N \setminus H$. If $u_{h_6}(\pi) \leq 6$ then $h_6$ has j-envy for $\alpha_{j_1}$ since $u_{h_6}(\pi) \leq 6 < 7 = u_{h_6}(\{ h_3, h_4 \})$, $v_{h_3}(\alpha_{j_1}) = 0 < 1 = v_{h_3}(h_6)$, and $v_{h_4}(\alpha_{j_1}) = 0 < 6 = v_{h_4}(h_6)$. It follows that $u_{h_6}(\pi) \geq 7$.  Since $v_{h_6}(\alpha_{j_2})=0$ it follows that $v_{h_6}(h_{i_3}) = u_{h_6}(\pi) \geq 7$, which is a contradiction.
    \item Suppose $v_{h_{i_1}}(h_{i_2}) = 4$. Assume without loss of generality that $\{ i_1, i_2 \} = \{ 2, 3 \}$. If $u_{h_4}(\pi) \leq 10$ then $h_4$ has j-envy for $\alpha_{j_1}$ since $u_{h_4}(\pi) \leq 10 < 11 = u_{h_4}(\{ h_2, h_3 \})$, $v_{h_2}(\alpha_{j_1}) = 0 < 6 = v_{h_2}(h_4)$, and $v_{h_3}(\alpha_{j_1}) = 0 < 5 = v_{h_3}(h_4)$. It follows that $u_{h_4}(\pi) \geq 11$ which, since $h_2 \notin \pi(h_4)$, is impossible. This contradicts our supposition that $v_{h_{i_1}}(h_{i_2}) = 4$.
    \item Suppose $v_{h_{i_1}}(h_{i_2}) = 3$. We assume without loss of generality that either $\{ i_1, i_2 \} = \{ 3, 11 \}$ or $\{ i_1, i_2 \} = \{ 2, 7 \}$.
\begin{itemize}
    \item Suppose $\{ i_1, i_2 \} = \{ 3, 11 \}$. If $u_{h_2}(\pi) \leq 8$ then $h_2$ has j-envy for $\alpha_{j_1}$ since $u_{h_2}(\pi) \leq 8 < 9 = u_{h_2}(\{ h_3, h_{11} \})$, $v_{h_3}(\alpha_{j_1}) = 0 < 4 = v_{h_3}(h_2)$, and $v_{h_{11}}(\alpha_{j_1}) = 0 < 5 = v_{h_{11}}(h_2)$. It follows that $u_{h_2}(\pi) \geq 9$. By the design of $H$, it follows that $\pi(h_2) = t_5$. Now consider $h_1$. Since $h_1 \notin t_5$ it must be that either $h_1 \in t_2$, $h_1 \in t_3$, or $h_1 \in t_4$. Assume without loss of generality that $h_1 \in t_2$ and $t_2 = \{ h_1, h_{i_3}, \alpha_{j_2} \}$ where $1\leq i_3 \leq 11$ and $\alpha_{j_2} \in N \setminus H$. Since $v_{h_1}(\alpha_{j_2})=0$ it follows that $u_{h_1}(\pi) = v_{h_1}(h_{i_3})$. By the design of $H$, it must be that $v_{h_1}(h_{i_3}) = 2$ so $u_{h_1}(\pi) = 2$. Now $h_1$ has j-envy for $\alpha_{j_1}$, since $u_{h_1}(\pi) = 2 < 4 = u_{h_1}(\{ h_3, h_{11} \})$, $v_{h_3}(\alpha_{j_1}) = 0 < 2 = v_{h_3}(h_1)$, and $v_{h_{11}}(\alpha_{j_1}) = 0 < 2 = v_{h_{11}}(h_1)$. This is a contradiction. 
    \item Suppose $\{ i_1, i_2 \} = \{ 2, 7 \}$. Consider $h_1$. As before, if $u_{h_1}(\pi) \leq 3$ then $h_1$ has j-envy for $\alpha_{j_1}$, since $u_{h_1}(\pi) \leq 3 < 4 = u_{h_1}(\{ h_2, h_7 \})$, $v_{h_2}(\alpha_{j_1}) = 0 < 2 = v_{h_2}(h_1)$, and $v_{h_7}(\alpha_{j_1}) = 0 < 2 = v_{h_7}(h_1)$. It follows that $u_{h_1}(\pi) \geq 4$. By the design of $H$, it must be that $\pi(h_1) = \{ h_1, h_{i_3}, h_{i_4} \}$ where $2\leq i_3, i_4 \leq 11$. It follows that $\pi(h_1) = t_5$. Now consider $h_4$. If $u_{h_4}(\pi) \leq 6$ then $h_4$ has j-envy for $\alpha_{j_1}$, since $u_{h_4}(\pi) \leq 6 < 7 = u_{h_1}(\{ h_2, h_7 \})$, $v_{h_2}(\alpha_{j_1}) = 0 < 6 = v_{h_2}(h_4)$, and $v_{h_7}(\alpha_{j_1}) = 0 < 1 = v_{h_7}(h_4)$. It follows that $u_{h_4}(\pi) \geq 7$ so, similarly, $\pi(h_4)$ must contain three agents in $H$ and thus $h_4 \in t_5$. Assume without loss of generality that $h_4 = h_{i_3}$ so $t_5 = \{ h_1, h_4, h_{i_4} \}$. Since $u_{h_4}(\pi) \geq 7$ and $v_{h_4}(h_1)=2$ it must be that $v_{h_4}(h_{i_4}) \geq 5$ and thus that either $i_4 = 3$ or $i_4 = 6$. Consider $h_{10}$. If $u_{h_{10}}(\pi) \leq 6$ then $h_{10}$ has j-envy for $\alpha_{j_1}$, since $u_{h_{10}}(\pi) \leq 6 < 7 = u_{h_{10}}(\{ h_2, h_7 \})$, $v_{h_2}(\alpha_{j_1}) = 0 < 6 = v_{h_2}(h_{10})$, and $v_{h_7}(\alpha_{j_1}) = 0 < 1 = v_{h_7}(h_{10})$. It follows that $u_{h_{10}}(\pi) \geq 7$. Since $h_{10} \notin t_5$ it must be that either $h_{10} \in t_2$, $h_{10} \in t_3$, or $h_{10} \in t_4$. Assume without loss of generality that $h_{10} \in t_2$ and that $t_2 = \{ h_{10}, h_{i_5}, \alpha_{j_2} \}$ where $1\leq i_5 \leq 11$ and $\alpha_{j_2} \in N \setminus H$. Since $v_{h_{10}}(\alpha_{j_2})=0$ it follows that $v_{h_{10}}(h_{i_5}) = u_{h_{10}}(\pi) \geq 7$, which is a contradiction.
\end{itemize}
\item Suppose $v_{h_{i_1}}(h_{i_2}) = 2$. It follows that either $i_1 = 1$ or $i_2 = 1$. Assume without loss of generality that $i_1 = 1$. Note that $2 \leq i_2 \leq 11$. Consider $h_{i_2}$. Note that since $v_{h_{i_2}}(\alpha_{j_1})=0$ it must be that $u_{h_{i_2}}(\pi) = v_{h_{i_2}}(h_1) = 2$. By the design of $H$, for each possible assignment of $i_2$, namely $2 \leq i_2 \leq 11$, there exist five agents $h_{i_3}, h_{i_4}, h_{i_5}, h_{i_6}, h_{i_7}$ such that $v_{h_{i_2}}(h_{i_k}) > 2$ for $3 \leq k \leq 7$. A counting argument shows that at least one of these five agents does not belong to $t_5$ and hence must belong to either $t_2$, $t_3$, or $t_4$. Assume without loss of generality that $h_{i_3} \in t_2$ and $t_2 = \{ h_{i_3}, h_{i_8}, \alpha_{j_2} \}$ where $2 \leq i_8 \leq 11$ and $\alpha_{j_2} \in N \setminus H$. Recall that $v_{h_{i_2}}(h_{i_3}) > 2$. By the design of $H$ it follows that $u_{h_{i_2}}(\{ h_{i_3}, h_{i_8} \}) > 3$. Now $h_{i_2}$ has j-envy for $\alpha_{j_2}$ since $u_{h_{i_2}}(\pi) = 2 < 3 < u_{h_{i_2}}(\{ h_{i_3}, h_{i_8} \})$, $v_{h_{i_3}}(\alpha_{j_2}) = 0 < 1 \leq v_{h_{i_3}}(h_{i_2})$, and $v_{h_{i_8}}(\alpha_{j_2}) = 0 < 1 \leq v_{h_{i_8}}(h_{i_2})$.
\item Suppose $v_{h_{i_1}}(h_{i_2}) = 1$. Without loss of generality we assume that either $\{ i_1, i_2 \} = \{ 2, 5 \}$ or $\{ i_1, i_2 \} = \{ 2, 6 \}$.
\begin{itemize}
    \item Suppose $\{ i_1, i_2 \} = \{ 2, 5 \}$. If $u_{h_{4}}(\pi) \leq 9$ then $h_{4}$ has j-envy for $\alpha_{j_1}$, since $u_{h_{4}}(\pi) \leq 9 < 10 = u_{h_{4}}(\{ h_2, h_5 \})$, $v_{h_2}(\alpha_{j_1}) = 0 < 6 = v_{h_2}(h_{4})$, and $v_{h_5}(\alpha_{j_1}) = 0 < 4 = v_{h_5}(h_{4})$. It follows that $u_{h_{4}}(\pi) \geq 10$. The only possibility is that $\pi(h_4) = \{ h_3, h_4, h_{6} \}$ and hence $\pi(h_4) = t_5$. Consider $h_1$. If $u_{h_1}(\pi) \leq 3$ then $h_1$ has j-envy for $\alpha_{j_1}$, since $u_{h_1}(\pi) \leq 3 < 4 = u_{h_1}(\{ h_2, h_5 \})$, $v_{h_2}(\alpha_{j_1}) = 0 < 2 = v_{h_2}(h_1)$, and $v_{h_5}(\alpha_{j_1}) = 0 < 2 = v_{h_5}(h_1)$. It follows that $u_{h_1}(\pi) \geq 4$. By the design of $H$, it follows that $\pi(h_1)$ must contain three agents in $H$, which is a contradiction since $h_1 \notin t_5$.
    \item Suppose $\{ i_1, i_2 \} = \{ 2, 6 \}$. It follows that $u_{h_4}(\pi) \leq 9$ so $h_4$ has j-envy for $\alpha_{j_1}$ since $u_{h_4}(\pi) \leq 9 < 12 = u_{h_4}(\{ h_2, h_6 \})$, $v_{h_2}(\alpha_{j_1}) = 0 < 6 = v_{h_2}(h_4)$, and $v_{h_6}(\alpha_{j_1}) = 0 < 6 = v_{h_6}(h_4)$.\qedhere
\end{itemize}
\end{itemize}
\end{proof}

\begin{lem}
\label{lem:threed_efr_as_jef_hopen}
If $(N, V)$ contains a j-envy-free partition into triples $\pi$ then $H$ has an open configuration in $\pi$.
\end{lem}
\begin{proof}
By definition, $4 \leq \sigma(H, \pi) \leq 11$. If $\sigma(H, \pi) \leq 5$ then $H$ has an open configuration in $\pi$, by Lemmas~\ref{lem:threed_efr_as_jef_3332_case} and~\ref{lem:threed_efr_as_jef_32222_case}. If $6 \leq \sigma(H, \pi) \leq 11$ then, by a counting argument, at least one triple in $\pi$ must contain exactly one agent in $H$. In other words, $H$ has an open configuration in $\pi$.
\end{proof}

We have shown, in Lemma~\ref{lem:threed_efr_as_jef_hopen}, that if $(N, V)$ contains a j-envy-free partition into triples $\pi$ then $H$ has an open configuration in $\pi$. By definition, some triple $t_{\beta}$ in $\pi$ contains exactly one agent in $H$. Since $|H|=11$, if $t_{\beta}$ is the only triple in $\pi$ to contain exactly one agent in $H$ then there must exist some triple in $\pi$ that contains exactly two agents in $H$. By Lemma~\ref{lem:threed_efr_as_jef_two_and_one_in_h}, this is a contradiction. It follows that at least two triples in $\pi$ exist that each contains exactly one agent in $H$. Suppose $t_{\beta}$ and $t_{\gamma}$ are two such triples where $t_{\beta} = \{ h_{a_1}, \alpha_{b_1}, \alpha_{b_2} \}$ and $t_{\gamma} = \{ h_{a_2}, \alpha_{b_3}, \alpha_{b_4} \}$.

\begin{lem}
\label{lem:threed_efr_as_jef_lequalsthefourisolated}
If $(N, V)$ contains a j-envy-free partition into triples then $\{ \alpha_{b_1},\allowbreak\alpha_{b_2}, \alpha_{b_3},\allowbreak\alpha_{b_4} \} = L$.
\end{lem}
\begin{proof}
Suppose for a contradiction that $\{ \alpha_{b_1}, \alpha_{b_2}, \alpha_{b_3}, \alpha_{b_4} \} \neq L$.

By definition, $\{ \alpha_{b_1}, \alpha_{b_2}, \alpha_{b_3}, \alpha_{b_4} \} \cap H = \varnothing$ and $\{ \alpha_{b_1}, \alpha_{b_2}, \alpha_{b_3}, \alpha_{b_4} \} \neq L$ it must be that at least one agent in $\{ \alpha_{b_1}, \alpha_{b_2}, \alpha_{b_3}, \alpha_{b_4} \}$ belongs to $C$. Assume without loss of generality that $\alpha_{b_1} \in C$.

We have already shown that $t_{\beta}$ contains exactly one agent in $H$. Since $v_{\alpha_{b_1}}(h_{a_1}) = 0$, by the design of the instance it must be that $u_{\alpha_{b_1}}(\pi) = v_{\alpha_{b_1}}(\alpha_{b_2}) \leq 3$. By the design of the instance $v_{\alpha_{b_1}}(\alpha_{b_3}) \geq 2$ and $v_{\alpha_{b_1}}(\alpha_{b_4}) \geq 2$ so $u_{\alpha_{b_1}}(\{ \alpha_{b_3}, \alpha_{b_4} \}) \geq 4$. Now $\alpha_{b_1}$ has j-envy for $h_{a_2}$ since $u_{\alpha_{b_1}}(\pi) \leq 3 < 4 \leq u_{\alpha_{b_1}}(\{ \alpha_{b_3}, \alpha_{b_4} \})$, $v_{\alpha_{b_3}}(h_{a_2}) = 0 < 2 \leq v_{\alpha_{b_3}}(\alpha_{b_1})$, and $v_{\alpha_{b_4}}(h_{a_2}) = 0 < 2 \leq v_{\alpha_{b_4}}(\alpha_{b_1})$.
\end{proof}

\begin{lem}
\label{lem:threed_efr_as_jef_structureofL}
If $(N, V)$ contains a j-envy-free partition into triples then $\{ \{ \alpha_{b_1},\allowbreak\alpha_{b_2} \}, \{ \alpha_{b_3},\allowbreak\alpha_{b_4} \} \} = \{ \{ l_1,\allowbreak l_2 \}, \{ l_3,\allowbreak l_4 \} \}$.
\end{lem}
\begin{proof}
By Lemma~\ref{lem:threed_efr_as_jef_lequalsthefourisolated}, $\{ \alpha_{b_1}, \alpha_{b_2}, \alpha_{b_3}, \alpha_{b_4} \} = L$. There are now three possibilities: first that $\{ \{ \alpha_{b_1}, \alpha_{b_2} \}, \{ \alpha_{b_3}, \alpha_{b_4} \} \} = \{ \{ l_1, l_3 \}, \{ l_2, l_4 \} \}$, second that $\{ \{ \alpha_{b_1}, \alpha_{b_2} \}, \{ \alpha_{b_3}, \alpha_{b_4} \} \} = \{ \{ l_1,l_4 \}, \{ l_2, l_3 \} \}$, and third that $\{ \{ \alpha_{b_1},\allowbreak\alpha_{b_2} \}, \{ \alpha_{b_3},\allowbreak\alpha_{b_4} \} \} = \{ \{ l_1,\allowbreak l_2 \}, \{ l_3,\allowbreak l_4 \} \}$.

First suppose $\{ \{ \alpha_{b_1}, \alpha_{b_2} \}, \{ \alpha_{b_3}, \alpha_{b_4} \} \} = \{ \{ l_1, l_3 \}, \{ l_2, l_4 \} \}$. Now $l_1$ has j-envy for $h_{a_2}$ since $u_{l_1}(\{ h_{a_1}, l_3 \}) = 1 < 3 \leq u_{l_1}(\{ l_2, l_4 \})$, $v_{l_2}(h_{a_2}) = 0 < 2 = v_{l_2}(l_1)$, and $v_{l_4}(h_{a_2}) = 0 < 1 \leq v_{l_4}(l_1)$.

Second suppose $\{ \{ \alpha_{b_1}, \alpha_{b_2} \}, \{ \alpha_{b_3}, \alpha_{b_4} \} \} = \{ \{ l_1, l_4 \}, \{ l_2, l_3 \} \}$. As before, $l_1$ has j-envy for $h_{a_2}$ since $u_{l_1}(\{ h_{a_1}, l_4 \}) = 1 < 3 \leq u_{l_1}(\{ l_2, l_3 \})$, $v_{l_2}(h_{a_2}) = 0 < 2 = v_{l_2}(l_1)$, and $v_{l_3}(h_{a_2}) = 0 < 1 \leq v_{l_3}(l_1)$.

It remains that $\{ \{ \alpha_{b_1}, \alpha_{b_2} \}, \{ \alpha_{b_3}, \alpha_{b_4} \} \} = \{ \{ l_1, l_2 \}, \{ l_3, l_4 \} \}$.
\end{proof}

By Lemma~\ref{lem:threed_efr_as_jef_structureofL}, either $\{ \alpha_{b_1}, \alpha_{b_2} \} = \{ l_1, l_2 \}$ or $\{ \alpha_{b_1}, \alpha_{b_2} \} = \{ l_3, l_4 \}$. Without loss of generality assume that $\{ \alpha_{b_1}, \alpha_{b_2} \} = \{ l_1, l_2 \}$.

\begin{lem}
\label{lem:threed_efr_as_jef_eachpigets6}
If $(N, V)$ contains a j-envy-free partition into triples then $u_{c_i}(\pi) = 6$ for each agent $c_i$ in $C$.
\end{lem}
\begin{proof}
Suppose to the contrary that some $1\leq i \leq 3q$ exists where $u_{c_i}(\pi) < 6$. Then $c_i$ has j-envy for $h_{a_1}$ since $u_{c_i}(\pi) \leq 5 < 6 \leq u_{c_i}(\{ l_1, l_2 \})$, $v_{l_1}(h_{a_1}) = 0 < 3 = v_{l_1}(c_i)$, and $v_{l_2}(h_{a_1}) = 0 < 3 = v_{l_2}(c_i)$.
\end{proof}

It is now straightforward to prove Lemma~\ref{lem:threed_efr_as_jef_second_direction}.

\lemthreedefrasjefseconddirection*
\begin{proof}
Suppose $(N, V)$ contains a j-envy-free partition into triangles $\pi$. Lemma~\ref{lem:threed_efr_as_jef_eachpigets6} shows that $u_{c_i}(\pi) = 6$ for each agent $c_i$ in $C$. By construction, it follows that $\pi(c_i)$ contains two agents $c_j, c_k$ such that $v_{c_i}(c_j) = v_{c_i}(c_k) = 3$. By construction, $c_j$ and $c_k$ therefore correspond to vertices $w_j$ and $w_k$ in $W$ where $\{ w_i, w_j \} \in E$ and $\{ w_i, w_k \} \in E$. It follows thus that there are exactly $q$ triples in $\pi$ each containing three agents $\{ c_i, c_j, c_k \}$, where the three corresponding vertices $w_i, w_j, w_k$ are pairwise adjacent in $G$. From these triples a partition into triangles $X$ can be easily constructed.
\end{proof}

\end{appendices}

\bibliography{mybib}


\end{document}